\documentclass[acmsmall,screen]{acmart}

\usepackage{pgfplots}
\usepackage{prettyref}
\usepackage{tikz}
\usepackage{xcolor,colortbl}
\usepackage{multirow}
\usepackage[ruled, linesnumbered, vlined, commentsnumbered]{algorithm2e}
\usepackage[section]{placeins}
\usepackage{booktabs}
\usepackage{threeparttable}
\usepackage{makecell}
\usepackage{graphicx}
\usepackage[listings,skins,breakable]{tcolorbox}
\usepackage{csquotes}
\usepackage{subcaption}
\usepackage{standalone}
\usepackage[skip=0.5\baselineskip]{caption}
\usepackage{wrapfig}
\usepackage{amsmath}
\usetikzlibrary{pgfplots.statistics}
\usepackage{tcolorbox}
\usepackage{pifont}
\usepackage{balance}
\usepackage{xfp}
\usepackage{enumitem}
\usepackage{hyperref}

\usepgfplotslibrary{fillbetween}

\usetikzlibrary{arrows.meta,arrows}

\newtcolorbox{quotebox}{colback=steel!10,boxrule=0.4pt,colframe=black,fonttitle=\bfseries,top=2pt,bottom=2pt}

\newcommand{\xmark}{\ding{53}}%

\SetKwInput{kwDeclare}{Declare}

\newcolumntype{x}[1]{>{\centering\arraybackslash\hspace{0pt}}p{#1}}

\mathchardef\mhyphen="2D
\newcommand{\vect}[1]{\boldsymbol{#1}}
\DeclareMathAlphabet\mathbfcal{OMS}{cmsy}{b}{n}

\newbox\aMark
\setbox\aMark\hbox{\begin{pgfpicture}\textcolor{red}{\pgfuseplotmark{o}}\end{pgfpicture}}

\newbox\bMark
\setbox\bMark\hbox{\begin{pgfpicture}\textcolor{red}{\pgfuseplotmark{star}}\end{pgfpicture}}

\newrefformat{fig}{Fig.~\ref{#1}}
\newrefformat{tab}{Table~\ref{#1}}
\newrefformat{sec}{Section~\ref{#1}}
\newrefformat{app}{Appendix~\ref{#1}}
\newrefformat{alg}{Algorithm~\ref{#1}}
\newrefformat{property}{Property~\ref{#1}}
\newrefformat{theorem}{Theorem~\ref{#1}}
\newrefformat{lemma}{Lemma~\ref{#1}}
\newrefformat{corollary}{Corollary~\ref{#1}}
\newrefformat{proposition}{Proposition~\ref{#1}}
\newrefformat{def}{Definition~\ref{#1}}
\newrefformat{eq}{equation~(\ref{#1})}

\definecolor{steel}{rgb}{0.16, 0.67, 0.8} 

\pgfplotsset{compat=newest}

\pgfplotsset{compat=1.11,
    /pgfplots/ybar legend/.style={
    /pgfplots/legend image code/.code={%
       \draw[##1,/tikz/.cd,yshift=-0.25em]
        (0cm,0cm) rectangle (3pt,0.8em);},
   },
}

\DeclareMathOperator*{\argmin}{argmin}

\def\signed #1{{\leavevmode\unskip\nobreak\hfil\penalty50\hskip2em
  \hbox{}\nobreak\hfil(#1)%
  \parfillskip=0pt \finalhyphendemerits=0 \endgraf}}

\newsavebox\mybox

   \newcommand{\squart}[4]{\begin{adjustbox}{max width=.1\textwidth}\begin{picture}(100,5)%1
    {\color{black}\put(0,5){\line(1,0){100}}\color{black}\put(0,5){\line(0,1){10}}\put(50,5){\line(0,1){10}}\put(100,5){\line(0,1){10}}\put(25,5){\line(0,1){5}}\put(75,5){\line(0,1){5}}\put(-2,-8){\LARGE$0$}\put(42,-8){\LARGE$0.5$}\put(96,-8){\LARGE$1$}}\end{picture}\end{adjustbox}}

\def\BibTeX{{\rm B\kern-.05em{\sc i\kern-.025em b}\kern-.08em
    T\kern-.1667em\lower.7ex\hbox{E}\kern-.125emX}}
    
\AtBeginDocument{%
  \providecommand\BibTeX{{%
    Bib\TeX}}}

\setcopyright{rightsretained}
\acmDOI{10.1145/3643751}
\acmYear{2024}
\copyrightyear{2024}
\acmSubmissionID{fse24main-p388-p}
\acmJournal{PACMSE}
\acmVolume{1}
\acmNumber{FSE}
\acmArticle{25}
\acmMonth{7}
\received{2023-09-27}
\received[accepted]{2024-01-23}

\begin{document}

%\title{Adapting Multi-objectivization for Software Configuration Tuning}
\title{Adapting Multi-objectivized Software Configuration Tuning}

\author{Tao Chen}
%\authornote{This work was conducted when visiting UESTC.}
\authornote{Tao Chen is the corresponding author (t.chen@bham.ac.uk).}

%\affiliation{%
%  \institution{UESTC}
  %\state{Birmingham}
%  \country{China}
%}
\affiliation{%
  \institution{University of Birmingham}
  \state{Birmingham}
  \country{United Kingdom}
}
  \email{t.chen@bham.ac.uk}

  \author{Miqing Li}
  \authornote{Both authors make commensurable contributions to this research.}
\affiliation{%
  \institution{University of Birmingham}
  \state{Birmingham}
  \country{United Kingdom}}
  \email{m.li.8@bham.ac.uk}

\begin{CCSXML}
<ccs2012>
<concept>
<concept_id>10011007.10011074.10011784</concept_id>
<concept_desc>Software and its engineering~Search-based software engineering</concept_desc>
<concept_significance>500</concept_significance>
</concept>
<concept>
<concept_id>10011007.10010940.10011003.10011002</concept_id>
<concept_desc>Software and its engineering~Software performance</concept_desc>
<concept_significance>500</concept_significance>
</concept>
   <concept>
       <concept_id>10011007.10011006.10011071</concept_id>
       <concept_desc>Software and its engineering~Software configuration management and version control systems</concept_desc>
       <concept_significance>300</concept_significance>
       </concept>
 </ccs2012>
\end{CCSXML}

\ccsdesc[500]{Software and its engineering~Search-based software engineering}
\ccsdesc[500]{Software and its engineering~Software performance}
\ccsdesc[300]{Software and its engineering~Software configuration management and version control systems}

\keywords{Configuration tuning, performance optimization, search-based software engineering, multi-objectivization}


\begin{abstract}
When tuning software configuration for better performance (e.g., latency or throughput), 
an important issue that many optimizers face is the presence of local optimum traps, compounded by a highly rugged configuration landscape and expensive measurements. To mitigate these issues, a recent effort has shifted to focus on the level of optimization model 
(called meta multi-objectivization or \textsc{MMO}) instead of designing better optimizers as in traditional methods. This is done by using an auxiliary performance objective, together with the target performance objective, 
to help the search jump out of local optima. 
While effective, 
\textsc{MMO} needs a fixed weight to balance the two objectives---a parameter that has been found to be crucial as there is a large deviation of the performance between the best and the other settings.
However, 
given the variety of configurable software systems, 
the ``sweet spot'' of the weight can vary dramatically in different cases and it is not possible to find the right setting without time-consuming trial and error. 
In this paper, 
we seek to overcome this significant shortcoming of \textsc{MMO} by proposing a weight adaptation method, 
dubbed \textsc{AdMMO}. Our key idea is to adaptively adjust the weight at the right time during tuning, such that a good proportion of the nondominated configurations can be maintained. Moreover, we design a partial duplicate retention mechanism to handle the issue of too many duplicate configurations without losing the rich information provided by the ``good'' duplicates. 

Experiments on several real-world systems, objectives, and budgets show that, 
for 71\% of the cases, 
\textsc{AdMMO} is considerably superior to \textsc{MMO} and a wide range of state-of-the-art optimizers while achieving generally better efficiency with the best speedup between $2.2\times$ and $20 \times$. 
\end{abstract}

\maketitle

\section{Introduction}
\label{sec:intro}
 
% [] report that, for \textsc{Storm}, a configuration can be up to $480\times$ worse than the optimal one on latency. 

\begin{figure}[t!]
	\centering
	\begin{subfigure}{.38\columnwidth}
		\includegraphics[width=\columnwidth]{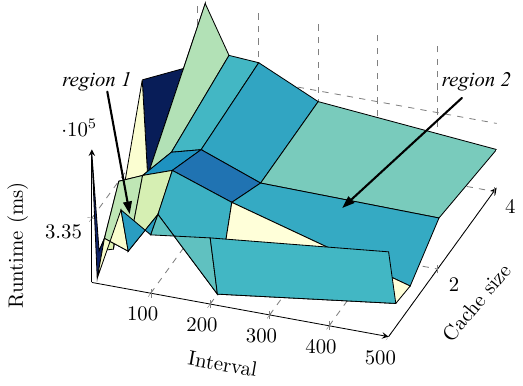}
		\subcaption{\textsc{MongoDB}}
	\end{subfigure}
	~\hspace{2cm}
	\begin{subfigure}{.38\columnwidth}
		%\vspace{0.2cm}
		\includegraphics[width=\columnwidth]{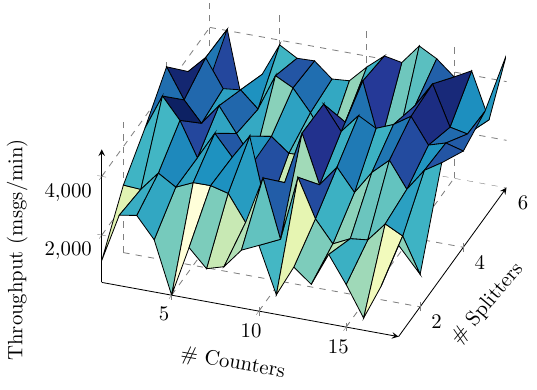}
		\subcaption{\textsc{Storm}}
	\end{subfigure}
	\caption{The projected configuration landscapes of two configurable software systems.}
	\label{fig:example}
 \vspace{-0.4cm}
\end{figure}

Have you ever struggled in configuring your complex software system? There is no need to worry as we have automatic software configuration tuning. Indeed, it has been shown that 59\% of the most severe performance bugs are caused by poor configuration~\cite{DBLP:conf/esem/HanY16}, making it one of the most dangerous threats to software quality. Over the past decade, software configuration tuning has been an important way to optimize the most important performance concern, such as runtime, throughput, or accuracy~\cite{DBLP:conf/icse/LiX0WT20,DBLP:conf/sigsoft/0001Chen21,DBLP:conf/sc/BehzadLHBPAKS13,Chen2018FEMOSAA,DBLP:conf/sigsoft/ShahbazianKBM20,DBLP:journals/peerj-cs/Silva-MunozFB21,DBLP:conf/kbse/LiXCT20,DBLP:conf/ae/CaceresPFS17,nair2018finding,DBLP:conf/icse/0003XC021,DBLP:journals/tsc/ChenB17,DBLP:conf/nips/BergstraBBK11,DBLP:journals/jmlr/ZuluagaKP16}. However, an unpleasant issue which an optimizer faces is that the tuning may easily be trapped at local optima~\cite{DBLP:conf/sigsoft/0001Chen21,DBLP:journals/corr/abs-2112-07303,DBLP:conf/icpads/DingLQ15,DBLP:conf/seams/Chen22,DBLP:conf/hpdc/LiZMTZBF14,DBLP:conf/cloud/ZhuLGBMLSY17}---some configurations that perform better than all the neighbors but being undesirably sub-optimal. While this issue may well be inherited and common in Search-Based Software Engineering (SBSE), there are some unique characteristics in software configuration tuning that make it particularly hard to tackle: (1) the measurement of configurations is expensive. Valov \textit{et al.}~\cite{DBLP:conf/wosp/ValovPGFC17} report that it takes 1,536 hours to explore merely 11 options of configuration for \textsc{x264}. (2) The configuration landscape is rather sparse and rugged, i.e., the close configurations can also have radically different performance~\cite{DBLP:conf/sigsoft/0001Chen21,DBLP:journals/corr/abs-2112-07303,nair2018finding,DBLP:conf/sigsoft/GongChen2023,DBLP:journals/tosem/ChenL23} (see Figure~\ref{fig:example}). This is because, for example, switching the data structures is merely a single digit change of an option, but that may affect the performance drastically~\cite{nair2018finding,DBLP:journals/tosem/ChenL23a}.

From the literature, numerous optimizers have been proposed to tune software configurations. Among others, most of those are mainly guided by direct measurement of the systems, such as Genetic Algorithm~\cite{DBLP:conf/sc/BehzadLHBPAKS13,Chen2018FEMOSAA,DBLP:conf/sigsoft/ShahbazianKBM20}, IRACE~\cite{lopez2016irace,DBLP:journals/peerj-cs/Silva-MunozFB21,DBLP:conf/ae/CaceresPFS17}, and Random Search~\cite{DBLP:journals/jmlr/BergstraB12}. Local optima are handled by search operators, e.g., increasing the
neighborhood size~\cite{DBLP:conf/sc/BehzadLHBPAKS13}, or by search strategies that balance exploration and exploitation, such as allowing sampling on less promising regions~\cite{lopez2016irace,DBLP:journals/peerj-cs/Silva-MunozFB21,DBLP:conf/ae/CaceresPFS17}. In contrast, surrogate-guided optimizers based on, e.g., Bayesian optimization, also exist~\cite{nair2018finding,DBLP:conf/icse/0003XC021,DBLP:conf/nips/BergstraBBK11}. 
Yet, 
a major limitation of those optimizers is that they may still easily get stuck at sparse regions of local optima (e.g., if the technique for handling local optima is not particularly effective) or cannot find better configurations efficiently 
(e.g., if the exploration is overemphasized)~\cite{DBLP:conf/sigsoft/0001Chen21,DBLP:journals/corr/abs-2112-07303,DBLP:conf/wcre/Chen22}. 

%Yet, a major limitation of those optimizers is that the best configuration (i.e., global optimum) may not be able to be easily found (e.g., if the technique for getting rid of local optima is not particularly effective), or to be found in a resource-efficient manner (e.g., if the exploration is overemphasized)~\cite{DBLP:conf/sigsoft/0001Chen21,DBLP:journals/corr/abs-2112-07303}. 

%Yet, a major limitation of those optimizers is that the local optima is either not considered or there is no ``incentive'' to traverse
%the wide search space and locating many local optima as possible while aiming to find the global optimum,
%thus finding the best one in a resource-efficient manner~\cite{DBLP:conf/sigsoft/0001Chen21,DBLP:journals/corr/abs-2112-07303}.

Recently, \citet{DBLP:conf/sigsoft/0001Chen21,DBLP:journals/corr/abs-2112-07303} 
proposed a meta multi-objectivization (MMO) model to tackle the above limitation. 
Unlike others who work on the design of optimizers, 
MMO focuses on the optimization model. 
This is achieved by transforming the tuning into a meta-objective space 
that involves linear combinations (by a weight) of two performance objectives: 
one is the target that is of concern 
and the other is an auxiliary that is of no interest to the software engineers for the considered system. 
Through constructing two conflicting meta-objectives,
MMO keeps a good tendency towards the best target performance objective while preserving high-quality dissimilar configurations, which helps to mitigate local optima. 

%Through constructing two conflicting meta-objectives using the target and auxiliary objectives

%Despite that the authors of~\cite{DBLP:conf/sigsoft/0001Chen21,DBLP:journals/corr/abs-2112-07303} have demonstrated the effectiveness of MMO, 

Despite the promising effectiveness of MMO, \citet{DBLP:conf/sigsoft/0001Chen21,DBLP:journals/corr/abs-2112-07303} also revealed that its performance can be highly sensitive to the weight---a key parameter that impacts the balances between finding 
better target-objective configuration and diversifying configurations during the tuning. 
A small weight puts an emphasis on optimizing the target performance objective (exploitation)
while a large weight encourages exploration in the search space. It has been shown that an inappropriate weight setting can lead to rather poor results~\cite{DBLP:conf/sigsoft/0001Chen21,DBLP:journals/corr/abs-2112-07303}. However,
finding a good weight on a given configurable software system is not easy. 
The ``sweet spot'' of the weight value, which achieves the ideal balance between exploitation and exploration of the tuning, varies dramatically on different systems or even the status of the tuning, 
due to the difference in their configuration landscape, scale, and types of options \textit{etc}. For example, Figure~\ref{fig:example} shows the different landscapes of two systems. Although both landscapes are rugged, relatively, a smaller weight that favors exploitation more might be preferred to stress finding better performance in Figure~\ref{fig:example}a since the landscape is smoother; in contrast, in Figure~\ref{fig:example}b, the weight value might need to be larger so that more exploration can be encouraged to overcome local optima due to the higher ruggedness. The same applied within one system, e.g., in Figure~\ref{fig:example}a, when the tuning is trapped at \textit{region 1} then a larger weight is needed to jump out from the area while a smaller weight is better when the tuning is located at \textit{region 2}, as the local landscape is relatively easier to tackle. As such, it is not possible to infer the right weight setting without time-consuming profiling beforehand. 

%Further, the original MMO work did not provide a sound theory as to why and how the weight can impact the tuning. 

To overcome the above limitation of \textsc{MMO} for software configuration tuning, 
in this work, we propose a weight adaptation method, 
dubbed Adaptive \textsc{MMO} or \textsc{AdMMO}. 
\textsc{AdMMO} does not require any prior knowledge or effort on setting the weight, 
and the weight is adjusted adaptively as the tuning proceeds. Specifically, our contributions are:

%It may not be possible to know a right setting of the weight without in-depth prior knowledge of the system or time-consuming profiling beforehand.

%Finding a good weight for a system is not easy as tuning is always expensive and a badly chosen one may lead to devastating outcome. Although in a subsequent work~\cite{DBLP:journals/corr/abs-2112-07303}, they propose a fix by patching a locally bounded normalization and constantly set $w=1$, but the fundamental problem remains unsolved because we entirely lose control over the contribution between target and auxiliary, leaving a risk of biased contribution towards one of them during the tuning (\textcolor{red}{we may need more convinced reasons}).

%This is measured by the proportion of unique nondominated configurations.
%What makes \textsc{AdMMO} unique is that it does not require any prior knowledge on setting the weight for MMO, but instead dynamically adapting it to reach the contribution between target and auxiliary performing objectives that leads to an unbiased proportion of nondominated configurations as the tuning proceeds. 

\begin{itemize}
	\item Through geometric transformation and empirical evidence, we explain how \textsc{MMO}'s weight~\cite{DBLP:conf/sigsoft/0001Chen21,DBLP:journals/corr/abs-2112-07303} can impact the tuning and why a fixed value is harmful.
	
	\item We propose a way to adapt the weight dynamically in \textsc{MMO} on the fly, aiming to maintain an appropriate proportion of the unique nondominated configurations\footnote{A configuration $\mathbf{\overline{a}}$ is dominated by $\mathbf{\overline{b}}$ if all objectives of $\mathbf{\overline{b}}$ are better or equivalent to those of $\mathbf{\overline{a}}$ while there is at least one objective of $\mathbf{\overline{b}}$ performs better than that of $\mathbf{\overline{a}}$. A configuration is nondominated if it cannot be dominated by others in the set.} throughout the tuning, which serves as a better indicator of the ideal balance between exploitation and exploration.
	%keep an ideal balance between exploitation and exploration preferred by the system hence 
	\item To avoid overfitting the weight, we design a progressive trigger that only enables weight adaptation when needed.
	
	\item To mitigate the detrimental effects caused by having too many duplicate configurations\footnote{By duplicate configurations, we refer to those with identical values in the configuration space rather than those of the same performance values, since it is still possible for different configurations to have identical performance.}, 
	we propose a novel partial duplicate retention mechanism that de-emphasizes the duplicate configurations but still allows some good ones to be preserved in the tuning.
	
\end{itemize}

% MMO explain
% MMO limitation
% contribution

Through extensive experiments on 14 system-objective pairs of commonly used real-world software systems and performance objectives~\cite{DBLP:journals/corr/abs-2112-07303,nair2018finding,DBLP:conf/sigsoft/0001Chen21,DBLP:conf/mascots/MendesCRG20,DBLP:conf/sigsoft/JamshidiVKS18,DBLP:conf/icse/SiegmundKKABRS12} along with different budget sizes, 
we compare \textsc{AdMMO} with the improved version of original MMO~\cite{DBLP:journals/corr/abs-2112-07303}, 
the plain multi-objectivization model used in other SBSE problems~\cite{DBLP:journals/tse/YuanB20,derakhshanfar2020good,DBLP:conf/ssbse/MkaouerKBC14,DBLP:conf/ssbse/SoltaniDPDZD18}, seven state-of-the-art optimziers for configuration tuning~\cite{DBLP:conf/ae/CaceresPFS17,DBLP:conf/sigsoft/ShahbazianKBM20,DBLP:journals/jair/HutterHLS09,nair2018finding,DBLP:conf/icse/0003XC021,DBLP:conf/lion/HutterHL11,DBLP:journals/jmlr/BergstraB12} and three of its variants. Empirically, we have also examined whether there exists a generally best proportion of nondominated configurations to maintain in \textsc{AdMMO}. The results are encouraging, 
from which we demonstrate that \textsc{AdMMO}:

\begin{itemize}
	\item is effective as it outperforms state-of-the-art optimizers, including the improved MMO, in 71\% of the cases with the best normalized improvements between 17\% and 100\%;
	\item is efficient, achieving considerable speedup over others with the best from $2.2\times$ to $20 \times$;
	\item can indeed benefit from the proposed progressive trigger and partial duplicate retention; 
	\item exists a generally best proportion of nondominated configurations to maintain according to this study, i.e., around 30\%. 
\end{itemize}

%To promote open science, we anonymously release all sources: \url{https://github.com/3c23/AdMMO}.

The rest of this paper is organized as follows. 
Section~\ref{sec:background} introduces some background information and related work. Section~\ref{sec:problem} explains the role of \textsc{MMO} weight, the limitation, and why it is challenging to address it. Section~\ref{sec:method} elaborates the designs of our \textsc{AdMMO}. 
Sections~\ref{sec:exp} and~\ref{sec:result} present our experiment methodology and a detailed analysis of the results, respectively. 
Section~\ref{sec:discussion} discusses the threats to validity followed by Section~\ref{sec:con} that concludes the paper and sheds light on future work.

% SCT
% SBSE for SCT
% (requirement) two types of Pareto search
% limitation
% finding 
% insight

\section{Background and Related Work}
\label{sec:background}

In this section, we present the background and discuss the existing methods compared in this work.

\subsection{Tuning Software Configuration}

%$i\in \{1,2...,n\}$ and 

A configurable software system often comes with a set of critical configuration options to tune, for example, \textsc{Storm} allows one to change the \texttt{num\_splitters} and \texttt{num\_counters} for better latency or throughput~\cite{nair2018finding,DBLP:conf/sigsoft/0001Chen21}. $x_i$ denotes the $i$th option, which can be either a binary or integer variable, among $n$ options for a software system. The goal we considered in this work is to search for better software configurations, from the space of $\mathbfcal{X}$, that optimize a single performance objective $f$\footnote{Without loss of generality, we assume minimizing scenarios which can be converted to maximizing via additive inverse.}:
\begin{align}
\argmin~f(\vect{x}),~~\vect{x} \in \mathbfcal{X}
\label{Eq:SOP}
\end{align}
where $\vect{x} = (x_1, x_2, ..., x_n)$. The measurement of $f$ depends on the target system and the performance attribute, for which we make no assumption about the characteristics in this work.

\subsection{Measurement-based Optimizer}

Measurement-based optimizers tune the configuration by solely measuring the software~\cite{DBLP:conf/sigsoft/0001Chen21}. In that regard, many optimizers that leverage different search algorithms exist~\cite{DBLP:conf/icpads/DingLQ15,DBLP:conf/hpdc/LiZMTZBF14,DBLP:conf/sc/BehzadLHBPAKS13,Chen2018FEMOSAA,DBLP:conf/sigsoft/ShahbazianKBM20}. Here, we briefly introduce some state-of-the-art optimizers which we compared against in our experiments:

\begin{itemize}
	
	%As a state-of-the-art optimizer for software configuration turning~\cite{lopez2016irace,DBLP:journals/peerj-cs/Silva-MunozFB21,DBLP:conf/ae/CaceresPFS17}, 

	\item \textbf{Iterated Racing (\textsc{IRACE})}: 
	\textsc{IRACE}~\cite{lopez2016irace,DBLP:journals/peerj-cs/Silva-MunozFB21,DBLP:conf/ae/CaceresPFS17} measures new configurations according to the distributions of configuration options, 
	aiming to jump out of local optima. At each iteration, the distributions are updated using the parent configuration selected and the iteration count, thereby focusing on the search around the best configuration found with more iterations. 
	
	%The best configuration is returned upon termination. 

	\item \textbf{Genetic Algorithm (\textsc{GA})}: \textsc{GA}~\cite{DBLP:conf/sc/BehzadLHBPAKS13,Chen2018FEMOSAA,DBLP:conf/sigsoft/ShahbazianKBM20} is a population-based optimizer that evolves configurations through reproduction. The mutation is used to overcome local optima. 
	
	\item \textbf{Random Search (\textsc{RS})}: In \textsc{RS}, a configuration is randomly formed and measured during each iteration. Theoretically, it is insensitive to the local optima issue. RS has been shown to be effective in the general configuration domain~\cite{DBLP:journals/jmlr/BergstraB12} and it serves as a baseline in this work.
	
	\item \textbf{\textsc{ParamILS}}: \textsc{ParamILS}~\cite{DBLP:journals/jair/HutterHLS09} iteratively conducts local search around the best configuration found so far but doing so with a probability to jump out from the likely local optima area.

	%GA is also a widely used and state-of-the-art optimizer for software configuration turning~\cite{DBLP:conf/sc/BehzadLHBPAKS13,Chen2018FEMOSAA,DBLP:conf/sigsoft/ShahbazianKBM20}.

\end{itemize}

Nevertheless, 
the above optimizers may not be effective in finding the best configuration (i.e., global optimum).
Some struggle to get rid of local optima
and some others fail to find the better configurations efficiently as reported~\cite{DBLP:conf/sigsoft/0001Chen21,DBLP:journals/corr/abs-2112-07303}.

\subsection{Model-based Optimizer}

To guide the tuning, model-based optimizers do not solely rely on the measurements, but also on a gradually updated surrogate~\cite{DBLP:conf/msr/GongC22,DBLP:journals/tse/ChenB17,DBLP:conf/sigsoft/0001L24} that can cheaply predict the performance~\cite{DBLP:conf/mascots/JamshidiC16,nair2018finding,DBLP:conf/icse/0003XC021,DBLP:conf/nips/BergstraBBK11,DBLP:journals/jmlr/ZuluagaKP16}. Typically, they follow the procedure of Bayesian Optimization or its variants. In this work, we experimentally examine three model-based optimizers, namely:

\begin{itemize}
	\item \textbf{\textsc{Flash}}: Published at TSE~\cite{nair2018finding}, \textsc{Flash} extends the Bayesian Optimization by using a regression tree as the surrogate and conducting acquisition without uncertainty. 
	
	%It updates the surrogate whenever a configuration is measured and returns the best found.
	
	\item \textbf{\textsc{BOCA}}: Another optimizer based on Bayesian Optimization from ICSE~\cite{DBLP:conf/icse/0003XC021}. \textsc{BOCA} uses Random Forest as the surrogate with Expected Improvement. Further, it prioritizes sampling of the important configuration options based on Gini importance from the surrogate.

	\item \textbf{\textsc{SMAC}}: As a general purpose optimizer published at LION~\cite{DBLP:conf/lion/HutterHL11}, \textsc{SMAC} improves \textsc{ParamILS} with a Random Forest model, hence making it a model-based optimizer.

	%Similar to FLASH, the surrogate is progressively updated and the best configuration is returned on termination.
\end{itemize}

%Indeed, other optimizers exist for general algorithmic configuration (such as TPE~\cite{DBLP:conf/nips/BergstraBBK11} and $\epsilon$-PAL~\cite{DBLP:journals/jmlr/ZuluagaKP16}), but we do not consider these in this work 

We selected the above because (1) they are state-of-the-art optimizers from both the software engineering and general optimization community~\cite{DBLP:conf/icse/0003XC021,DBLP:journals/corr/abs-2112-07303,DBLP:conf/sigsoft/0001Chen21,bartz2021tuning}. (2) It has been reported that, for software configuration tuning, they outperform other optimizers, e.g., \textsc{BOCA} is better than \textsc{TPE}~\cite{DBLP:conf/icse/0003XC021} and \textsc{Flash} is superior to $\epsilon$\textsc{-PAL}~\cite{nair2018finding}. However, the model-based optimizers suffer from local optima, and prior study has shown that the inaccuracy of the surrogate can cause serious issue~\cite{DBLP:conf/cloud/ZhuLGBMLSY17}.

\subsection{Plain Multi-objectivization (PMO)}
\label{sec:pmo}

% in order to introduce incomparability between currently best configurations (by Pareto dominance relation),  thus preventing the search from being trapped into local optima

%Unlike the optimizers that rely on the single-objective model,  multi-objectivization is an alternative way to address the local optimum issue for software configuration tuning~\cite{DBLP:conf/sigsoft/0001Chen21,DBLP:journals/corr/abs-2112-07303}. 
%but useful for mitigating local optima. 

As an alternative way to address the local optimum issue in software configuration tuning for a single performance objective, multi-objectivization assumes two performance objectives to be considered~\cite{DBLP:conf/sigsoft/0001Chen21,DBLP:journals/corr/abs-2112-07303}: 
a target performance objective $f_t$ that is of concern (e.g., runtime) and another auxiliary performance objective $f_a$ that is generally of no interest to the software engineer (e.g., CPU load) on the given system (at least on the occasion of the corresponding tuning round).
In other SBSE problems~\cite{DBLP:journals/tse/YuanB20,derakhshanfar2020good,DBLP:conf/ssbse/MkaouerKBC14,DBLP:conf/ssbse/SoltaniDPDZD18}, 
a natural way of multi-objectivization would be to optimize both $f_t$ and $f_a$ simultaneously, hence leveraging the incomparability of Pareto dominance relation to jump out of the local optima. This, denoted as plain multi-objectivization (PMO), can be expressed as: 
\begin{align}
\begin{split}
\text{minimize}
\begin{cases}
f_a(\vect{x})\\
f_t(\vect{x})\\
\end{cases}
\end{split}
\label{Eq:PMOmodel}
\end{align}

Since PMO is an optimization model, it is agnostic to the underlying multi-objective optimizer. However, a major issue with PMO is that, 
a configuration with a poor $f_t$ 
but a good $f_a$ will also be regarded as a good configuration, 
despite the fact that $f_a$ is of no interest to the software engineer. 
This does not fit our purpose and 
can lead to undesired consumption of the tuning budget 
(due to the unnecessary efforts of optimizing $f_a$), 
especially when the measurements are expensive.

%To further confirm this point, we also evaluate PMO based on NSGA-II~\cite{Deb2002}---an overwhelmingly used multi-objective measurement-based optimizer from SBSE---in this work. 

%which is an overwhelmingly used optimizer in SBSE
%Note that since PMO is an optimization model, it can be both measurement- or model-based, but we examine the measurement-based PMO as this rules out the possible errors caused by the surrogate. 

\subsection{Meta Multi-objectivization (MMO)}

%MMO works at the level of optimization model as opposed to the level of optimizers, which means it can be either measurement- or model-based and we use the measurement-based version in this work as~\cite{DBLP:conf/sigsoft/0001Chen21} did. 

%MMO is also an optimizer-independent optimization model and we again use the measurement-based variant with NSGA-II.

%guide the tuning and optimize for our concerned performance objective

\citet{DBLP:journals/corr/abs-2112-07303,DBLP:conf/sigsoft/0001Chen21} propose a new way of tuning software configuration, dubbed MMO, firstly appeared at FSE'21~\cite{DBLP:conf/sigsoft/0001Chen21}. Like PMO, MMO is an optimization model and it can also be paired with any multi-objective optimizers, although Chen and Li used NSGA-II~\cite{Deb2002}---an extremely common multi-objective optimizer in SBSE---as the default. Yet, it does not treat $f_t$ and $f_a$ equally. 
That is, 
MMO optimizes $f_t$ whilst diversifying the values of $f_a$ during the tuning via the following model\footnote{We use the simplest linear form of MMO, as \citet{DBLP:conf/sigsoft/0001Chen21} showed that different forms perform similarly.}:
\begin{align}
\begin{split}
\text{minimize}
\begin{cases}
g_1(\vect{x}) = f_t(\vect{x}) + wf_a(\vect{x})\\
g_2(\vect{x}) = f_t(\vect{x}) - wf_a(\vect{x})\\
\end{cases}
\end{split}
\label{Eq:model}
\end{align}
whereby $g_1(\vect{x})$ and $g_2(\vect{x})$ are the meta-objectives, each of which shares the same $f_t$, 
but differs (effectively being opposite) 
on $f_a$. 
$w$ is a weight parameter that controls the relative contribution. As such, MMO transforms the tuning from searching in the original objectives into the space of meta-objectives, namely the MMO space. Since the performance objectives may come with radically different scales, both $f_t(\vect{x})$ and $f_a(\vect{x})$ need to be normalized.

Essentially,
MMO aims to seek a diverse set of configurations with two characteristics: 
(1) the configurations all have fairly high performance on $f_t$,
and (2) the configurations are of clear dissimilarity on $f_a$. 
The first characteristic is as the result of both meta-objectives in Eq.~(\ref{Eq:model}) optimizing $f_t$~\cite{DBLP:conf/sigsoft/0001Chen21},
which differs from the PMO, 
where both $f_t$ and $f_a$ are optimized together.
The second characteristic is the result of the two meta-objectives having opposite terms on $f_a$, making configurations incomparable (i.e., nondominated)~\cite{DBLP:conf/sigsoft/0001Chen21}---the key to mitigate local optima in the highly rugged and sparse configuration landscape, which allows MMO to outperform other optimizers~\cite{DBLP:journals/corr/abs-2112-07303,DBLP:conf/sigsoft/0001Chen21}.

However, \citet{DBLP:journals/corr/abs-2112-07303,DBLP:conf/sigsoft/0001Chen21} have also revealed 
that in MMO the weight $w$ is a highly sensitive parameter to the performance. Even with the improved MMO\footnote{Throughout the paper, we use MMO refers to both the original~\cite{DBLP:conf/sigsoft/0001Chen21} and improved MMO~\cite{DBLP:journals/corr/abs-2112-07303}, as they only differ on whether fixing $w=1$ and the normalization. Yet, we apply the improved MMO for all experiments and as the basis of \textsc{AdMMO}.} under a new improved normalization~\cite{DBLP:journals/corr/abs-2112-07303}, the resolution is simply fixing $w=1$ as opposed to adapt it (which is our aim). 

%This is due to the lack of a sound theoretical analysis on how and why $w$ can impact the performance of MMO.
%While they subsequently proposed a way to fix the weight to be 1 via a locally bounded normalization that achieves stable performance~\cite{DBLP:journals/corr/abs-2112-07303}, it also loses the opportunity to fine tune the contributions between $f_t(\vect{x})$ and $f_a(\vect{x})$ for boosting the outcome of configuration tuning. 
%This is what this paper tries to address.

%Next,we will explain what role the weight plays and why its adaptation in the tuning is beneficial.  

\section{What is Wrong with \textsc{MMO}?}
\label{sec:problem}
Here, we justify what role the weight plays in \textsc{MMO} and why its adaptation in tuning is beneficial.  

%we articulate the main issue with the original/improved MMO for configuration tuning. 

%and why it is a non-trivial challenge to overcome.

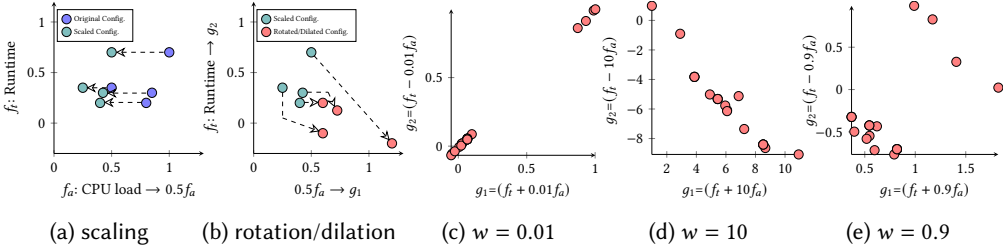
\begin{figure}[t!]
	\centering
 \begin{subfigure}{.2\columnwidth}
		\newcommand{\equal}{=}
\begin{tikzpicture}
    \begin{axis}[
        xmin=0,
        xmax=1.3,
        ymin=-0.3,
        ymax=1.2,
      width=5cm,
    height=5cm,
        axis x line  = bottom,
    axis y line  = left,
        ylabel=$f_t$: Runtime,
         xlabel=$f_a$: CPU load $\rightarrow 0.5f_a$,
       ylabel style={at={(-0.2,0.5)}},
        legend cell align={left},
       legend style={at={(0.02,0.95),font=\fontsize{5}{5}\selectfont},
      anchor=north west,legend columns=1}]
      
    \addplot[only marks,mark=*,fill=blue!50,draw=black,mark size=3pt] plot coordinates {
(0.8,0.2)
(0.85,0.3)
(0.5,0.35)
(1.0,0.7)

};

%w=0.5
    \addplot[only marks,mark=*,fill=teal!50,draw=black,mark size=3pt] plot coordinates {
(0.4,0.2)
(0.425,0.3)
(0.25,0.35)
(0.5,0.7)

};

\addlegendentry{Original Config.}
\addlegendentry{Scaled Config.}

       \end{axis}
       \draw[dashed,-{Stealth[open,length=2mm, width=2mm]}] (2.1,1.36) -- (1.22,1.36);
         \draw[dashed,-{Stealth[open,length=2mm, width=2mm]}] (2,1.15) -- (1.15,1.15);
          \draw[dashed,-{Stealth[open,length=2mm, width=2mm]}] (2.5,2.28) -- (1.42,2.28);
            \draw[dashed,-{Stealth[open,length=2mm, width=2mm]}] (1.2,1.48) -- (0.75,1.48);
%\node[align=left,anchor=west,black] (source) at (0,3.8){\footnotesize MMO$+$ for \textsc{Storm}};
    \end{tikzpicture}
		\subcaption{scaling}
	\end{subfigure}
 \hspace{-0.3cm}
 \begin{subfigure}{.2\columnwidth}
		\newcommand{\equal}{=}
\begin{tikzpicture}
    \begin{axis}[
        xmin=0,
        xmax=1.3,
        ymin=-0.3,
        ymax=1.2,
      width=5cm,
    height=5cm,
        axis x line  = bottom,
    axis y line  = left,
      ylabel=$f_t$: Runtime $\rightarrow g_2$,
         xlabel=$0.5f_a\rightarrow g_1$,
  %   ylabel=$g_2$\equal$(f_t-0.5f_a)$,
    %     xlabel=$g_1$\equal$(f_t+0.5f_a)$,
        ylabel style={at={(-0.2,0.5)}},
             legend cell align={left},
       legend style={at={(0.02,0.95),font=\fontsize{5}{5}\selectfont},
      anchor=north west,legend columns=1}]

%w=0.5
    \addplot[only marks,mark=*,fill=teal!50,draw=black,mark size=3pt] plot coordinates {
(0.4,0.2)
(0.425,0.3)
(0.25,0.35)
(0.5,0.7)

};

    \addplot[only marks,mark=*,fill=red!50,draw=black,mark size=3pt] plot coordinates {
(0.6,0.2)
(0.725,0.125)
(0.6,-0.1)
(1.2,-0.2)

};

\addlegendentry{Scaled Config.}
\addlegendentry{Rotated/Dilated Config.}

       \end{axis}
%\node[align=left,anchor=west,black] (source) at (0,3.8){\footnotesize MMO$+$ for \textsc{Storm}};
  \draw[dashed,-{Stealth[open,length=2mm, width=2mm]}] (1.15,1.15) -- (1.48,1.15);
   \draw[dashed,-{Stealth[open,length=2mm, width=2mm]}] (1.38,2.2) -- (3.1,0.3);
    \draw[dashed] (1.22,1.38) -- (1.7,1.38);
       \draw[dashed,-{Stealth[open,length=2mm, width=2mm]}] (1.7,1.38) -- (1.85,1.05);
       
         \draw[dashed] (0.66,1.38) -- (0.66,0.8);
           \draw[dashed,-{Stealth[open,length=2mm, width=2mm]}] (0.66,0.8) -- (1.5,0.5);
    \end{tikzpicture}
		\subcaption{rotation/dilation}
	\end{subfigure}
 \hspace{-0.3cm}
	\begin{subfigure}{.2\columnwidth}
		\newcommand{\equal}{=}
\begin{tikzpicture}
    \begin{axis}[
      width=5cm,
    height=5cm,
        axis x line  = bottom,
    axis y line  = left,
        ylabel=$g_2$\equal$(f_t-0.01f_a)$,
         xlabel=$g_1$\equal$(f_t+0.01f_a)$,
        ylabel style={at={(-0.2,0.5)}},
       legend style={at={(0.2,0.95),font=\fontsize{7}{8}\selectfont},
      anchor=north west,legend columns=1}]
      
    \addplot[only marks,mark=*,fill=red!50,draw=black,mark size=3pt] plot coordinates {
(0.06598190060649702,0.055213186594980715)
(0.06561276631836457,0.048681384360591065)
(0.006080074376199617,-0.006080074376199617)
(0.026723044295916983,0.019029306292078215)
(0.06561276631836457,0.048681384360591065)
(0.09906638540969145,0.08734475633099664)
(0.9890055481302034,0.9890055481302034)
(0.9327136877082577,0.9127136877082577)
(0.8729638045325954,0.8609748110105224)
(1.0019050803742802,0.9980949196257197)
(0.026723044295916983,0.019029306292078215)
(-0.04552898752751633,-0.05545812020505951)
(-0.05214433814643511,-0.06673844803127196)
(0.026723044295916983,0.019029306292078215)
(-0.025537032953425517,-0.03775887196494183)
(0.021207401194478042,0.003898410071637347)
(0.06598190060649702,0.055213186594980715)
(0.06598190060649702,0.055213186594980715)
(0.06561276631836457,0.048681384360591065)
(0.06561276631836457,0.048681384360591065)

};

       \end{axis}
%\node[align=left,anchor=west,black] (source) at (0,3.8){\footnotesize MMO$+$ for \textsc{Storm}};
    \end{tikzpicture}
		\subcaption{$w=0.01$}
	\end{subfigure}
 \hspace{-0.3cm}
	\begin{subfigure}{.2\columnwidth}
		\newcommand{\equal}{=}
\begin{tikzpicture}
    \begin{axis}[
      width=5cm,
    height=5cm,
        axis x line  = bottom,
    axis y line  = left,
        ylabel=$g_2$\equal$(f_t-10f_a)$,
         xlabel=$g_1$\equal$(f_t+10f_a)$,
      ylabel style={at={(-0.2,0.5)}},
       legend style={at={(0.2,0.95),font=\fontsize{7}{8}\selectfont},
      anchor=north west,legend columns=1}]
      
    \addplot[only marks,mark=*,fill=red!50,draw=black,mark size=3pt] plot coordinates {
(5.444954549358893,-5.3237594621574145)
(8.52283805422623,-8.408543903547276)
(6.080074376199617,-6.080074376199617)
(3.869745177213382,-3.823992826625387)
(8.52283805422623,-8.408543903547276)
(5.954020110217751,-5.767608968477063)
(0.9890055481302034,0.9890055481302034)
(10.922713687708258,-9.077286312291742)
(6.861466068808026,-5.127527453264909)
(2.90508037428023,-0.9050803742802298)
(3.869745177213382,-3.823992826625387)
(4.914072784905305,-5.015059892637882)
(7.237613549329572,-7.356496335507279)
(3.869745177213382,-3.823992826625387)
(6.0792715532989705,-6.142567458217338)
(8.667048467053405,-8.64194265578729)
(5.444954549358893,-5.3237594621574145)
(5.444954549358893,-5.3237594621574145)
(8.52283805422623,-8.408543903547276)
(8.52283805422623,-8.408543903547276)

};

       \end{axis}
%\node[align=left,anchor=west,black] (source) at (0,3.8){\footnotesize MMO$+$ for \textsc{Storm}};
    \end{tikzpicture}
		\subcaption{$w=10$}
	\end{subfigure}
 \hspace{-0.3cm}
	\begin{subfigure}{.2\columnwidth}
		\newcommand{\equal}{=}
\begin{tikzpicture}
    \begin{axis}[
      width=5cm,
    height=5cm,
        axis x line  = bottom,
    axis y line  = left,
        ylabel=$g_2$\equal$(f_t-0.9f_a)$,
         xlabel=$g_1$\equal$(f_t+0.9f_a)$,
        ylabel style={at={(-0.2,0.5)}},
       legend style={at={(0.2,0.95),font=\fontsize{7}{8}\selectfont},
      anchor=north west,legend columns=1}]
      
    \addplot[only marks,mark=*,fill=red!50,draw=black,mark size=3pt] plot coordinates {
(0.5451896741189727,-0.423994586917495)
(0.8190592634392857,-0.70476511276033)
(0.5472066938579655,-0.5472066938579655)
(0.3690943854667422,-0.323342034878747)
(0.8190592634392857,-0.70476511276033)
(0.6206788794116108,-0.4342677376709227)
(0.9890055481302034,0.9890055481302034)
(1.8227136877082577,0.022713687708257657)
(1.406474016264841,0.32746459927827676)
(1.1714572336852207,0.8285427663147793)
(0.3690943854667422,-0.323342034878747)
(0.3963174166231555,-0.4973045243557313)
(0.5972935517288047,-0.7161763379065118)
(0.3690943854667422,-0.323342034878747)
(0.5183348030590502,-0.5816307079774177)
(0.7914575061608888,-0.7663516948947735)
(0.5451896741189727,-0.423994586917495)
(0.5451896741189727,-0.423994586917495)
(0.8190592634392857,-0.70476511276033)
(0.8190592634392857,-0.70476511276033)

};

       \end{axis}
%\node[align=left,anchor=west,black] (source) at (0,3.8){\footnotesize MMO$+$ for \textsc{Storm}};
    \end{tikzpicture}
		\subcaption{$w=0.9$}
	\end{subfigure}
	\caption{The theoretical analysis and empirical evidence on the role of \textsc{MMO} weights for system \textsc{MariaDB} with normalized target and auxiliary performance objectives ($f_t$ is \textit{runtime} and $f_a$ is \textit{CPU load}). (a) shows the scaling on $f_a$ by a factor of $w=0.5$; (b) is the $45^{\circ}$ rotation and dilation on both axes by a factor of $\sqrt{2}$ afterwards. (c), (d), and (e) are the empirical results on configurations in an iteration with distinct $w$ thereof.}
	\label{fig:pre-analysis1}
 \vspace{-0.3cm}
\end{figure}

\subsection{Theory: Interpreting the Weight in \textsc{MMO}}
\label{sec:theory}

 Compared with \textsc{PMO}, \textsc{MMO} essentially does two main geometric operations to transform the original space of two performance objectives into a meta-objective space: (1) it scales (stretches or shrinks) the configurations along the $f_a$ axis by a factor of $w$ (Figure~\ref{fig:pre-analysis1}a); (2) it rotates the scaled configurations by $45^{\circ}$ clockwise and then dilates them on both $f_t$ and $f_a$ by a factor of $\sqrt{2}$ (Figure~\ref{fig:pre-analysis1}b). Geometrically, \textsc{MMO} can be decomposed via the following transformation metrics in linear algebra:
\begin{align}
\begin{bmatrix}
g_1(\vect{x})\\
g_2(\vect{x})
\end{bmatrix}
=
\overbrace{
\sqrt{2}
\begin{bmatrix}
\cos{{{\pi}\over{4}}} & \sin{{{\pi}\over{4}}}\\
-\sin{{{\pi}\over{4}}} & \cos{{{\pi}\over{4}}}
\end{bmatrix}
}^{\text{rotation/dilation matrix}}
\overbrace{
\begin{bmatrix}
w & 0\\
0 & 1
\end{bmatrix}
}^{\text{scaling matrix}}
\overbrace{
\begin{bmatrix}
f_a(\vect{x}) \\
f_t(\vect{x}) 
\end{bmatrix}
}^{\text{original space}}
=
\overbrace{
\begin{bmatrix}
f_t(\vect{x}) + wf_a(\vect{x})\\
f_t(\vect{x}) - wf_a(\vect{x})
\end{bmatrix}
}^{\text{MMO space}}
\label{Eq:MMO-detail}
\end{align}
The scaling based on weight $w$ in \textsc{MMO}, together with the rotation/dilation, impacts the balance between enhancing $f_t$ 
and diversifying configurations via $f_a$ during tuning. Since $w$ can be arbitrarily predefined, its value plays a more critical role therein. A smaller $w$ means a bigger shrink on $f_a$, thus more configurations become comparable after the rotation/dilation, putting more emphasis on optimizing $f_t$, i.e., \textbf{\textit{exploitation}}. 
In the extreme case when $w=0$,
$f_a$ is completely ruled out, leaving the two meta-objectives identical, and as such all configurations are comparable provided that they differ on $f_t$.
On the other hand,
a larger $w$ suggests a bigger stretch on $f_a$, making more of the configurations become incomparable following the rotation/dilation, which encourages the \textbf{\textit{exploration}} in the search space to find more diverse configurations. 
In the extreme case where $w=\infty$,
the differences between configurations on $f_a$ become infinitely large, rendering the two meta-objectives linearly conflicted,
hence all configurations are incomparable if they differ on $f_a$.

%which is fixed and predefined, specifies how the configurations should be stretched/shrunk along the axis of $f_a$. As such, it 

Figure~\ref{fig:pre-analysis1}c to ~\ref{fig:pre-analysis1}e demonstrate that distinct weight values can lead to different states of the configurations during the tuning even with the improved \textsc{MMO}~\cite{DBLP:journals/corr/abs-2112-07303}.
When $w=0.01$ (Figure~\ref{fig:pre-analysis1}c),
all configurations are comparable on the two meta-objectives. In this case, the problem degenerates to a single-objective problem,
thus the search may easily be trapped in local optima.
On the other extreme, when $w=10$ (Figure~\ref{fig:pre-analysis1}d),  
almost all configurations are incomparable (i.e. nondominated) in terms of the two meta-objectives, meaning that there is no selection pressure (discriminative power) driving the search toward the Pareto front. In contrast, when $w=0.9$ (Figure~\ref{fig:pre-analysis1}e),  
configurations are of more variety in terms of their quality---they can be superior, inferior, or incomparable between superior/inferior ones, which tends to favor exploitation while still maintaining a certain level of exploration. This also suggests that an ideal proportion of nondominated configurations also implies an appropriate balance between exploitation and exploration

%a good balance between exploitation and exploration balance.  

\begin{figure}[t!]
	\centering
	\begin{subfigure}{.37\columnwidth}
		\includegraphics[width=\columnwidth]{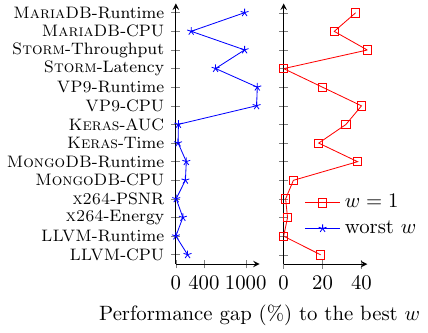}
			\subcaption{gap to best $w$}
	\end{subfigure}
	\hfill
	\begin{subfigure}{.35\columnwidth}
		\includegraphics[width=\columnwidth]{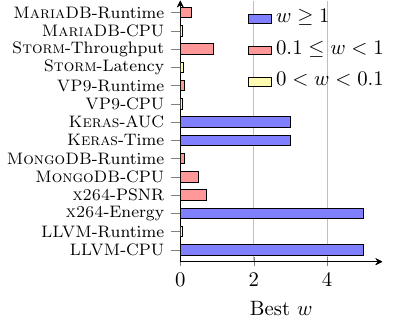}
				\subcaption{best $w$ per system-objective pair}
	\end{subfigure}
	\hfill
	\begin{subfigure}{.255\columnwidth}
		\includegraphics[width=\columnwidth]{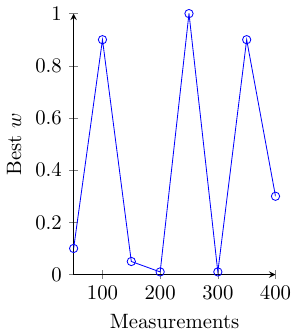}
		\subcaption{best $w$ per budget}	
	\end{subfigure}
	%\hspace{-0.2cm}
	\caption{The sensitivity of (improved) \textsc{MMO} to weight $w$ over 50 runs on different system-objective pairs. (a) shows the \% improvement that can be made by the best $w$. (b) illustrates the best fixed $w$. (c) demonstrates the best $w$ throughout the tuning for system \textsc{MariaDB} using \textit{runtime} as the target. The best $w$ (among $\{0.01,0.03,...,10\}$) is identified by Scott-Knott test~\cite{scott1974cluster} and the average performance results.}
	%as used by Chen and Li~\cite{DBLP:conf/sigsoft/0001Chen21}
	\label{fig:pre-analysis2}
 \vspace{-0.2cm}
\end{figure}

\subsection{Limitation of the Fixed Weight in \textsc{MMO} and Why It is Challenging}

% In multi-objective search,
% the proportion of nondominated solutions (i.e., configurations here) in the population significantly affects the search performance~\cite{Li2014}.
% If all the solutions in the population are comparable to each other 
% (i.e., either dominating or being dominated),
% like the case of Figure~\ref{fig:pre-analysis1}a,
% the problem degenerates to a single-objective problem,
% thus the search may easily be trapped in local optima.
% If all the solutions in the population are incomparable to each other
% (i.e., nondominated),
% like the case of Figure~\ref{fig:pre-anazlysis1}b,
% there is no selection pressure (discriminative power) driving the search toward the Pareto front. 
% Therefore,
% maintaining a certain proportion of nondominated solutions---neither too high nor too low (Figure~\ref{fig:pre-analysis1}c)---in the population during the search is critical for balancing exploitation and exploration. 

The above analysis explains what makes the weight in \textsc{MMO}~\cite{DBLP:conf/sigsoft/0001Chen21,DBLP:journals/corr/abs-2112-07303} important: its value determines the proportion of nondominated configurations in the tuning, which reflects the relationship between exploitation and exploration in the tuning hence can severely influence the performance of \textsc{MMO}. In subsequent work, \citet{DBLP:journals/corr/abs-2112-07303} have acknowledged such and attempted to fix the issue by removing the weight from \textsc{MMO} with an improved normalization scheme, i.e., setting $w=1$. However, there is still no guarantee that $w=1$ can help to achieve the reasonable proportion of nondominated configuration for all systems/performance objectives, hence blurring the full potential of \textsc{MMO}. Indeed, from Figure~\ref{fig:pre-analysis2}a (we use the improved MMO with the new normalization from~\cite{DBLP:journals/corr/abs-2112-07303}), we see that the performance gap between the worst and best $w$ can be rather high---up to 1167\%. Even when setting $w=1$, the truly best setting can still lead to considerable improvement in general with up to 43\% (throughput on \textsc{Storm}).

Resolving the weight setting issue is not easy, because for different systems/performance objectives,
the ``sweet spot'' (best setting) of the weight, which achieves a reasonably balanced proportion of nondominated configurations during the tuning, can be very disparate 
due to the difference of configuration landscape and scale, \textit{etc}~\cite{DBLP:conf/sigsoft/0001Chen21,DBLP:journals/corr/abs-2112-07303}. From Figure~\ref{fig:pre-analysis2}b, we note that the optimal weight value varies dramatically, 
e.g., the best $w$ value for the \textit{runtime} of \textsc{MongoDB} and the \textit{CPU load} of \textsc{LLVM} are $0.05$ and $5$, respectively---a $100\times$ difference. This brings a big difficulty to the attempt to set the right weight via trial and error, 
which itself is time-consuming due to the expensive measurement of configurations.

%\textcolor{red}{[Tao, could you add one sentence to explain the figure like ``in xxx system, the optimal setting is 0.1 while in xxx, the optimal setting is...'']}

On top of that, 
even for the same system/performance objective,
different budgets/stages of the tuning can have distinct optimal weight settings. When the configurations concentrate in a small region of the search space,
a large weight can be helpful to jump out of the local optimum. When the configurations scatter widely over the space,
a small weight can be beneficial to steer the tuning towards the right direction 
(i.e., better target performance).
Figure~\ref{fig:pre-analysis2}c gives the optimal weight settings during different tuning budgets/stages 
(every 50 measurements) for the system \textsc{MariaDB}.
As can be seen,
when the tuning consumes 200 measurements, 
a tiny weight value is the best,
whereas after 50 measurements at a budget of 250, 
the weight being $1$ is the best.

The above suggests that 
a fixed weight setting for all cases and throughout the tuning is not ideal,
hence an adaptive weight, 
which can change in line with the tuning, is needed for dynamically maintaining a good proportion of nondominated configurations---a neither too high nor too low weight value that indicates a good balance of exploitation and exploration, e.g., in Figure~\ref{fig:pre-analysis1}e.

\section{Software Configuration Tuning with Adaptive \textsc{MMO}}
\label{sec:method}

{
\begin{algorithm}[t!]
	\DontPrintSemicolon
	\footnotesize
	
	\caption{Pseudo-code of \textsc{AdMMO} (with NSGA-II~\cite{Deb2002} as the base optimizer)}
	\label{alg:main}
	\KwIn{Configuration space $\mathbfcal{V}$; system $\mathbfcal{S}$; initial weight in MMO $w=1$; budget $B$; offset $T=1$; cut-off point $C=0.5$; expected proportion of nondominated configuration $p$}

	\KwOut{the best configuration on $f_t$}
	%\KwOut{$\mathbfcal{S}_{best}$ the best adaptation plan on $f_t(\vect{x})$}
	%\kwDeclare{bound vectors $\mathbf{\overline{z}_{max}}$ and $\mathbf{\overline{z}_{min}}$}
	Randomly initialize a population of $n$ configurations $\mathbfcal{P}$\\
    %\tcc{\textcolor{blue}{measuring $f_t$ and $f_a$.}}
	\textsc{measure($\mathbfcal{P},\mathbfcal{S}$)} \tcc*[h]{\textcolor{blue}{measuring $f_t$ and $f_a$.}}\\
	 \textsc{normalize}($\mathbfcal{P}$)\\
	 
	 $\mathbfcal{U}\leftarrow${\textsc{computeMMO($\mathbfcal{P}, w$)}} \tcc*[h]{\textcolor{blue}{computing MMO meta-objectives $g_1$ and $g_2$.}}\\
    $b = b + n$\\
	\While{b < B}
	{  
		
		$\mathbfcal{P'}=\emptyset$\\
		
		\While{$\mathbfcal{P'}<n$}
		{ 
		
			$\{p_x,p_y\}\leftarrow$\textsc{mating($\mathbfcal{P}$)}
			\tcc*[h]{\textcolor{blue}{mating based on $g_1$ and $g_2$.}}\\
			$\{s_x,s_y\}\leftarrow$\textsc{doCrossoverAndMutation($\mathbfcal{V}, p_x,p_y$)}\\

		\For{$\forall s_i \in \{s_x,s_y\}$ that is new}{
		\textsc{measure($s_i,\mathbfcal{S}$)} \tcc*[h]{\textcolor{blue}{new configurations are measured; use previous measurement otherwise.}}\\
		   	$b = b+1$\\
		}
		
%		   \If{$o_x$ is new} {
%		   	\textsc{measure($o_x,\mathbfcal{S}$)}\\
%		   	$b = b+1$\\
%		   }
%		   \If{$o_y$ is new} {
%		   	\textsc{measure($o_y,\mathbfcal{S}$)}\\
%		   	 $b = b+1$\\
%		   }

			$\mathbfcal{P'}=\mathbfcal{P'}\bigcup\{s_x,s_y\}$\\
		}
		
		\If{a new best configuration on $f_t$ found in $\mathbfcal{P'}$}{
		 o = 0
		}\Else {
		o = o + 1
		}
	 
		$\mathbfcal{U}=\mathbfcal{P}\bigcup\mathbfcal{P'}$\\
		%\tcc{\textcolor{blue}{normalizing $f_t$ and $f_a$ using local bounds in $\mathbfcal{U}$.}}
	    \textsc{normalize}($\mathbfcal{U}$) \tcc*[h]{\textcolor{blue}{the new local normalization as in the improved MMO~\cite{DBLP:journals/corr/abs-2112-07303}}}\\
		
	\If(\tcc*[h]{\textcolor{blue}{triggering adaptation via Equation~\ref{eq:prob}.}}){\textsc{isTrigger}($o,T,b,B,C$)}{
		
		$\mathbfcal{U}\leftarrow${\textsc{runMMOwithAdaptiveWeight($\mathbfcal{U}, w, p$)}} \tcc*[h]{\textcolor{blue}{adapting $w$ to compute $g_1$ and $g_2$.}}\\
	}\Else {
		
	$\mathbfcal{U}\leftarrow${\textsc{runMMO($\mathbfcal{U}, w$)}} \tcc*[h]{\textcolor{blue}{computing $g_1$ and $g_2$ with current $w$.}}\\
	}

		$\mathbfcal{P}\leftarrow$\textsc{MMOwithPartialDuplicates($\mathbfcal{U}$)}\tcc*[h]{\textcolor{blue}{survival selection with partial duplicate retention.}}\\

	}
	
	\Return the configuration with the best $f_t$ in $\mathbfcal{P}$\\
	
\end{algorithm}
\vspace{0.2cm}
} 
%\subsection{Adapting the Weight in \textsc{MMO}}

%This work presents a weight adaptive method for \textsc{MMO},  called \textsc{AdMMO}. Its 

We now delineate the designs of \textsc{AdMMO}. As in Algorithm~\ref{alg:main}, similar to \textsc{MMO}, 
\textsc{AdMMO} also uses NSGA-II~\cite{Deb2002} as the underlying multi-objective optimizer and 
the tuning runs in the transformed \textsc{MMO} space. 
The extensions of \textsc{AdMMO} to the original/improved \textsc{MMO} are three-fold: 

\begin{itemize}
    \item We maintain a good proportion of unique nondominated configurations by dynamically adapting the weight in \textsc{MMO} (line 22);
    \item but doing so only when it is necessary (line 21);
    \item while preserving the promising configurations with partially retained duplicates (line 25).
\end{itemize}

In what follows, we elaborate on these designs in detail.

%i.e., the best configuration does not change while the tuning has made some reasonable progresses. 

\subsection{Progressive Trigger for Adapting $w$ in MMO during Tuning}

Constantly adapting the weight during the tuning is unnecessary, or can even be harmful,
as it takes a few consecutive iterations to gather the tuning state.
In \textsc{AdMMO}, 
we design a \textit{progressive trigger} to determine when a weight adaptation is needed (at line 21 in Algorithm~\ref{alg:main}). 
In general, our trigger design is derived from two observations from tuning software configuration:

\begin{itemize}
    \item The less the measurements (i.e., at an earlier stage of tuning), 
    the less likely that the weight adaptation is needed as there is a smaller chance that the tuning has stuck (either due to local optima or limited discriminative power).
    %converged reasonably
    \item If a better configuration can still be discovered, then it is less desirable to adapt the weight since this implies that the current proportion of nondominated configurations can still be effective in guiding the tuning.
\end{itemize}

% \begin{align}
% 	prob = 1 - \exp{(-{{\text{max}(0,o-T)}\over{\sigma^2}}})
% \end{align}
% \begin{align}
% 	\sigma^2 = -{{S^2}\over{\log{C}}}
% \end{align}

\begin{figure}[t!]
	\centering
	\begin{subfigure}{.32\columnwidth}
		\includegraphics[width=\columnwidth]{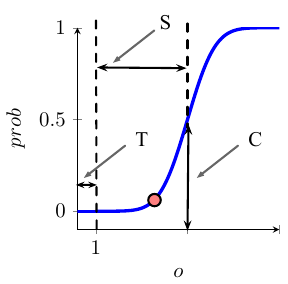}
		%\subcaption{$T=1$, $C=0.5$, $S=8$ ($B=400$; $b=50$)}
	\end{subfigure}
	~\hspace{2cm}
	\begin{subfigure}{.32\columnwidth}
		\vspace{0.2cm}
		\includegraphics[width=\columnwidth]{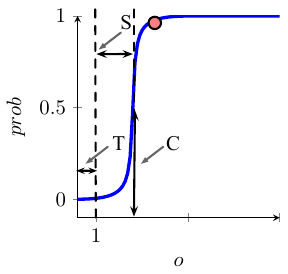}
		%\subcaption{$T=1$, $C=0.5$, $S=4$ ($B=400$; $b=100$)}
	\end{subfigure}

~\hspace{1cm}
 \begin{subfigure}{.4\columnwidth}
		\subcaption{$T=1$, $C=0.5$, $S=8$ ($B=400$; $b=50$)}
	\end{subfigure}
	~\hspace{0.8cm}
  \begin{subfigure}{.4\columnwidth}
		\subcaption{$T=1$, $C=0.5$, $S=4$ ($B=400$; $b=100$)}
	\end{subfigure}
	\caption{The probability distribution in progressive trigger of \textsc{AdMMO}. The red dot denotes $o=5$: 5 consecutive iterations in which the best configuration has not been changed. Clearly, even with the same $o$, there is a higher probability of trigger in (b) than (a) as the former consumes more measurements.}
	\label{fig:trigger}
\end{figure}

As a result, at each iteration of the tuning, we compute the probability of triggering weight adaptation, $prob$, using the following function (with an illustration in Figure~\ref{fig:trigger}):
\begin{align}
	prob = 1 - \exp{({{\ln{C} \times \text{max}(0,o-T)}\over{S^2}}})
 \label{eq:prob}
\end{align}
whereby $o$ is the number of consecutive iterations that no better configuration is found and $T$ is a given offset, indicating the tolerance level of $o$ (which we set the most restricted value of $T=1$ in this work). 
Hence the more iterations that the tuning gets stuck in a configuration, 
the more likely that we need to adapt the weight. 
$S$ determines the slope of the function, which is calculated by the ratio between the budget ($B$) and the current number of measurements ($b$): $S={B \over b}$. As such, $S$ decreases as more configurations are measured, meaning that the steeper the slope and hence the more likely it is to trigger weight adaptation for the same $o$ value. $C$ is a cut-off probability that also influences the slope of the shape and we use the most pragmatic setting of 0.5 (50\% probability). In this way, we progressively increase the probability of adapting the weight proportional to the number of measurements (by $S$) and the iteration count with no better configuration found (by $o$).

\begin{figure}[t!]
\centering
\begin{subfigure}{.32\columnwidth}
\includegraphics[width=\columnwidth]{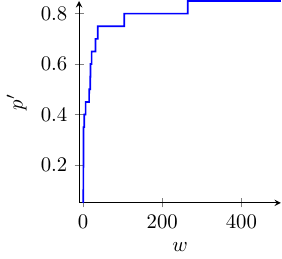}
\subcaption{\textsc{Storm}}
\end{subfigure}
\hspace{2cm}
\begin{subfigure}{.32\columnwidth}
\includegraphics[width=\columnwidth]{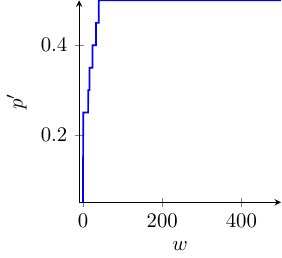}
\subcaption{\textsc{MongoDB}}
\end{subfigure}
\caption{Correlation between the actual proportion of nondominated configurations $p'$ and the weight $w$ under the (improved) MMO in one run (population size of 10 and without duplicates). $f_t$ and $f_a$ of \textsc{Storm} are throughput and latency, receptively; for \textsc{MongoDB}, the $f_t$ is runtime while $f_a$ is CPU load.}
\label{fig:weight-change}
\end{figure}

\begin{algorithm}[t!]
	\DontPrintSemicolon
	\footnotesize
	
	\caption{\textsc{runMMOwithAdaptiveWeight}}
	\label{alg:adapt}
	\KwIn{Union of current population and the generated offsprings $\mathbfcal{U}$; current weight $w$; expecting proportion of nondominated configurations $p$}
    \KwOut{Configurations with MMO meta-objectives under adapted $w$}
	\kwDeclare{$\Delta=0.1$; bounds of weight: $w_{max}=10^3$; $w_{min}=0$}

	\While{$p' \neq p$ and $w_{min} < w < w_{max}$}
	{
	    {\textsc{computeMMO($\mathbfcal{U}, w$)}}\\
	    $\mathbfcal{U'}\leftarrow$\textsc{removeDuplicates($\mathbfcal{U}$)}\\
		$\mathbfcal{F}\leftarrow$\textsc{nondominatedSorting($\mathbfcal{U'}$)}\\
		$p'$ = \textsc{getProportionOfNondominatedConfigurations}($\mathbfcal{F}$)\\
		
		\If{$p' < p$}{
		    \If{$w + \Delta \geq 0.1$}{
		        $\Delta = 0.1$\\
		    }
		    $w = w + \Delta$\\
		   % \If{$w \geq w_{max}$}{
		     %  \Return $\mathbfcal{U}$\\
		    %}
		}\ElseIf{$p' > p$}{
	    	\If{$w - \Delta < 0.1$}{
		        $\Delta =  0.1 \times 10^{-3}$\\
		    }
		    $w = w - \Delta$\\
		   %  \If{$w \leq w_{min}$}{
		      % \Return $\mathbfcal{U}$\\
		    %}
		}
	}

    \Return $\mathbfcal{U}$\\
	
\end{algorithm}
%\vspace{-1cm}

\subsection{Maintaining Nondominated Configurations in the MMO Space}

As discussed in Section~\ref{sec:problem}, the proportion of nondominated configurations is controllable via the weights in \textsc{MMO}, hence under each trigger, we seek to adapt the weights such that the proportion, denoted as $p'$, reaches a given level $p$ (line 22 in Algorithm~\ref{alg:main}). To this end, our design was derived from the theoretical analysis in Section~\ref{sec:problem} and some important properties for tuning software configuration in that regard, as illustrated by the exampled systems from Figure~\ref{fig:weight-change}: 

\begin{itemize}
    \item Clearly, when the weight increases, the proportion would never become smaller, and similarly,
     decrements of the weight would never make the proportion larger (recall Section~\ref{sec:theory}).
    \item Although changing \textsc{MMO} weight can control the proportion of nondominated configurations, the sensitivity of which can differ depending on the given system.
\end{itemize}

Bearing the above in mind, \textsc{AdMMO} adapts the weight via the following steps (Algorithm~\ref{alg:adapt}):

\begin{enumerate}
    \item Examine the current proportion of unique nondominated configurations\footnote{The divisor is less important in the calculation as it can be with or without duplicate configurations; we use the one without duplicates for its interpretability. The most important factor is the count of nondominated configurations.} $p'={n_{d}\over n_{u}}$ under $w$ by using the count of unique nondominated configurations ($n_{d}$) together with the size of the current unique population and offsprings $n_{u}$ (lines 2-5 and see Section~\ref{sec:handling}).
    
    \item If $p' < p$, we increase $w$ by $\Delta$; if $p' > p$, we decrease $w$ by $\Delta$. The $\Delta$ is updated depending on two situations, which we found appropriate (lines 6-9 and 10-13):
    \begin{align}
     \begin{split}
      \Delta =
      \begin{cases}
      0.1 \text{, if } p' < p \text{ and } w + \Delta \geq 0.1 \\
      0.1 \times 10^{-3} \text{, if } p' > p \text{ and } w - \Delta < 0.1\\
    %     \begin{cases}
    %   0.1 \text{, if } p' < p \text{ and } w + \Delta \geq 0.1 \text{ or } p' > p \text{ and } w + \Delta < 0.1\\
    %   0.1 \times 10^{-3} \text{, if } p' > p \text{ and } w - \Delta < 0.1 \text{ or } p' < p \text{ and } w - \Delta \geq 0.1 \\
    %   \end{cases}
      \end{cases}
      \end{split}
     \end{align}
   When $p'=p$, the adaptation terminates. 
   
    % 0.1 \text{, if } p' < p \text{ while } w + \Delta \geq 0.1 \text{ or } w \geq 0.1\\
    %   0.1 \times 10^{-3} \text{, if } p' > p \text{ while } w - \Delta < 0.1 \text{ or } w < 0.1\\
   
   \item Repeat from step 1 with the updated $w$. To prevent adapting forever, we set bounds as $w_{max}=10^{3}$ and $w_{min}= 0$ to force the adaptation to stop once either of them is reached, as we found that they produce the most stable results on all systems/performance objectives. 
   
   %Note that when $w=0$, the \textsc{MMO} degenerates to the original single-objective model.

\end{enumerate}

%It is difficult to confirm what proportion level is appropriate, other than knowing too small or too large would be detrimental. However, 

As we will show in Section~\ref{sec:rq4}, empirically we are able to conclude that $p=0.3$ (30\% proportion) is generally the best and most stable level to maintain across the cases. 

%The above step-wise adaptation of the weight is important, as we found that, although changing \textsc{MMO} weight can control the proportion of nondominated configurations when tuning software configuration, the sensitivity of which can differ depending on the given system, as shown in Figure~\ref{fig:weight-change}.

\subsection{Partial Duplicate Retention for Preserving Promising Configurations}
\label{sec:handling}

Since \textsc{MMO} preserves the tendency toward the best target performance objective and there is often a high sparsity for configurable software systems~\cite{DBLP:conf/sigsoft/0001Chen21,nair2018finding,DBLP:conf/sigsoft/GongChen2023}, \textsc{MMO} will likely to accumulate many duplicate configurations. Yet, having many duplicate configurations can harm the weight adaptation as the duplicates might also be nondominated. As such, the proportion of nondominated configurations may appear to be appropriate but in fact, most (if not all) of those nondominated ones can be duplicates. Hence, when examining the actual proportion of nondominated configurations $p'$, we need to consider only the unique configurations (line 3 in Algorithm~\ref{alg:adapt}).

% When directly adapting the \textsc{MMO} weight with \textsc{AdMMO}, one issue we noticed was that, during the tuning, it is easy for the population to contain a large number of duplicate configurations. This is possible since \textsc{MMO} preserves the tendency toward the best target performance objective and there is often a high sparsity for configurable software systems~\cite{DBLP:conf/sigsoft/0001Chen21,nair2018finding}. However, having many duplicate configurations is detrimental to weight adaptation, 
% because those duplicates are also regarded as nondominated according to the notion of Pareto dominance. 
% This means that, when calculating the proportion of nondominated configurations, it may appear to be appropriate but in fact most (if not all) of those nondominated ones are duplicate. Hence, when examining the actual proportion $p'$, we consider only the unique configurations (line 3 in Algorithm~\ref{alg:adapt}).

%Similarly, too many duplicate configurations can also be harmful in the standard survival selection of NSGA-II that determines which one to keep for the next iteration (\textcolor{red}{Some more explanation why?}).

%The computed proportion (without duplicates) needs to be reflected in the survival selection of NSGA-II that determines which one to keep for the next iteration, otherwise, too many duplicate configurations can also be harmful. \textcolor{red}{[Tao, I do not quite understand what this means here.]} 

Next, in the survival selection---a procedure that preserves promising configurations for the next iteration---of the optimizer that underpins \textsc{AdMMO} (i.e., NSGA-II~\cite{Deb2002}), we also need to distinguish the duplicates or otherwise the selection will lose the guidance since there could be too many nondominated yet duplicate configurations. {However, unlike the calculation of $p'$, what makes it more challenging is that simply removing all but one duplicate (which is a common method~\cite{DBLP:conf/gecco/FortinP13}) can also be harmful, as poor configurations dominated by some removed (and better) duplicates may be preserved, thereby undermining the tuning performance.} Indeed, our experiments suggest that considering only the unique configurations in the selection is detrimental (see Section~\ref{sec:ablation}).

% In addition, 
% to align with the calculated proportion of nondominated configurations, 
% duplicate configurations also need to be handled in the survival selection of the optimizer that underpins \textsc{AdMMO} (i.e., NSGA-II~\cite{Deb2002}). 
% To that end, 
% a naive way would be to again simply remove all duplicates as before. 
% However, 
% simply removing all duplicate configurations means that configurations dominated by some removed duplicates may be selected. 
% This may undermine the tuning towards better target performance.

%However, there are two situations in one of which such a simple method can be problematic:
%\begin{itemize}
%
%    \item If the number of unique nondominated configurations is greater than (or equivalent to) the size of the population, we can indeed remove all duplicates before performing the nondominated sorting selection in NSGA-II since only the configuration in the first front would be considered anyway.
%
%    \item Otherwise, simply removing all duplicate configurations means that unique configurations dominated by some removed duplicates would be selected, 
%    even though they are inferior to the target performance objective. This may be undesired.
%\end{itemize}

To overcome the above, we design a partial duplicate retention mechanism in the survival selection of \textsc{AdMMO}, ensuring that the duplicates are distributed amongst different fronts and the better ones will be with higher-ranked fronts (line 25 in Algorithm~\ref{alg:main}). As such, high-quality duplicates (i.e., the good ones) still have a chance to survive when there are fewer unique nondominated configurations than the population size (but not always). As shown in Algorithm~\ref{alg:dup}, we perform the steps below to determine which configurations can be preserved to the next iteration:

\begin{enumerate}

\item Run nondominated sort in NSGA-II to produce a vector of fronts $\mathbfcal{F}=\{\vect{F}_0,\vect{F}_1,...,\vect{F}_n\}$ ($\vect{F}_0$ is the nondominated front) where duplicate configurations will be in the same front (line 1).
%Start from the best one in $\mathbfcal{F}$
\item For a front $\vect{F}_i$ where $0 \leq i \leq n-1$, we preserve only the unique configurations in $\vect{F}_i$ and move all the remaining duplicates $\vect{D}$ to $\vect{F}_{i+1}$ (lines 3-6). 
%\item Repeat Step 2 for $\vect{F}_{i+1}$ until $\vect{F}_{n-1}$.

%The last front $\vect{F}_{n}$ will have all the duplicates.

\item Preserve configurations to survive from $\vect{F}_i$ (lines 7-8). 
If not all configurations in $\vect{F}_i$ can be fitted in the next population, select them based on crowding distance (lines 9-11). 
This is the same as in the original NSGA-II.

\item Return the population if it is full. Otherwise, repeat from step 2 until the population is full.

%\item Return the population if it is full. Otherwise, repeat from step 2 for front $\vect{F}_{i+1}$ until the population becomes full. 

%or all the fronts have been visited.

\begin{algorithm}[t!]
	\DontPrintSemicolon
	\footnotesize
	
	\caption{\textsc{MMOwithPartialDuplicates}}
	\label{alg:dup}
	\KwIn{Union of current population and the generated offsprings $\mathbfcal{U}$}

	\kwDeclare{The set of duplicate configurations $\vect{D}$}
	\KwOut{New population of configurations $\mathbfcal{P}$}

	$\mathbfcal{F}$ = \textsc{nondominatedSorting($\mathbfcal{U}$)}\\
	\For{$\forall \vect{F}_i \in \mathbfcal{F}$ when $\mathbfcal{P}$ is not full}{

	     \If{$\vect{F}_{i+1}$ exists}{
	     $\vect{D}$ = \textsc{getDuplicate}($\vect{F}_i$)\\
	     $\vect{F}_i$ = \textsc{removeDuplicateConfigurations}($\vect{F}_i$,$\vect{D}$)\\
	     $\vect{F}_{i+1} = \vect{F}_{i+1} \bigcup \vect{D}$\\
	     }
	     
	     %$n = \mathbfcal{P} - \vect{F}_i$\\
	         \If{the remaining size of $\mathbfcal{P} \geq$ the size of $\vect{F}_i$}{
	            $\mathbfcal{P} = \mathbfcal{P} \bigcup \vect{F}_i$\\
	         }\Else{
	            $\vect{F'}_i$ = \textsc{sortByCrowdingDistance}($\vect{F}_i$)\\
	            $\mathbfcal{P} = \mathbfcal{P} \bigcup$ top $k$ configurations from $\vect{F'}_i$ until $\mathbfcal{P}$ is full\\
	         }
	     
	   %  \If{$\mathbfcal{P}$ is full}{
	     %   \Return $\mathbfcal{P}$\\
	     %}

	}

	\Return $\mathbfcal{P}$
	
\end{algorithm}
%\vspace{-1cm}

\end{enumerate}

\begin{figure}[t!]
\centering
\begin{subfigure}{.32\columnwidth}
\includegraphics[width=\columnwidth]{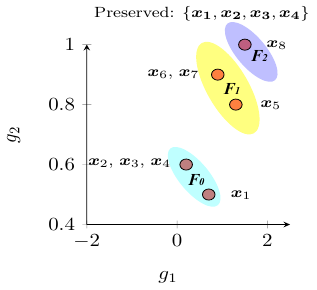}
\subcaption{without distinguish duplicates}
\end{subfigure}
\hfill
\begin{subfigure}{.32\columnwidth}
\includegraphics[width=\columnwidth]{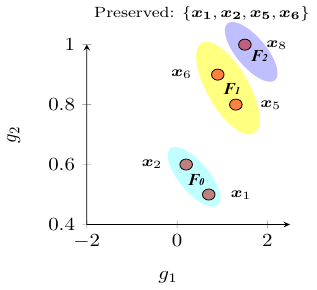}
\subcaption{remove extra duplicates}
\end{subfigure}
\hfill
\begin{subfigure}{.32\columnwidth}
\includegraphics[width=\columnwidth]{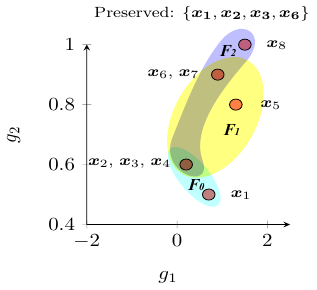}
%\vspace{-0.49cm}
\subcaption{partial duplicate retention}
\end{subfigure}
\caption{The results from different ways of handling duplicates when preserving configurations for the next iteration under the \textsc{MMO} meta-objectives space. $\vect{x_2}$, $\vect{x_3}$, and $\vect{x_4}$ are duplicate; $\vect{x_6}$ and $\vect{x_7}$ are also duplicate.}
\label{fig:dup-exp}
\vspace{-0.4cm}
\end{figure}

Figure~\ref{fig:dup-exp} gives an example. 
Suppose that the population size is 4 and after the normal nondominated sorting, we have three fronts: $\vect{F}_0 = \{\vect{x_1},\vect{x_2},\vect{x_3},\vect{x_4}\}$, $\vect{F}_{1}=\{\vect{x_5},\vect{x_6},\vect{x_7}\}$, and $\vect{F}_{2}=\{\vect{x_8}\}$, which contains a rather high number of nondominated yet duplicate configurations. Clearly, when computing $p'$ without duplicates, the number of nondominated configurations considered is $2$. 

In this case, if we do not distinguish the duplicate nondominated configurations in the selection (Figure~\ref{fig:dup-exp}a), then there would be 4 nondominated configurations, which is inconsistent with the calculation of $p'$, and the preserved ones are $\{\vect{x_1},\vect{x_2},\vect{x_3},\vect{x_4}\}$ with three duplicates. If we simply remove all duplicates in the selection (Figure~\ref{fig:dup-exp}b), then indeed we can have 2 nondominated configurations, but the preserved ones would be $\{\vect{x_1},\vect{x_2},\vect{x_5},\vect{x_6}\}$, which is not ideal since both $\vect{x_5}$ and $\vect{x_6}$ are configurations that have been dominated by others. With partial duplicate retention (Figure~\ref{fig:dup-exp}c), we can have $\vect{F}_0 = \{\vect{x_1},\vect{x_2}\}$ and it would certainly be preserved. Now, we move $\vect{x_3}$ and $\vect{x_4}$ to $\vect{F}_{1}$ which would become $\vect{F}_{1}=\{\vect{x_5},\vect{x_6},\vect{x_7},\vect{x_3},\vect{x_4}\}$. After step 2 at this front, we would have $\vect{F}_{1} = \{\vect{x_5},\vect{x_6},\vect{x_3}\}$, leaving $\vect{x_4}$ and $\vect{x_7}$ again to $\vect{F}_{2}$ and having $\vect{F}_{2}=\{\vect{x_4},\vect{x_7},\vect{x_8}\}$. Since we only need two more configurations, we stick with $\vect{F}_{1}$ to preserve the top 2 most diverse ones from $\{\vect{x_5},\vect{x_6},\vect{x_3}\}$, which will be $\vect{x_3}$ and $\vect{x_6}$ according to the crowding distance. As such, there would still be $2$ nondominated configurations and it is consistent with the calculation of $p'$, while good configurations ($\vect{x_2}$) and their duplicates ($\vect{x_3}$ and $\vect{x_4}$) are distributed across different fronts, some of which are higher-ranked. Here, the preserved ones will be $\{\vect{x_1},\vect{x_2},\vect{x_3},\vect{x_6}\}$, which is more ideal than the outcome of ignoring the duplicates in terms of diversity since only $\vect{x_2}$ and $\vect{x_3}$ are duplicate, while also clearly better than the result of simply removing all duplicates as $\vect{x_3}$ dominates $\vect{x_5}$. 

Notably, the above retention mechanism is tailored for \textsc{AdMMO} to use an underlying multi-objective optimizer that is compatible with nondominated sorting, therefore it will directly work with those in the same family of NSGA-II, e.g., SMS-EMOA~\cite{DBLP:journals/eor/BeumeNE07} and GrEA~\cite{DBLP:journals/tec/YangLLZ13} (the diversity distance therein is flexibly changeable). Although the idea/purpose behind retention is generic, adopting the mechanism (and \textsc{AdMMO}) to pair other families of optimizers, e.g., IBEA~\cite{DBLP:conf/ppsn/ZitzlerK04}, requires some amendments, which we leave for future work. 

%Compared with NSGA-II, IBEA belongs to a different family as it uses a single/continuous indicator in selection. 

%the bottom number when calculating the proportion is less important

%than, e.g., that of $\vect{x_8}$, and hence its duplicates $\vect{x_3}$ are still more likely to be preserved than $\vect{x_8}$.}

%\textcolor{red}{[Tao, this actually is slightly different from the approach I thought previously.]}

\section{Evaluation}
\label{sec:exp}
In this section, we elaborate on the experiment settings. All experiments were run on a dedicated High-Performance Computing server with 64-core Intel Xeon 2.6GHz and 256GB DDR RAM.

\subsection{Research Questions}
\label{sec:rq}

%To understand the usefulness of the \textsc{MMO} model, 

Our experimental evaluation seeks to answer four research questions (RQs):
%in software configuration tuning:

\begin{itemize}
    \item \textbf{RQ1:} How does \textsc{AdMMO} perform against the state-of-the-art optimizers for tuning?
    \item \textbf{RQ2:} How efficient is the \textsc{AdMMO} in utilizing resources over the state-of-the-art optimizers?
    \item \textbf{RQ3:} How beneficial are the progressive trigger and partial duplicate retention? %for \textsc{AdMMO}?
    \item \textbf{RQ4:} What is the sensitivity of \textsc{AdMMO} to the maintained proportion of nondominated configurations?
\end{itemize}

\textbf{RQ1} helps us to understand the ability of \textsc{AdMMO} to tune the configurations. However, doing so by consuming a large amount of resources (the number of measurements) is clearly undesirable. Therefore, we ask \textbf{RQ2} to assess whether \textsc{AdMMO} is efficient in finding promising configurations. \textbf{RQ3} allows us to conduct ablation studies of \textsc{AdMMO}, hence the usefulness of our adaptation designs can be examined. We use \textbf{RQ4} to study the sensitivity of \textsc{AdMMO} to $p$.

%\textbf{RQ1} helps us to understand the ability of \textsc{AdMMO}, i.e., whether it can lead to better configurations compared with a variety of other optimizers under the same budget. However, doing so by consuming a large amount of resources (the number of measurements) is clearly undesirable. Therefore, we ask \textbf{RQ2} to assess whether \textsc{AdMMO} is efficient in achieving promising configurations. \textbf{RQ3} allows us to conduct ablation analysis of \textsc{AdMMO}, hence the usefulness of our designs can be examined. We use \textbf{RQ4} to study how the results can change under different proportions of nondominated configurations for multi-objectivizing software configuration tuning. 

%\subsection{MMO instances}

%Clearly, {MMO-Linear} builds a linear effect while the other two are nonlinear.

\subsection{Configurable Software Systems and State-of-the-art Optimizers}

In this work, we consider the highly configurable software systems that have been widely used in existing work~\cite{DBLP:journals/corr/abs-2112-07303,nair2018finding,DBLP:conf/mascots/MendesCRG20,DBLP:conf/sigsoft/JamshidiVKS18,DBLP:conf/icse/SiegmundKKABRS12}, including the same sets of performance objectives, options and possible configurations. Since evaluating all systems is unrealistic, we select them according to three criteria:
 
 \begin{itemize}
      \item The systems have been measured on at least two performance objectives.
     \item The systems have both binary and categorical/numeric options or more than 10 options.
    % \item We do not use simple systems, i.e., those with less than 10 configuration options, unless they involve a mix of binary and categorical/numeric options.
     \item The systems should have clear instructions on how they are deployed.
     \item For systems with multiple benchmarks, we use the one with the most deviated performance.

 \end{itemize}

 % As shown in Table~\ref{tb:sys}, the systems studied come from diverse domains while having different performance objectives (two each system), number of options, and search space.
 
 As shown in Table~\ref{tb:sys}, the systems studied are of a diverse nature. To examine \textsc{AdMMO}, we use each of the two performance objectives of a system as the target performance objective in turn while the other serves as the auxiliary performance objective according to Chen and Li~\cite{DBLP:conf/sigsoft/0001Chen21,DBLP:journals/corr/abs-2112-07303}. This gives us 14 system-objective pairs in total. All systems are tuned using the benchmarks setting from prior work (e.g., \textsc{WordCount} for \textsc{Storm})~\cite{DBLP:journals/corr/abs-2112-07303,nair2018finding,DBLP:conf/sigsoft/0001Chen21,DBLP:conf/mascots/MendesCRG20,DBLP:conf/sigsoft/JamshidiVKS18,DBLP:conf/icse/SiegmundKKABRS12}. 

%and \textsc{Coffee} dataset for the neural network in \textsc{Keras}

%Note that a single measurement can be expensive, e.g., it can take up to 341 seconds to measure a configuration on \textsc{MongoDB}. To reduce noise, each measurement is repeated at least twice until the standard deviation of a resulted performance is less than 10\% and the mean is used --- the same method adopted in existing work~\cite{DBLP:journals/corr/abs-2106-02716}.  

We compare all optimizers in Section~\ref{sec:background}, including the improved \textsc{MMO}~\cite{DBLP:journals/corr/abs-2112-07303}.

%~\cite{DBLP:journals/corr/abs-2112-07303,DBLP:conf/sigsoft/0001Chen21}

\begin{table}[t!]
\caption{Configurable software systems and performance objectives studied. $|\mathbfcal{O}|$ and $|\mathbfcal{S}|$ denote the number of options and search space, respectively.}
\label{tb:sys}
\centering
%\footnotesize

  \begin{adjustbox}{max width = 0.8\columnwidth}

\begin{tabular}{llllrr}

\toprule
\textbf{System}&\textbf{Language}&\textbf{Domain}&\textbf{The Two Performance Objectives}&\textbf{$|\mathbfcal{O}|$}&\textbf{$|\mathbfcal{S}|$}\\

\midrule

{\textsc{MariaDB}}&C/C$++$/Perl&{SQL database}&Runtime and CPU load&{10}&{864}\\

{\textsc{Storm}}&Java/Clojure&{Stream process}&Throughput and Latency&{6}&{2,880}\\

{\textsc{VP9}}&C&{Video encoding}&Runtime and CPU load&{12}&{3,008}\\

{\textsc{Keras}}&Python&{Deep learning}&AUC and Inference time&{13}&{12,288}\\

{\textsc{MongoDB}}&C$++$&{Non-SQL database}&Runtime and CPU load&{15}&{6,840}\\

{\textsc{x264}}&C&{Video encoding}&PSNR and Energy consumption&{17}&{53,662}\\

{\textsc{LLVM}}&C$++$&{Compiler}&Runtime and CPU load&{16}&{65,436}\\

% \textsc{MariaDB}~\cite{DBLP:journals/corr/abs-2112-07303}&SQL database&\textsc{O1:} runtime& \textsc{O2:} CPU load&10&864\\

% \textsc{Storm}~\cite{nair2018finding,DBLP:conf/mascots/JamshidiC16,DBLP:conf/sigsoft/0001Chen21}&stream process&\textsc{O1:} throughput& \textsc{O2:} latency&6&2,880\\

% \textsc{VP9}~\cite{DBLP:journals/corr/abs-2112-07303}&video encoding&\textsc{O1:} runtime& \textsc{O2:} CPU load&12&3,008\\

% \textsc{MongoDB}~\cite{DBLP:journals/corr/abs-2112-07303}&no-SQL database&\textsc{O1:} runtime& \textsc{O2:} CPU load&15&6,840\\

% \textsc{Keras}~\cite{DBLP:conf/mascots/MendesCRG20,DBLP:conf/sigsoft/JamshidiVKS18}&deep learning&\textsc{O1:} AUC& \textsc{O2:} inferred time&13&12,288\\

% \textsc{x264}~\cite{nair2018finding,DBLP:conf/icse/SiegmundKKABRS12,DBLP:conf/sigsoft/0001Chen21}&video encoding&\textsc{O1:} PSNR& \textsc{O2:} energy usage&17&53,662\\

% \textsc{LLVM}~\cite{nair2018finding,DBLP:journals/corr/abs-2112-07303}&compiler&\textsc{O1:} runtime& \textsc{O2:} CPU load&16&65,436\\
\bottomrule

%1.92$\times 10^{10}$

\end{tabular}
  \end{adjustbox}
  \vspace{-0.35cm}
\end{table}

\subsection{Tuning Budget}

To avoid noises from the implementation tricks and the nature of different programming languages underpinning the compared optimizers, we measure the ``speed'' of tuning by the number of measurements (a language-independent feature) as suggested in prior work~\cite{nair2018finding}. Notably, only the measurements of distinct configurations can consume the budgets. 

Since the measurement is often expensive, e.g., it can take up to 166 minutes to measure one configuration on \textsc{MariaDB}, the possible budgets are often not unrealistically high compared with some other domains and SBSE problems~\cite{DBLP:conf/sigsoft/0001Chen21,DBLP:journals/corr/abs-2112-07303,DBLP:conf/sc/BehzadLHBPAKS13,DBLP:conf/sigsoft/ShahbazianKBM20}. To match with the realistic settings, we examine different budget sizes including 100, 200, 300, and 400 measurements, denoted as $S_{100}$, $S_{200}$, $S_{300}$, and $S_{400}$, respectively. Note that even with $S_{100}$, in reality, systems like \textsc{MongoDB} will still require more than two full weeks to complete all the unique measurements. As such, the budgets serve as balanced choices between the convergence of optimizers and the limits of available resources in the practical scenarios.

%11 hours

Each measurement is run three times and the average is used. Like prior work~\cite{DBLP:conf/sigsoft/0001Chen21}, we stored the measurements as datasets for reuse (and for expediting the experiments). 
%That is, only the measurements of distinct configurations can consume the budgets. 

\subsection{Evaluation Metrics}

It is worth noting that, since we are interested in optimizing a single performance attribute that is of the most concern, the key evaluation metric would be the achieved performance when the tuning terminates (see Table~\ref{tb:sys}) rather than any multi-objective quality metrics~\cite{DBLP:journals/tse/LiCY22}. We also measure the efficiency of the optimizers in terms of the resources required to achieve a promising performance result, the details of which can be found in Section~\ref{sec:eff}.

% For comparing with the measurement-based optimizers, we set a tuning budget of 400 measurements, which is the smallest budget used among the existing work~\cite{DBLP:conf/sigsoft/0001Chen21,DBLP:journals/corr/abs-2112-07303,DBLP:conf/sc/BehzadLHBPAKS13,DBLP:conf/sigsoft/ShahbazianKBM20}. In contrast to some prior work~\cite{DBLP:conf/sigsoft/0001Chen21,DBLP:journals/corr/abs-2112-07303}, such a relatively small budget is a disadvantage to optimizers based on multi-objectivization like \textsc{AdMMO} due to the need to take an additional objective into account. However, as we will show, \textsc{AdMMO} is efficient even under this constrained setting.

%In contrast to the prior work where the budget that can lead to a reasonable convergence under a system is used~\cite{DBLP:conf/sigsoft/0001Chen21,DBLP:journals/corr/abs-2112-07303},

% As for the evaluation with the model-based optimizers, we use 50 measurements as a tuning budget, where the first 30 measurements are used for training. This is the exact setting used in the work of FLASH~\cite{nair2018finding}, which is one of the optimizers we compare in this paper. It is worth noting that, thanks to the surrogate, a small budget can often be beneficial to the model-based optimizers~\cite{DBLP:conf/mascots/JamshidiC16,nair2018finding,DBLP:conf/icse/0003XC021,DBLP:conf/nips/BergstraBBK11,DBLP:journals/jmlr/ZuluagaKP16}.

\subsection{Optimizer Settings}

Since \textsc{AdMMO} leverages NSGA-II~\cite{Deb2002}---a common multi-objective measurement-based optimizer from SBSE, we apply the binary tournament for mating selection and a population size of 10, together with the boundary mutation and uniformed crossover under the rates of 0.1 and 0.9, respectively, as prior work~\cite{Chen2018FEMOSAA,DBLP:conf/sigsoft/ShahbazianKBM20,DBLP:journals/infsof/ChenLY19}. Those settings are often recommended~\cite{DBLP:conf/icse/0003XC021} and we seek to relieve the dependency on specific components of these optimizers. We set $p=0.3$ (unless otherwise stated), meaning that \textsc{AdMMO} seeks to maintain 30\% of the unique nondominated configurations as this is a generally best setting (we will show in Section~\ref{sec:rq4}); other (less important) parameters are fixed as stated in Section~\ref{sec:method}. \textsc{AdMMO} is implemented using jMetal~\cite{DBLP:journals/aes/DurilloN11} and Opt4J~\cite{DBLP:conf/gecco/LukasiewyczGRT11}. 

For \textsc{GA}, \textsc{MMO} (in the experiments, we compare with the latest improved MMO~\cite{DBLP:journals/corr/abs-2112-07303} that uses $w=1$ and the improved normalization scheme, as it performs better than the original version~\cite{DBLP:conf/sigsoft/0001Chen21}), and \textsc{PMO}, we set the same settings as those for \textsc{AdMMO} while both \textsc{MMO} and \textsc{PMO} are also paired with NSGA-II. For other optimizers, we use exactly the same setting for \textsc{IRACE} as reported by \citet{lopez2016irace} and those for \textsc{ParamILS} from the work of \citet{DBLP:journals/jair/HutterHLS09}. While \textsc{Flash} has no other parameters except for the budgets, \textsc{BOCA} and \textsc{SMAC} require a few settings, for which we use the identical values as those reported in their work~\cite{nair2018finding,DBLP:conf/icse/0003XC021}. A sampling budget of 30 is used for initializing the model in model-based optimizers as recommended by \citet{nair2018finding}.

%The only exception is the top $K$ important configuration options considered in \textsc{BOCA}, for which it was set as $K=8$. This is because we study configurable systems with rather different scales compared with those studied by Chen \textit{et al.}, e.g., \textsc{Storm} has less than 8 options. Therefore, in this work, we set $K$ as 50\% of the number of configuration options for a system, which works well according to our preliminary analysis.

To mitigate stochastic bias, all experiments are repeated 50 runs.

\subsection{Statistical Validation}

We use the recommended non-parametric {Wilcoxon test}~\cite{Wilcoxon1945IndividualCB} with $a=0.05$ to verify the significance of pairwise comparisons between \textsc{AdMMO} and its counterpart over the 50 runs~\cite{ArcuriB11}. Further, we use $\mathbf{\hat{A}_{12}}$~\cite{Vargha2000ACA} to examine the effect size. According to Vargha and Delaney~\cite{Vargha2000ACA}, $0.56\leq \mathbf{\hat{A}_{12}}<0.64$ (or $0.36 < \mathbf{\hat{A}_{12}} \leq 0.44$) indicates a small yet non-trivial effect size while $0.64 \leq \mathbf{\hat{A}_{12}} < 0.71$ (or $0.29 < \mathbf{\hat{A}_{12}} \leq 0.36$) and $\mathbf{\hat{A}_{12}} \geq 0.71$ (or $\mathbf{\hat{A}_{12}} \leq 0.29$) mean a medium and a large effect size, respectively. 

In this work, we say the difference is statistically significant only when $\mathbf{\hat{A}_{12}} \geq 0.56$ (or $\mathbf{\hat{A}_{12}} \leq 0.44$) and $p$-$value <0.05$; otherwise the deviation in the comparison is trivial.

%We use the following methods to interpret the statistical significance of the results: 

% \begin{itemize}

%     \item\textbf{Wilcoxon rank-sum test~\cite{Wilcoxon1945IndividualCB}:} This is used to investigate the statistical significance of the performance objective comparisons over 50 runs under $a=0.05$. It is a non-parametric and non-paired test that is often recommended due to its strong statistical power on pair-wise comparisons~\cite{ArcuriB11}. If the \textit{p-value} $<0.05$, we say that the magnitude of differences is significant.
    
%     %To ensure the resulted differences are not generated from a trivial effect,
%     %To prevent drawing conclusions based on trivial effects in the samples,
%     \item\textbf{$\mathbf{\hat{A}_{12}}$ effect size~\cite{Vargha2000ACA}:} Additionally, we use $\mathbf{\hat{A}_{12}}$ to verify the effect size over 50 runs. When comparing \textsc{AdMMO} with an existing optimizer in this work, $\mathbf{\hat{A}_{12}}>0.5$ denotes that \textsc{AdMMO} wins and it is better for more than 50\% of the runs; $\mathbf{\hat{A}_{12}}<0.5$ means otherwise (\textsc{AdMMO} loses) and an exact $0.5$ denotes a tie. According to Vargha and Delaney, $0.56\leq \mathbf{\hat{A}_{12}}<0.64$ indicates a small yet non-trivial effect size while $0.64 \leq \mathbf{\hat{A}_{12}} < 0.71$ and $\mathbf{\hat{A}_{12}} \geq 0.71$ mean a medium and a large effect size, respectively. 

% \end{itemize}

\section{Results and Analysis}
\label{sec:result}

We now present and discuss the experiment results.
%\input{tables/rq1-1}
%\input{tables/rq1-2}

% CS Table for RQ2
\begin{table*}[t!]
\caption{Comparing the optimizers on the average of normalized target performance over 50 runs (smaller value is preferred). The bottom-right shows the \% of statistically significant comparisons (and $\hat{A}_{12}$) against \textsc{AdMMO} over 56 cases. The \setlength{\fboxsep}{1.5pt}\colorbox{steel!30}{blue cells} highlight the best in a case. The summarized raw results of all cases can be found at our repository: \url{https://github.com/ideas-labo/admmo/blob/main/supp.pdf}.}
    \label{tb:rq1}
   % \footnotesize
    \setlength{\tabcolsep}{1mm}
  \begin{center}
    \begin{adjustbox}{max width = 1\textwidth}

   \begin{tabular}{l||llll||llll||llll||llll}

     \toprule
     
            \cellcolor[gray]{1}\multirow{1}{*}{\textbf{Optimizer}}  & \cellcolor[gray]{1}\textbf{$S_{100}$} & \cellcolor[gray]{1}\textbf{$S_{200}$} & \cellcolor[gray]{1}\textbf{$S_{300}$} & \cellcolor[gray]{1}\textbf{$S_{400}$} & \cellcolor[gray]{1}\textbf{$S_{100}$} & \cellcolor[gray]{1}\textbf{$S_{200}$} & \cellcolor[gray]{1}\textbf{$S_{300}$} & \cellcolor[gray]{1}\textbf{$S_{400}$} & \cellcolor[gray]{1}\textbf{$S_{100}$} & \cellcolor[gray]{1}\textbf{$S_{200}$} & \cellcolor[gray]{1}\textbf{$S_{300}$} & \cellcolor[gray]{1}\textbf{$S_{400}$} & \cellcolor[gray]{1}\textbf{$S_{100}$} & \cellcolor[gray]{1}\textbf{$S_{200}$} & \cellcolor[gray]{1}\textbf{$S_{300}$} & \cellcolor[gray]{1}\textbf{$S_{400}$}  \\
            \cline{1-17}
            %&
            %\multicolumn{4}{c|}{\textsc{MariaDB-O1}}&
            %\multicolumn{4}{c|}{\textsc{MariaDB-O2}}&
            %\multicolumn{4}{c|}{\textsc{Storm-O1}}&
            %\multicolumn{4}{c}{\textsc{Storm-O2}}\\
            %\hline
\cellcolor{black!10}&\multicolumn{4}{c||}{\cellcolor{black!10}\textsc{MariaDB}-Runtime}&\multicolumn{4}{c||}{\cellcolor{black!10}\textsc{MariaDB}-CPU}&\multicolumn{4}{c||}{\cellcolor{black!10}\textsc{Storm}-Throughtput}&\multicolumn{4}{c}{\cellcolor{black!10}\textsc{Storm}-Latency}\\
\cline{1-17}

\textsc{IRACE}&0.4774&0.3543&0.2813&0.2670&0.4933&0.3122&0.1673&0.1085&1.0000&0.6713&0.4944&0.3125&0.2317&0.1384&0.0932&0.0644\\
\textsc{GA}&0.5061&0.1138&0.0867&0.0702&0.2857&0.0715&0.0352&0.0091&0.1671&0.0741&0.0571&0.0429&\cellcolor{steel!30}0.0124&\cellcolor{steel!30}0.0024&\cellcolor{steel!30}0.0000&\cellcolor{steel!30}0.0000\\
\textsc{RS}&0.5172&0.3473&0.2468&0.1849&0.4280&0.2812&0.2207&0.1343&0.7768&0.4745&0.3254&0.2803&0.1792&0.1169&0.0922&0.0692\\
\textsc{ParamILS}&1.0000&0.5762&0.3904&0.1650&1.0000&0.7029&0.5779&0.5326&0.7240&0.4206&0.1829&0.0665&1.0000&0.7186&0.3986&0.2560\\

\textsc{MMO}&0.2166&0.1045&0.0712&0.0477&\cellcolor{steel!30}0.2761&\cellcolor{steel!30}0.0589&0.0268&0.0052&\cellcolor{steel!30}0.1173&0.0584&0.0435&0.0329&0.0427&0.0016&\cellcolor{steel!30}0.0000&\cellcolor{steel!30}0.0000\\

\textsc{PMO}&0.5124&0.1573&0.0834&0.0515&0.4505&0.3112&0.2365&0.1544&0.6520&0.1887&0.0914&0.0480&0.2813&0.0740&0.0216&0.0108\\

\textsc{Flash}&0.4543&0.2270&0.1815&0.1657&0.4979&0.0941&\cellcolor{steel!30}0.0033&\cellcolor{steel!30}0.0000&0.8734&0.0481&0.0213&0.0058&0.4423&0.0913&0.0913&0.0913\\

\textsc{BOCA}&0.5545&0.3208&0.2437&0.1924&0.4230&0.2808&0.2018&0.1897&0.3323&0.1725&0.0928&0.0485&0.2510&0.1325&0.0982&0.0814\\

\textsc{SMAC}&0.5573&0.3956&0.3155&0.2699&0.5258&0.3187&0.2214&0.1871&0.9983&0.6810&0.4461&0.2944&0.2198&0.1040&0.0768&0.0659\\

\textsc{AdMMO}&\cellcolor{steel!30}0.2122&\cellcolor{steel!30}0.0531&\cellcolor{steel!30}0.0093&\cellcolor{steel!30}0.0000&0.2823&0.1193&0.0654&0.0268&0.1237&\cellcolor{steel!30}0.0299&\cellcolor{steel!30}0.0173&\cellcolor{steel!30}0.0000&0.0416&0.0048&\cellcolor{steel!30}0.0000&\cellcolor{steel!30}0.0000\\

\hline
\cellcolor{black!10}&\multicolumn{4}{c||}{\cellcolor{black!10}\textsc{VP9}-Runtime}&\multicolumn{4}{c||}{\cellcolor{black!10}\textsc{VP9}-CPU}&\multicolumn{4}{c||}{\cellcolor{black!10}\textsc{Keras}-AUC}&\multicolumn{4}{c}{\cellcolor{black!10}\textsc{Keras}-Time}\\
\cline{1-17}

\textsc{IRACE}&0.1069&0.0665&0.0436&0.0296&0.0486&0.0216&0.0181&0.0162&0.7580&0.6093&0.4898&0.3819&0.0696&0.0424&0.0366&0.0309\\

\textsc{GA}&0.1315&0.0565&0.0246&0.0060&0.0537&0.0230&0.0120&0.0075&0.7959&0.6997&0.6501&0.6239&0.0082&0.0033&0.0022&0.0019\\

\textsc{RS}&0.1625&0.0625&0.0281&0.0240&0.0587&0.0310&0.0237&0.0188&0.8017&0.6706&0.5685&0.4927&0.0799&0.0387&0.0281&0.0242\\

\textsc{ParamILS}&1.0000&0.9901&0.9711&0.8899&1.0000&0.9833&0.9773&0.8772&1.0000&0.9913&0.9913&0.9854&1.0000&0.6231&0.4191&0.3367\\

\textsc{MMO}&0.1925&0.0512&0.0251&0.0004&0.0831&0.0531&0.0425&0.0351&0.8455&0.7638&0.6880&0.6122&0.0068&0.0030&0.0020&0.0013\\

\textsc{PMO}&0.1791&0.1021&0.0722&0.0519&0.0820&0.0585&0.0438&0.0371&\cellcolor{steel!30}0.2711&\cellcolor{steel!30}0.0000&\cellcolor{steel!30}0.0000&\cellcolor{steel!30}0.0000&0.0128&0.0023&0.0017&0.0013\\

\textsc{Flash}&0.1504&0.0647&0.0399&0.0162&0.0503&0.0290&0.0208&0.0164&0.9708&0.9708&0.9708&0.9708&0.0799&0.0202&0.0186&0.0186\\

\textsc{BOCA}&0.1627&0.0600&0.0459&0.0237&0.0748&0.0428&0.0264&0.0172&0.7580&0.6006&0.5510&0.4898&0.0479&0.0280&0.0232&0.0193\\

\textsc{SMAC}&0.1746&0.1004&0.0489&0.0259&0.0530&0.0285&0.0211&0.0190&0.7522&0.5685&0.5219&0.4490&0.0601&0.0391&0.0334&0.0293\\

\textsc{AdMMO}&\cellcolor{steel!30}0.0860&\cellcolor{steel!30}0.0451&\cellcolor{steel!30}0.0059&\cellcolor{steel!30}0.0000&\cellcolor{steel!30}0.0447&\cellcolor{steel!30}0.0193&\cellcolor{steel!30}0.0034&\cellcolor{steel!30}0.0000&0.7085&0.5569&0.4402&0.4169&\cellcolor{steel!30}0.0059&\cellcolor{steel!30}0.0012&\cellcolor{steel!30}0.0003&\cellcolor{steel!30}0.0000\\

\hline
\cellcolor{black!10}&\multicolumn{4}{c||}{\cellcolor{black!10}\textsc{MongoDB}-Runtime}&\multicolumn{4}{c||}{\cellcolor{black!10}\textsc{MongoDB}-CPU}&\multicolumn{4}{c||}{\cellcolor{black!10}\textsc{x264}-PSNR}&\multicolumn{4}{c}{\cellcolor{black!10}\textsc{x264}-Energy}\\
\cline{1-17}

\textsc{IRACE}&0.5129&0.2621&0.1853&0.1013&0.4202&0.1947&0.1079&0.0809&0.3351&0.3351&0.3351&0.3351&0.0077&0.0077&0.0077&0.0077\\

\textsc{GA}&0.5056&0.2795&0.1965&0.1041&0.4777&0.2792&0.1536&0.0519&0.1559&\cellcolor{steel!30}0.0487&0.0325&0.0179&\cellcolor{steel!30}0.0029&0.0018&0.0015&0.0015\\

\textsc{RS}&0.4724&0.2829&0.1900&0.1354&0.4741&0.2624&0.1837&0.0642&1.0000&1.0000&1.0000&1.0000&0.0961&0.0961&0.0961&0.0961\\

\textsc{ParamILS}&1.0000&1.0000&1.0000&1.0000&1.0000&0.9936&0.9936&0.9936&1.0000&1.0000&1.0000&1.0000&1.0000&1.0000&1.0000&1.0000\\

\textsc{MMO}&0.5541&0.3459&0.1746&0.0685&0.4844&0.2831&0.1862&0.1185&0.1918&0.0708&0.0214&\cellcolor{steel!30}0.0000&0.0044&0.0016&0.0014&0.0011\\

\textsc{PMO}&0.5140&0.2656&0.1681&0.0804&0.4675&0.2236&0.1475&0.0604&0.2510&0.1096&0.0574&0.0043&0.0051&0.0021&0.0011&0.0007\\

\textsc{Flash}&0.5821&0.3013&0.2191&0.1442&0.4782&0.2191&0.1154&0.0803&0.3351&0.3351&0.3351&0.3351&0.0077&0.0077&0.0077&0.0077\\

\textsc{BOCA}&0.6089&0.4453&0.2763&0.1810&0.4570&0.2776&0.1724&0.1157&0.3351&0.3351&0.3351&0.3351&0.0077&0.0077&0.0077&0.0077\\

\textsc{SMAC}&0.5709&0.3439&0.2264&0.1195&0.5364&0.2590&0.1605&0.0931&0.3351&0.3351&0.3351&0.3351&0.0077&0.0077&0.0077&0.0077\\

\textsc{AdMMO}&\cellcolor{steel!30}0.4332&\cellcolor{steel!30}0.1317&\cellcolor{steel!30}0.0361&\cellcolor{steel!30}0.0000&\cellcolor{steel!30}0.3829&\cellcolor{steel!30}0.1611&\cellcolor{steel!30}0.0269&\cellcolor{steel!30}0.0000&\cellcolor{steel!30}0.1424&0.0781&\cellcolor{steel!30}0.0194&0.0042&0.0032&\cellcolor{steel!30}0.0012&\cellcolor{steel!30}0.0001&\cellcolor{steel!30}0.0000\\

%\cline{0-10}
\hline

\cellcolor{black!10}&\multicolumn{4}{c||}{\cellcolor{black!10}\textsc{LLVM}-Runtime}&\multicolumn{4}{c||}{\cellcolor{black!10}\textsc{LLVM}-CPU}&\multicolumn{2}{l}{\textbf{vs. \textsc{AdMMO}}}&\multicolumn{2}{c}{\textbf{trivial}}&\multicolumn{1}{c}{\textbf{small}}&\multicolumn{2}{c}{\textbf{medium}}&\multicolumn{1}{c}{\textbf{large}} \\

\cline{1-17}

\textsc{IRACE}&0.5210&0.4124&0.3804&0.3511&0.6465&0.4050&0.2831&0.2099&\multicolumn{2}{l}{\textsc{IRACE}}&\multicolumn{2}{c}{7\%}&\multicolumn{1}{c}{2\%}&\multicolumn{2}{c}{2\%}&\multicolumn{1}{c}{89\%} \\

\textsc{GA}&\cellcolor{steel!30}0.0774&\cellcolor{steel!30}0.0000&\cellcolor{steel!30}0.0000&\cellcolor{steel!30}0.0000&0.4918&0.2615&0.1821&0.1608&\multicolumn{2}{l}{\textsc{GA}}&\multicolumn{2}{c}{17\%}&\multicolumn{1}{c}{9\%}&\multicolumn{2}{c}{14\%}&\multicolumn{1}{c}{60\%} \\

\textsc{RS}&0.5998&0.4709&0.4278&0.3937&0.8072&0.5448&0.3880&0.2557&\multicolumn{2}{l}{\textsc{RS}}&\multicolumn{2}{c}{2\%}&\multicolumn{1}{c}{5\%}&\multicolumn{2}{c}{2\%}&\multicolumn{1}{c}{91\%} \\

\textsc{ParamILS}&1.0000&0.9729&0.9510&0.9307&1.0000&0.9191&0.8688&0.8171&\multicolumn{2}{l}{\textsc{ParamILS}}&\multicolumn{2}{c}{0\%}&\multicolumn{1}{c}{0\%}&\multicolumn{2}{c}{4\%}&\multicolumn{1}{c}{96\%} \\

\textsc{MMO}&0.0915&\cellcolor{steel!30}0.0000&\cellcolor{steel!30}0.0000&\cellcolor{steel!30}0.0000&0.4876&0.2396&0.1680&0.1164&\multicolumn{2}{l}{\textsc{MMO}}&\multicolumn{2}{c}{16\%}&\multicolumn{1}{c}{9\%}&\multicolumn{2}{c}{21\%}&\multicolumn{1}{c}{54\%} \\

\textsc{PMO}&0.2613&0.0956&0.0276&0.0155&0.5450&0.3317&0.2534&0.2112&\multicolumn{2}{l}{\textsc{PMO}}&\multicolumn{2}{c}{4\%}&\multicolumn{1}{c}{5\%}&\multicolumn{2}{c}{11\%}&\multicolumn{1}{c}{80\%} \\

\textsc{Flash}&0.4282&\cellcolor{steel!30}0.0000&\cellcolor{steel!30}0.0000&\cellcolor{steel!30}0.0000&\cellcolor{steel!30}0.4651&0.2760&0.1306&0.0277&\multicolumn{2}{l}{\textsc{Flash}}&\multicolumn{2}{c}{14\%}&\multicolumn{1}{c}{2\%}&\multicolumn{2}{c}{4\%}&\multicolumn{1}{c}{80\%} \\

\textsc{BOCA}&0.3971&0.3287&0.2961&0.2720&0.7422&0.4786&0.3340&0.2130&\multicolumn{2}{l}{\textsc{BOCA}}&\multicolumn{2}{c}{5\%}&\multicolumn{1}{c}{2\%}&\multicolumn{2}{c}{0\%}&\multicolumn{1}{c}{93\%} \\

\textsc{SMAC}&0.5058&0.3888&0.3204&0.2910&0.7112&0.4817&0.3210&0.2156&\multicolumn{2}{l}{\textsc{SMAC}}&\multicolumn{2}{c}{5\%}&\multicolumn{1}{c}{2\%}&\multicolumn{2}{c}{2\%}&\multicolumn{1}{c}{91\%} \\

\textsc{AdMMO}&0.1156&\cellcolor{steel!30}0.0000&\cellcolor{steel!30}0.0000&\cellcolor{steel!30}0.0000&0.5216&\cellcolor{steel!30}0.2378&\cellcolor{steel!30}0.0984&\cellcolor{steel!30}0.0000&\multicolumn{8}{c}{\textbf{\% of significant cases vs. \textsc{AdMMO} over all 56 cases.}}\\

\bottomrule
%\cline{0-10}

        \end{tabular}
        
  \end{adjustbox}
   \end{center}
   \vspace{-0.5cm}
\end{table*}

\subsection{Effectiveness}

\subsubsection{Method}

For \textbf{RQ1}, we compare \textsc{AdMMO} against all optimizers specified in Section~\ref{sec:background}. Since there are 14 system-objective pairs and 4 different budgets, we obtain 56 cases of comparisons. 

%Statistical tests are used for every comparison between AdMMO and a state-of-the-art optimizer.

\subsubsection{Findings}

As can be seen from Table~\ref{tb:rq1}, it is clear that \textsc{AdMMO} achieves considerably better performance over the state-of-the-art optimizers, as it is ranked the best for 71\% of the cases (40/56). In contrast to measurement-based optimizers such as \textsc{IRACE}, \textsc{GA}, \textsc{RS}, and \textsc{ParamILS}, the best improvement of \textsc{AdMMO} ranges from 21\% to 100\%. We also see that \textsc{RS} and \textsc{ParamILS} perform badly compared with the others, this is because the former has little heuristic to guide the tuning for better performance despite that it is good at handling local optima, while the latter suffers severely from the local optima issues due to the nature of its local search. Compared with the model-based optimizers, \textsc{AdMMO} again has superior results between 17\% and 45\% best improvements. Those optimizers appear to be also largely affected by the local optima issue as seen from cases of, e.g., on the \textit{PSNR} for \textsc{x264}. When directly comparing with \textsc{MMO}, \textsc{AdMMO} also obtain much better results with up to 20\% improvement. This evidences that the adaptive weight is indeed beneficial and necessary. With no surprise, \textsc{PMO} is much more inferior than \textsc{AdMMO}, except for one system-objective pair: i.e., \textit{AUC} for \textsc{Keras}. Despite being rare, this is indeed possible: in this case, lower inference time often means better AUC (but not vice versa), hence optimizing the former can also benefit the latter. As such, the drawback of \textsc{PMO} we discussed in Section~\ref{sec:pmo} would become blurred.

%talk about trace

All the above results are concluded with high statistical significance, as can be seen at the bottom-right of Table~\ref{tb:rq1}: majority of the cases have significant comparison ($\mathbf{\hat{A}_{12}} \geq 0.56$ or $\mathbf{\hat{A}_{12}} \leq 0.44$ while $p$-$value <0.05$) and exhibit medium to large effect size, i.e., $\mathbf{\hat{A}_{12}} \geq 0.64$ (or $\mathbf{\hat{A}_{12}} \leq 0.36$).

\begin{figure}[t!]
\centering
\begin{subfigure}{.32\columnwidth}
\includegraphics[width=\columnwidth]{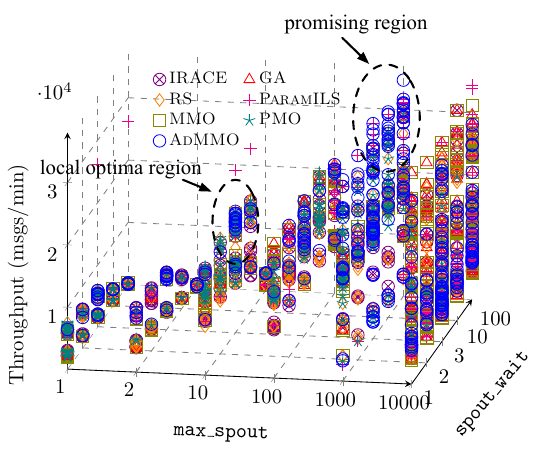}
\vspace{0.1cm}
\subcaption{explored configurations (measurement-based optimizers)}
\end{subfigure}
%\hspace{-0.25cm}
\begin{subfigure}{.32\columnwidth}
%\vspace{-1.8cm}
\includegraphics[width=\columnwidth]{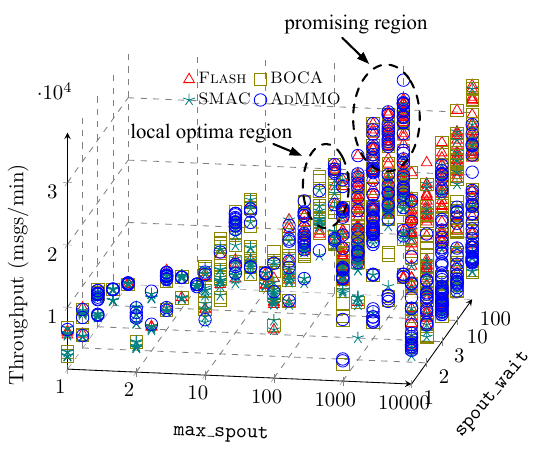}
\vspace{0.1cm}
\subcaption{explored configurations (model-based optimizers)}
\end{subfigure}
\hspace{-0.2cm}
\begin{minipage}[b]{0.35\columnwidth}
    \centering
     \footnotesize
      % \begin{adjustbox}{max width = \columnwidth}
   
        \begin{subfigure}{\columnwidth}
\includestandalone[width=\columnwidth]{figures/detail-weight}
\vspace{-0.5cm}
\subcaption{trajectory of $w$}
\end{subfigure}

\begin{subfigure}{\columnwidth}
\includestandalone[width=\columnwidth]{figures/detail-p}
\vspace{-0.5cm}
\subcaption{trajectory of $p'$}
\end{subfigure}
    % \end{adjustbox}
    \end{minipage}
% \begin{subfigure}{.3\columnwidth}
% \includestandalone[width=\columnwidth]{figures/detail-weight}
% \subcaption{trajectory of $w$}
% \end{subfigure}
% %\hspace{-0.1cm}
% \begin{subfigure}{.3\columnwidth}
% \includestandalone[width=\columnwidth]{figures/detail-p}
% \subcaption{proportion}
% \end{subfigure}
\vspace{-0.1cm}
\caption{Example run for system \textsc{Storm}. (a) and (b) are the projected landscape of explored configurations by all optimizers; (c) and (d) are the trajectories of $w$ and the actual proportion of nondominated configurations $p'$, respectively. The dashed line in (d) represents the expected proportion of nondominated configurations.}
\label{fig:details}
\vspace{-0.4cm}
\end{figure}
%(some optimizers are omitted as they are much more inferior than the others)

%To take a closer look, Figure~\ref{fig:details} shows all explored configurations for a run as well as the actual weight value and proportion of nondominated configurations.

To take a closer look, Figure~\ref{fig:details} shows all explored configurations together with how the $w$ and $p'$ change for a run. From Figures~\ref{fig:details}a and~\ref{fig:details}b, we see that \textsc{AdMMO} often successfully explores points around the promising regions, while the others can easily be trapped as some locally undesired areas. Unlike \textsc{MMO}, in Figures~\ref{fig:details}c and~\ref{fig:details}d, it is clear that \textsc{AdMMO} dynamically adapts the weights throughout the tuning and manages to maintain the proportion of nondominated configurations near a certain level, i.e., 30\% in this case ($p=0.3$). 

%This is the key reason behind its superiority.

%Therefore, we say:

\begin{quotebox}
   \noindent
   \textit{\textbf{RQ1:} \textsc{AdMMO} is effective as it outperforms \textsc{MMO} and other optimizers in 71\% cases with considerable improvements and high statistical significance.}
   %\textit{\textbf{RQ1:} AdMMO significantly outperforms the original MMO and other state-of-the-art optimizers, as adapting the weights helps to find considerably better configurations in general.}
\end{quotebox}

\subsection{Efficiency}
\label{sec:eff}

\subsubsection{Method} 

To answer \textbf{RQ2}, we examine the efficiency of \textsc{AdMMO} by comparing how much less (or more) resources it needs to reach the best result obtained by the other optimizers. Similar to Chen and Li~\cite{DBLP:conf/sigsoft/0001Chen21}, we follow the steps below:

\begin{enumerate}

\item Set a baseline, $b$, as the smallest number of measurements (up to 400) that one other optimizer needs to reach its best average (over 50 runs) of the target performance objective (says $T$).

\item For \textsc{AdMMO}, find the smallest number of measurements, $m$, at which the mean of the target performance objective (over 50 runs) is at least the same as $T$.

\item Following the metric used by Gao \textit{et al.}~\cite{DBLP:conf/icse/GaoZ0LY21}, the speedup of \textsc{AdMMO} over its counterpart, i.e., $s = {b \over m}$, is reported. 

%The percentage saving of resource, i.e., $s=(1-{m \over b}) \times 100\%$, is reported. 

\end{enumerate}

\begin{figure}[t!]
\centering
\includegraphics[width=0.85\columnwidth]{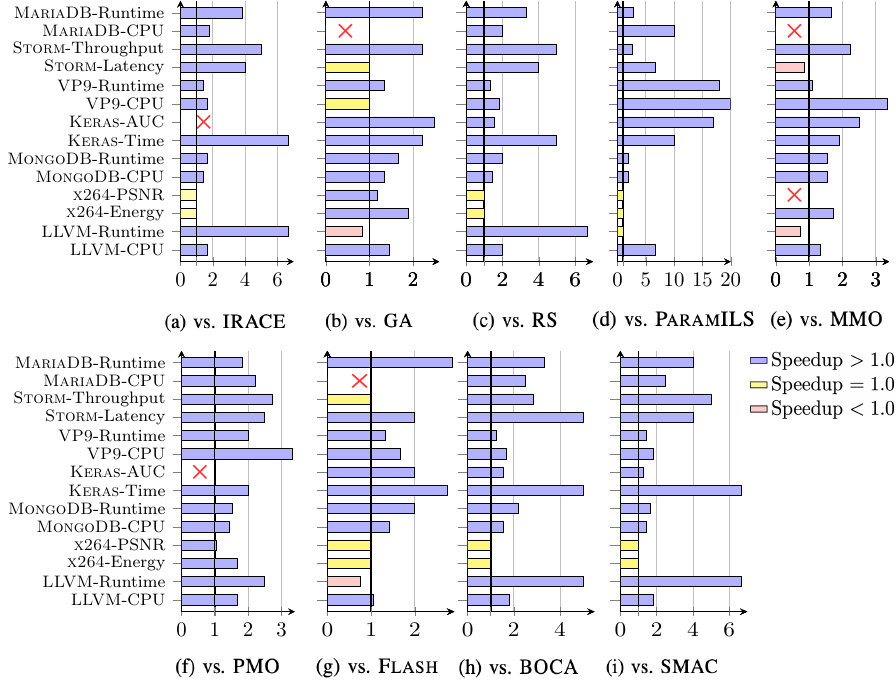}
\caption{The speedup ($s$) of \textsc{AdMMO} over the state-of-the-art optimizers on the budget of 400. \textcolor{red}{\xmark}~denotes a case where \textsc{AdMMO} cannot achieve the same result when the budget runs out.}
\label{fig:resources}
\vspace{-0.45cm}
\end{figure}

As such, for \textsc{AdMMO} to be efficient, we expect $s \geq 1$.

\subsubsection{Findings}

Figure~\ref{fig:resources} plots the results, from which we observe that compared with measurement-based optimizers (e.g., \textsc{IRACE}), \textsc{AdMMO} has $s \geq 1$ in the majority of the system-objective pairs (at least 12 out of 14), and it is at most of the time with $s > 1$ (at least 10 cases). Remarkably, the best speedup ranges from $2.2\times$ to $20 \times$ over the others. In contrast to the model-based \textsc{Flash}, \textsc{BOCA}, and \textsc{SMAC}, even without a surrogate, \textsc{AdMMO} can still obtain $s \geq 1$ for 12 to 14 pairs and at least 10 pairs of $s > 1$ with up to $6.7 \times$ speedup. \textsc{AdMMO} also significantly boosts \textsc{MMO} and \textsc{PMO}, achieving $s>1$ on 10 and 13 out of 14 pairs, respectively. This means that the weight adaptation also enables more efficient resource consumption in the tuning. 

%Overall, we conclude that:

% \begin{itemize}
%     \item Compared with measurement-based optimizers, AdMMO has $s \geq 1$ in the majority of the cases (at least 12 out of 14), and it is at most of the time with $s > 1$ (at least 10 cases). Remarkably, the best speedup ranges from $2.2\times$ to $20 \times$ over the others.
%     \item AdMMO also significantly boosts MMO, as it achieves $s>1$ on 10 out of 14 cases. This means that the weight adaptation also enables more efficient resource consumption in the tuning.
%     \item In contrast to the model-based FLASH and BOCA, even without a surrogate, AdMMO can still obtain $s \geq 1$ for 11 or 12 cases and at least 7 cases of $s > 1$ with up to $1.8 \times$ speedup for both.
% \end{itemize}

%\input{tables/rq3}

\begin{quotebox}
   \noindent
   \textit{\textbf{RQ2:} \textsc{AdMMO} is efficient, achieving $s > 1$ on at least 10 out of 14 system-objective pairs with the best speedup between $2.2\times$ and $20 \times$ against the state-of-the-art optimizers.}
\end{quotebox}

\subsection{Ablation Analysis}
\label{sec:ablation}

\subsubsection{Method}

To further study what parts in \textsc{AdMMO} can help to achieve the above results, in \textbf{RQ3}, we conduct an ablation analysis that converts \textsc{AdMMO} into the following variants: 

%each with one contribution of this paper being switched off:

\begin{itemize}
    \item \textbf{\textsc{AdMMO}$_i$}: A variant that \underline{\textbf{i}}ndistinguishes duplicate configurations in selection (Figure~\ref{fig:dup-exp}a). 
    \item \textbf{\textsc{AdMMO}$_r$}: A variant that \underline{\textbf{r}}emove all but one duplicates in selection (Figure~\ref{fig:dup-exp}b). 
    \item \textbf{\textsc{AdMMO}$_c$}: A variant that \underline{\textbf{c}}onstantly runs weight adaptation throughout the tuning, i.e., without the progressive trigger.
    
   % without progressive trigger, i.e., the weight adaptation runs constantly throughout the tuning.

    %Both the duplicate removal in the weight adaptation and the duplicate-aware selection are switched off.
    
    % and hence they are included in the normal survival selection of NSGA-II under MMO
\end{itemize}
 
Their performance is compared with the original \textsc{AdMMO}.

\subsubsection{Findings}

From Figure~\ref{fig:ablation}, we see that all variants are clearly much inferior to \textsc{AdMMO} regardless of the budgets. This is also supported by the statistical test and $\mathbf{\hat{A}_{12}}$, where the majority of the comparisons are statistically significant with medium to large effect sizes.

Interestingly, we observe that \textsc{AdMMO}$_i$ performs better than \textsc{AdMMO}$_r$ till around 200 budget size because, before this point, the number of duplicates might still be acceptable while it can be more harmful to preserve configurations dominated by those high-quality duplicate configurations that have been omitted, as we discussed in Section~\ref{sec:handling}. Yet, \textsc{AdMMO}$_i$ worsens faster with a larger budget. This makes sense since the more measurements, the higher the likelihood of involving too many duplicates as the tuning converges, hence hindering the effectiveness of weight adaptation. 

%As a result, we say: 

%removing all but one duplicate configuration would ignore some high-quality ones, leading to the preservation of some undesired configurations, which is more harmful as we discussed in Section~\ref{sec:handling}. 

% The results are illustrated in Table~\ref{tb:rq3}, from which we see that:

% \begin{itemize}
%     \item AdMMO$_{c}$ loses to AdMMO for 12 out of 14 cases (statistically significant on 9 of them) with considerable degradation. This means that overfitting the weight in MMO would often lead to harmful outcomes. 
%     \item AdMMO$_d$ is inferior to AdMMO on 9 out of the 14 cases (8 are statistically significant) with significantly degraded outcomes. This suggests that leaving duplicates unmanaged is detrimental to the weight adaptation and the tuning process.
% \end{itemize}

\begin{quotebox}
   \noindent
   \textit{\textbf{RQ3:} Both progressive trigger and partial duplicate retention can greatly improve \textsc{AdMMO} across different systems, objectives and budgets.}
\end{quotebox}  

\begin{figure}[t!]
\centering

 \begin{minipage}[b]{0.32\columnwidth}
   \begin{tikzpicture}
  \begin{axis}[
  width = 5cm,
  height = 5cm,
  xtick={1,2,3,4},
xticklabels={$S_{100}$,$S_{200}$,$S_{300}$,$S_{400}$},
  %  xlabel=\% of nondominated configurations,
  %xlabel=$p$,
    ylabel={Normalized performance},
      axis x line  = bottom,
    axis y line  = left  ,
    legend cell align={left},
    legend style={draw=none,fill=none,at={(1,1.05)}},             
  ]

    \addplot [mark=diamond,black,name path=K] plot coordinates {

(1,0.7530142857142855)
(2,0.2890714285714286)
(3,0.17395)
(4,0.1461857142857143)
 };

   \addplot [mark=square,red,name path=G] plot coordinates {
(1,0.8039357142857143)
(2,0.3486142857142857)
(3,0.17667857142857143)
(4,0.09372857142857142)
 };

    \addplot [mark=o,teal,name path=J] plot coordinates {
(1,0.9618642857142856)
(2,0.4510642857142857)
(3,0.24452142857142858)
(4,0.1537714285714286)
 };

   \addplot [mark=star,blue,name path=C] plot coordinates {
(1,0.6700357142857143)
(2,0.22134999999999996)
(3,0.06299999999999999)
(4,0.013935714285714286)

 };

\addplot [no marks,black!20,name path=M] plot coordinates {
(1,0.8022374052378064)
(2,0.332648616046317)
(3,0.2104524862865811)
(4,0.18605388717979782)

 };
 \addplot [no marks,black!20,name path=N] plot coordinates {
(1,0.7037911661907646)
(2,0.24549424109654014)
(3,0.1374475137134189)
(4,0.10631754139163078)
 };

\addplot [no marks,red!20,name path=E] plot coordinates {
(1,0.837701024656009)
(2,0.39115095399810934)
(3,0.20627147973957066)
(4,0.11884157811746211)
 };
 \addplot [no marks,red!20,name path=F] plot coordinates {
(1,0.7701704039154196)
(2,0.3060776174304621)
(3,0.1470856631175722)
(4,0.06861556473968074)
 };

 \addplot [no marks,teal!20,name path=H] plot coordinates {
(1,0.988221025175277)
(2,0.5138484800784197)
(3,0.2847849510540849)
(4,0.19124281960208817)
 };
 \addplot [no marks,teal!20,name path=I] plot coordinates {
(1,0.9355075462532942)
(2,0.38828009135015173)
(3,0.20425790608877228)
(4,0.11630003754076901)
 };

\addplot [no marks,blue!20,name path=A] plot coordinates {
(1,0.717838908405729)
(2,0.24926120349630668)
(3,0.07923938271672934)
(4,0.023459195508797263)
 };
 \addplot [no marks,blue!20,name path=B] plot coordinates {
(1,0.6222325201656996)
(2,0.19343879650369325)
(3,0.046760617283270633)
(4,0.004412233062631308)
 };

   %% filling
   \addplot[blue!20,opacity=0.5] fill between[of=B and C];
    \addplot[blue!20,opacity=0.5] fill between[of=A and C];
    
       \addplot[red!20,opacity=0.5] fill between[of=E and G];
    \addplot[red!20,opacity=0.5] fill between[of=F and G];

        \addplot[teal!20,opacity=0.5] fill between[of=H and J];
    \addplot[teal!20,opacity=0.5] fill between[of=I and J];

        \addplot[teal!20,opacity=0.5] fill between[of=M and K];
    \addplot[teal!20,opacity=0.5] fill between[of=N and K];

\addlegendentry{\textsc{AdMMO}$_i$}
\addlegendentry{\textsc{AdMMO}$_r$}
\addlegendentry{\textsc{AdMMO}$_c$}
\addlegendentry{\textsc{AdMMO}}
%\addlegendentry{\textsc{Flash}}
    
  \end{axis}

\end{tikzpicture}
\subcaption{overall performance}
\end{minipage}
%\hspace{0.8cm}
%\begin{subfigure}{.51\columnwidth}
% \begin{minipage}[b]{0.47\columnwidth}
 \begin{minipage}[b]{0.65\columnwidth}
    \centering
     %\footnotesize
     %  \begin{adjustbox}{max width = 0.6\columnwidth}
     \begin{adjustbox}{max width = 1\columnwidth}
     \setlength{\tabcolsep}{0.85mm}
   %  \begin{tabular}{llccc}
   %  \multicolumn{5}{c}{\textbf{\% of statistically significant cases vs. \textsc{AdMMO}}}\\ \hline
   %  &  & \textbf{\textsc{AdMMO}$_i$} & \textbf{\textsc{AdMMO}$_r$}  & \textbf{\textsc{AdMMO}$_c$} \\ \hline
   % \multirow{4}{*}{$S_{100}$}& trivial   &  14\% & 7\% &  0\% \\
   %  & small   &  7\% & 21\% & 7\% \\
   %  & medium &  21\% & 29\% &  0\% \\
   %  & large  &  58\% & 43\% &  93\% \\ \hline
   % \multirow{4}{*}{$S_{200}$}& trivial   &  14\% & 22\% &  14\% \\
   %  & small   &  22\% & 14\% &  7\% \\
   %  & medium &  14\% & 14\% &  7\% \\
   %  & large  &  50\% & 50\% &  72\% \\ \hline
   % \multirow{4}{*}{$S_{300}$}& trivial   &  14\% & 21\%  &  14\%\\
   %  & small   &  7\% & 0\%  &  14\%\\
   %  & medium &  0\% & 21\% &  7\% \\
   %  & large  &  79\% & 58\% &  65\% \\ \hline
   % \multirow{4}{*}{$S_{400}$}& trivial   &  14\% & 14\% &  14\% \\
   %  & small   &  7\% & 0\% &  21\% \\
   %  & medium &  0\% & 21\% &  0\% \\
   %  & large  &  88\% & 65\% &  65\% \\  \hline

   %    \end{tabular}

   \begin{tabular}{llccc||llccc}
 
    \multicolumn{10}{c}{\textbf{\% of statistically significant cases vs. \textsc{AdMMO}}}\\ \\   \toprule
    &  & \textbf{\textsc{AdMMO}$_i$} & \textbf{\textsc{AdMMO}$_r$}  & \textbf{\textsc{AdMMO}$_c$}& &  & \textbf{\textsc{AdMMO}$_i$} & \textbf{\textsc{AdMMO}$_r$}  & \textbf{\textsc{AdMMO}$_c$} \\ \midrule
   \multirow{4}{*}{$S_{100}$}& trivial   &  14\% & 7\% &  0\% &  \multirow{4}{*}{$S_{300}$}& trivial   &  14\% & 21\%  &  14\%\\
    & small   &  7\% & 21\% & 7\% & & small   &  7\% & 0\%  &  14\% \\
    & medium &  21\% & 29\% &  0\% & & medium &  0\% & 21\% &  7\%  \\
    & large  &  58\% & 43\% &  93\% & & large  &  79\% & 58\% &  65\% \\ \hline
   \multirow{4}{*}{$S_{200}$}& trivial   &  14\% & 22\% &  14\% &  \multirow{4}{*}{$S_{400}$}& trivial   &  14\% & 14\% &  14\% \\
    & small   &  22\% & 14\% &  7\% & & small   &  7\% & 0\% &  21\% \\
    & medium &  14\% & 14\% &  7\% &  & medium &  0\% & 21\% &  0\% \\
    & large  &  50\% & 50\% &  72\% & & large  &  79\% & 65\% &  65\% \\ 
   \bottomrule

      \end{tabular}
        \end{adjustbox}
       \vspace{0.48cm}
      \subcaption{statistical significance}
    \end{minipage}
%\end{subfigure}
\caption{Comparing \textsc{AdMMO} variants over 50 runs across all systems/objectives. (a) and (b) show the average/deviation of normalized target performance and the \% of statistically significant comparisons (also classified based on $\hat{A}_{12}$) against \textsc{AdMMO} on 14 system-objective pairs, respectively.}
\label{fig:ablation}
\vspace{-0.2cm}
\end{figure}

\subsection{Sensitivity of AdMMO to Maintained Proportion of Nondominated Configurations $p$}
\label{sec:rq4}

\subsubsection{Method}

In \textbf{RQ4} we seek to empirically examine whether there exists a generally best $p$. To that end, we verify on different settings of $p$: $\{0.05,0.1,0.3,0.5,0.7,0.9,1\}$. 

%While we cannot theoretically justify exactly what proportion can be the best,

\begin{figure}[t!]
\centering
\begin{subfigure}{.4\columnwidth}
   \begin{tikzpicture}
  \begin{axis}[
  width = 6cm,
  height = 6cm,
  xmin=0,
  xmax=1,
  ytick={1,2,3,4,5,6,7,8,9,10,11,12,13,14},
    yticklabel style = {font=\footnotesize},
    yticklabels={\textsc{LLVM}-CPU,\textsc{LLVM}-Runtime,\textsc{x264}-Energy,\textsc{x264}-PSNR,\textsc{MongoDB}-CPU,\textsc{MongoDB}-Runtime,\textsc{Keras}-Time,\textsc{Keras}-AUC,\textsc{VP9}-CPU,\textsc{VP9}-Runtime,\textsc{Storm}-Latency,\textsc{Storm}-Throughput,\textsc{MariaDB}-CPU,\textsc{MariaDB}-Runtime},
        xlabel=Best $p$,
%yticklabels={$I_{14}$,$I_{13}$,$I_{12}$,$I_{11}$,$I_{10}$,$I_{9}$,$I_{8}$,$I_{7}$,$I_{6}$,$I_{5}$,$I_{4}$,$I_{3}$,$I_{2}$,$I_{1}$},   xlabel=Best $p$,
   % ylabel={Normalized performance},
      axis x line  = bottom,
    axis y line  = left  ,
    legend cell align={left},
    legend style={draw=none,fill=none,at={(0.95,1.05)}},             
  ]

   \addplot [mark=diamond,black] plot coordinates {

(0.3,1)
(0.3,2)
(0.3,3)
(0.3,4)
(0.3,5)
(0.3,6)
(0.3,7)
(0.3,8)
(0.3,9)
(0.3,10)
(0.05,11)
(0.3,12)
(0.5,13)
(0.3,14)

 };

   \addplot [mark=square,red] plot coordinates {

(0.3,1)
(0.3,2)
(0.3,3)
(0.3,4)
(0.1,5)
(0.3,6)
(0.9,7)
(0.3,8)
(0.3,9)
(0.3,10)
(0.3,11)
(0.3,12)
(0.05,13)
(0.3,14)

 };

   \addplot [mark=o,teal] plot coordinates {

(0.3,1)
(0.3,2)
(0.3,3)
(0.3,4)
(0.3,5)
(0.3,6)
(0.9,7)
(0.3,8)
(0.05,9)
(0.3,10)
(0.3,11)
(0.3,12)
(0.1,13)
(0.3,14)

 };

   \addplot [mark=star,blue] plot coordinates {

(0.3,1)
(0.3,2)
(0.3,3)
(0.3,4)
(0.3,5)
(0.3,6)
(0.3,7)
(0.7,8)
(0.3,9)
(0.3,10)
(0.05,11)
(0.3,12)
(0.1,13)
(0.3,14)

 };

\addlegendentry{$S_{100}$}
\addlegendentry{$S_{200}$}
\addlegendentry{$S_{300}$}
\addlegendentry{$S_{400}$}
    
  \end{axis}

\end{tikzpicture}
%\vspace{-0.45 cm}
\subcaption{best $p$ across systems-objective pairs}
\end{subfigure}
\hspace{1.5cm}
\begin{subfigure}{.33\columnwidth}
   \begin{tikzpicture}
  \begin{axis}[
  width = 6cm,
  height = 6cm,
  xtick={1,2,3,4,5,6,7},
xticklabels={0.05,0.1,0.3,0.5,0.7,0.9,1},
    xlabel=$p$,
    ylabel={Normalized performance},
      axis x line  = bottom,
    axis y line  = left  ,
    legend cell align={left},
    legend style={draw=none,fill=none,at={(0.4,1.05)}},             
  ]

   \addplot [mark=diamond,black,name path=F] plot coordinates {

(1,0.42857142857142855)
(2,0.4371428571428572)
(3,0.3692857142857143)
(4,0.5228571428571429)
(5,0.5228571428571429)
(6,0.6092857142857143)
(7,0.6392857142857142)

 };

    \addplot [mark=square,red,name path=I] plot coordinates {
(1,0.205)
(2,0.18357142857142858)
(3,0.1457142857142857)
(4,0.2492857142857143)
(5,0.3514285714285714)
(6,0.5392857142857143)
(7,0.6278571428571429)

 };

    \addplot [mark=o,teal,name path=L] plot coordinates {
(1,0.12571428571428572)
(2,0.10071428571428573)
(3,0.04)
(4,0.1457142857142857)
(5,0.2942857142857143)
(6,0.5185714285714286)
(7,0.612142857142857)

 };

    \addplot [mark=star,blue,name path=C] plot coordinates {

(1,0.07928571428571429)
(2,0.050714285714285726)
(3,0.009285714285714286)
(4,0.10071428571428573)
(5,0.24714285714285714)
(6,0.5157142857142858)
(7,0.6049999999999999)

 };

\addplot [no marks,black!20,name path=D] plot coordinates {

(1,0.5049756779915103)
(2,0.515117398850513)
(3,0.43144469647699046)
(4,0.6120624127403013)
(5,0.6047569607746884)
(6,0.6955959300934024)
(7,0.7173258941887635)
 };
 \addplot [no marks,black!20,name path=E] plot coordinates {

(1,0.35216717915134677)
(2,0.35916831543520145)
(3,0.3071267320944381)
(4,0.43365187297398455)
(5,0.4409573249395974)
(6,0.5229754984780263)
(7,0.561245534382665)
 };

\addplot [no marks,red!20,name path=G] plot coordinates {

(1,0.2546698282333923)
(2,0.22843638629443197)
(3,0.17445866024314602)
(4,0.3026946515271487)
(5,0.4195809902160871)
(6,0.6144214030102918)
(7,0.7037745158482172)
 };
 \addplot [no marks,red!20,name path=H] plot coordinates {

(1,0.15533017176660768)
(2,0.1387064708484252)
(3,0.1169699111854254)
(4,0.19587677704427991)
(5,0.28327615264105577)
(6,0.46415002556113666)
(7,0.5519397698660686)
 };

\addplot [no marks,teal!20,name path=J] plot coordinates {
(1,0.16182231927281057)
(2,0.13732397950295722)
(3,0.050642617446965243)
(4,0.1840297630276283)
(5,0.36565507750152915)
(6,0.6023462551862537)
(7,0.7254613734247266)
 };
 \addplot [no marks,teal!20,name path=K] plot coordinates {

(1,0.08960625215576087)
(2,0.06410459192561425)
(3,0.029357382553034758)
(4,0.10739880840094312)
(5,0.2229163510698995)
(6,0.4347966019566034)
(7,0.5388243408609874)
 };

\addplot [no marks,blue!20,name path=A] plot coordinates {

(1,0.10784757492857025)
(2,0.07421991434360503)
(3,0.015631553812545353)
(4,0.12886229622349973)
(5,0.32492085473614174)
(6,0.6096949300834057)
(7,0.7053043836916167)
 };
 \addplot [no marks,blue!20,name path=B] plot coordinates {

(1,0.05072385364285833)
(2,0.027208657084966423)
(3,0.0029398747588832196)
(4,0.07256627520507172)
(5,0.1693648595495725)
(6,0.4217336413451659)
(7,0.504695616308383)
 };

   %% filling
   \addplot[blue!20,opacity=0.5] fill between[of=B and C];
    \addplot[blue!20,opacity=0.5] fill between[of=A and C];
    
       \addplot[black!20,opacity=0.5] fill between[of=D and F];
    \addplot[black!20,opacity=0.5] fill between[of=E and F];
    
           \addplot[red!20,opacity=0.5] fill between[of=H and I];
    \addplot[red!20,opacity=0.5] fill between[of=G and I];
    
         \addplot[teal!20,opacity=0.5] fill between[of=J and L];
    \addplot[teal!20,opacity=0.5] fill between[of=K and L];
   
%\addlegendentry{\textsc{Flash}$_\textsc{mmo}$}
%\addlegendentry{\textsc{Flash}}
    \addlegendentry{$S_{100}$}
\addlegendentry{$S_{200}$}
\addlegendentry{$S_{300}$}
\addlegendentry{$S_{400}$}
    
  \end{axis}

\end{tikzpicture}
\subcaption{impact of $p$ on performance}
\end{subfigure}

\caption{The role of $p$ in \textsc{AdMMO} over 50 runs. (a) plots the best $p$ on each system-objective pair; (b) shows the sensitivity of normalized target performance (mean and standard error) to $p$ on all 56 cases. }
\label{fig:sen}
\vspace{-0.3cm}
\end{figure}
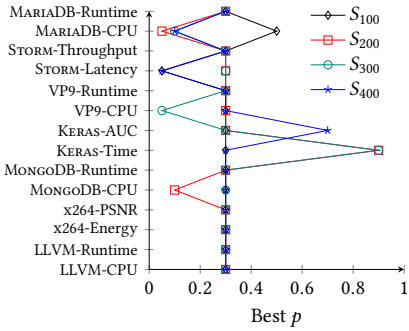
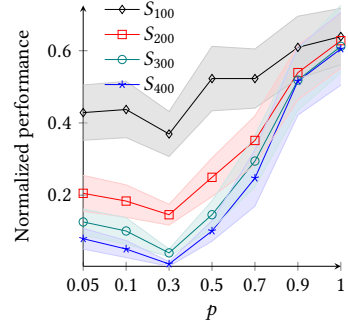

\subsubsection{Findings}

From Figure~\ref{fig:sen}a. It is clear that except for a few cases, most of the time $p=0.3$ tends to achieve the best performance regardless of the systems, objectives, and budgets. Additionally, from Figure~\ref{fig:sen}b, we note that, for all budgets and across the systems/objectives, both too small (e.g.,  $p=0.05$) or too large $p$ (e.g.,  $p=1$) can be detrimental, as the former suffers the local optima issues and the latter loses the discriminative power. In particular, we found that a too-large proportion is even more dangerous than a too-small one, 
as a small $p$ still retains the selection pressure (despite being overpressurized) whereas a large $p$ loses it completely. 

%In summary, we conclude that:

%  %This means that a proportion of 30\% is overall the best and more stable value than the others.

% \begin{itemize}
%     \item Both too small (e.g., 5\%) or too large $p$ (e.g., 100\%) can be detrimental, as the former suffers the local optima issues and the latter loses the discriminative power.
%     \item A too large proportion is even more dangerous than a too small one, 
%     as the latter still retains the selection pressure (despite being overpressurized) whereas the former loses it completely. 
%     %\item A proportion of 30\% is generally the best and more stable value than the others.
% \end{itemize}

\begin{quotebox}
   \noindent
   \textit{\textbf{RQ4:} Unlike the weight for which the best setting can largely vary case by case (recall Figure~\ref{fig:pre-analysis2}), maintaining a proportion of the nondominated configurations as around 30\% ($p=0.3$) can be overwhelmingly the best level according to all the cases in our study.}
\end{quotebox}

\section{Threats to Validity}
\label{sec:discussion}

Threats to \textbf{internal validity} can be related to the tuning budget. To tackle this, we examine different budget sizes and assess the efficiency of utilizing resources. The parameters of optimizers can also harm internal validity, hence for the state-of-the-art optimizers, we set their values as identical to what have been fine-tuned in existing work. For \textsc{AdMMO}, we use the most pragmatic settings of the parameters that are less significant (e.g., $T$ in the progressive trigger); most importantly, we confirm that there exists a generally best value for the most crucial parameter, i.e., $p$, for which we have used throughout the experiments. To mitigate bias, we repeated 50 experiment runs for each case. Indeed, we cannot completely rule out the impacts of some unusual parameter settings.

%or confirmed as the generally best one by our empirical study (i.e., the $p$ in AdMMO) unless inapplicable. 

% for comparing with measurement-based optimizers, we use the smallest budget used to stress test AdMMO; for comparing with model-based optimizers, we leverage exactly the same setting as that of FLASH~\cite{nair2018finding}. 

% All other parameters are set as identical to existing work or confirmed as the generally best one by our empirical study (e.g., the $p$ in AdMMO) unless inapplicable. To mitigate bias, we repeated 50 experiment runs under each case.
% %We have also examined the sensitivity of the MMO to its internal weight.

To reduce threats to \textbf{construct validity}, we compare the results of target performance objectives as that of \citet{DBLP:conf/sigsoft/0001Chen21}. As recommended by \citet{nair2018finding} and \citet{DBLP:conf/icse/GaoZ0LY21}, we also use the number of measurements (a language-independent feature) required to converge to the same result as an indicator of efficiency. To perform statistical validation, we leverage the non-parametric Wilcoxon test and $\mathbf{\hat{A}_{12}}$. The use of NSGA-II (and the operators) under \textsc{AdMMO} might be a threat, as it is merely a pragmatic choice, but it is not difficult to migrate \textsc{AdMMO} to the other multi-objective optimizers that are of the same family (i.e., compatible with nondominated sorting). We leave a more thorough study of diverse optimizers and operators in future work.

Threats to \textbf{external validity} can be raised from the subjects studied. We mitigated this by using seven systems that are of different scales and performance attributes, together with different budgets and nine state-of-the-art optimizers/models from diverse technical foundations. Nonetheless, if resources permit, we agree that using more systems and budgets may prove fruitful. It is worth noting that, in this work, we do not aim to optimize multiple performance objectives but use a tailored and improved idea/model of multi-objective search to solve a single objective SBSE problem. Indeed, there are cases where optimizing multiple performance objectives is desirable, yet we believe that this work serves as a first step and \textsc{AdMMO} can well be extended for those cases.

\section{Conclusion and Future Work}
\label{sec:con}

This paper presents a significant improvement on \textsc{MMO} for tuning software configuration, namely \textsc{AdMMO}. 
We contribute to a weight adaptation method that is capable of maintaining an unbiased proportion of nondominated configurations, 
together with a progressive trigger and a partial duplicate retention mechanism. 
Experiments on 14 system-objective pairs reveal that \textsc{AdMMO}:

\begin{itemize}

\item considerably outperforms the (improved) \textsc{MMO} and other optimizers for the tuning;
\item and achieves so with a significant speedup in most cases while;
%\item can be benefited the progressive trigger and partial duplicates retention mechanism while;
\item maintaining around 30\% of the nondominated configurations leads to the best outcome.
\end{itemize}

In future work, 
we hope to perform further analysis on the impact of the auxiliary performance objective chosen along with any characteristics, together with the role of different multi-objective optimizers and their operators. Understanding the suitability of \textsc{AdMMO} and the broader idea of multi-objectivization for other SBSE problems is also an interesting direction.

%\section*{Data Availability}

\textbf{Data Availability:} To promote open science, all source code, data, and supplementary materials can be accessed at our repository: \url{https://github.com/ideas-labo/admmo}.

%may significantly impact the results and the the proportion $p$ to maintain

% Although many aspects of \textsc{AdMMO} are specifically designed based on the properties/observations of software configuration tuning, 
% which may or may not be suitable for other SBSE problems, 
% it could be possible to tailor it to fit a given domain in software engineering. For example, for other problems with less sparse landscapes, there could be much fewer duplicates and hence it may make less sense to (partially) eliminate them.

%\textcolor{red}{[Tao, I am wondering if we could state more specifically here, for example for another problem, how to tailor the AdMMO for? at least some insights/directions are desirable to provide.]}

\begin{acks}
This work was supported by a UKRI Grant (10054084) and a NSFC Grant (62372084).
\end{acks}

\balance
\bibliographystyle{ACM-Reference-Format}
\bibliography{reference}


\begin{thebibliography}{54}

%%% the Web address.

\ifx \showCODEN    \undefined \def \showCODEN     #1{\unskip}     \fi
\ifx \showDOI      \undefined \def \showDOI       #1{#1}\fi
\ifx \showISBNx    \undefined \def \showISBNx     #1{\unskip}     \fi
\ifx \showISBNxiii \undefined \def \showISBNxiii  #1{\unskip}     \fi
\ifx \showISSN     \undefined \def \showISSN      #1{\unskip}     \fi
\ifx \showLCCN     \undefined \def \showLCCN      #1{\unskip}     \fi
\ifx \shownote     \undefined \def \shownote      #1{#1}          \fi
\ifx \showarticletitle \undefined \def \showarticletitle #1{#1}   \fi
\ifx \showURL      \undefined \def \showURL       {\relax}        \fi
% invisible to TeX
\providecommand\bibfield[2]{#2}
\providecommand\bibinfo[2]{#2}
\providecommand\natexlab[1]{#1}
\providecommand\showeprint[2][]{arXiv:#2}

\bibitem[Arcuri and Briand(2011)]%
        {ArcuriB11}
\bibfield{author}{\bibinfo{person}{Andrea Arcuri} {and}
  \bibinfo{person}{Lionel~C. Briand}.} \bibinfo{year}{2011}\natexlab{}.
\newblock \showarticletitle{A practical guide for using statistical tests to
  assess randomized algorithms in software engineering}. In
  \bibinfo{booktitle}{\emph{ICSE'11: Proc. of the 33rd International Conference
  on Software Engineering}}. \bibinfo{publisher}{{ACM}},
  \bibinfo{pages}{1--10}.
\newblock


\bibitem[Bartz-Beielstein et~al\mbox{.}(2021)]%
        {bartz2021tuning}
\bibfield{author}{\bibinfo{person}{Thomas Bartz-Beielstein},
  \bibinfo{person}{Frederik Rehbach}, {and} \bibinfo{person}{Margarita
  Rebolledo}.} \bibinfo{year}{2021}\natexlab{}.
\newblock \showarticletitle{Tuning Algorithms for Stochastic Black-Box
  Optimization: State of the Art and Future Perspectives}.
\newblock \bibinfo{journal}{\emph{Black Box Optimization, Machine Learning, and
  No-Free Lunch Theorems}} (\bibinfo{year}{2021}), \bibinfo{pages}{67--108}.
\newblock


\bibitem[Behzad et~al\mbox{.}(2013)]%
        {DBLP:conf/sc/BehzadLHBPAKS13}
\bibfield{author}{\bibinfo{person}{Babak Behzad}, \bibinfo{person}{Huong
  Vu~Thanh Luu}, \bibinfo{person}{Joseph Huchette}, \bibinfo{person}{Surendra
  Byna}, \bibinfo{person}{Prabhat}, \bibinfo{person}{Ruth~A. Aydt},
  \bibinfo{person}{Quincey Koziol}, {and} \bibinfo{person}{Marc Snir}.}
  \bibinfo{year}{2013}\natexlab{}.
\newblock \showarticletitle{Taming parallel {I/O} complexity with auto-tuning}.
  In \bibinfo{booktitle}{\emph{International Conference for High Performance
  Computing, Networking, Storage and Analysis, SC'13, Denver, CO, {USA} -
  November 17 - 21, 2013}}, \bibfield{editor}{\bibinfo{person}{William Gropp}
  {and} \bibinfo{person}{Satoshi Matsuoka}} (Eds.). \bibinfo{publisher}{{ACM}},
  \bibinfo{pages}{68:1--68:12}.
\newblock
\urldef\tempurl%
\url{https://doi.org/10.1145/2503210.2503278}
\showDOI{\tempurl}


\bibitem[Bergstra et~al\mbox{.}(2011)]%
        {DBLP:conf/nips/BergstraBBK11}
\bibfield{author}{\bibinfo{person}{James Bergstra}, \bibinfo{person}{R{\'{e}}mi
  Bardenet}, \bibinfo{person}{Yoshua Bengio}, {and}
  \bibinfo{person}{Bal{\'{a}}zs K{\'{e}}gl}.} \bibinfo{year}{2011}\natexlab{}.
\newblock \showarticletitle{Algorithms for Hyper-Parameter Optimization}. In
  \bibinfo{booktitle}{\emph{Advances in Neural Information Processing Systems
  24: 25th Annual Conference on Neural Information Processing Systems 2011.
  Proceedings of a meeting held 12-14 December 2011, Granada, Spain}},
  \bibfield{editor}{\bibinfo{person}{John Shawe{-}Taylor},
  \bibinfo{person}{Richard~S. Zemel}, \bibinfo{person}{Peter~L. Bartlett},
  \bibinfo{person}{Fernando C.~N. Pereira}, {and} \bibinfo{person}{Kilian~Q.
  Weinberger}} (Eds.). \bibinfo{pages}{2546--2554}.
\newblock
\urldef\tempurl%
\url{https://proceedings.neurips.cc/paper/2011/hash/86e8f7ab32cfd12577bc2619bc635690-Abstract.html}
\showURL{%
\tempurl}


\bibitem[Bergstra and Bengio(2012)]%
        {DBLP:journals/jmlr/BergstraB12}
\bibfield{author}{\bibinfo{person}{James Bergstra} {and}
  \bibinfo{person}{Yoshua Bengio}.} \bibinfo{year}{2012}\natexlab{}.
\newblock \showarticletitle{Random Search for Hyper-Parameter Optimization}.
\newblock \bibinfo{journal}{\emph{J. Mach. Learn. Res.}}  \bibinfo{volume}{13}
  (\bibinfo{year}{2012}), \bibinfo{pages}{281--305}.
\newblock
\urldef\tempurl%
\url{http://dl.acm.org/citation.cfm?id=2188395}
\showURL{%
\tempurl}


\bibitem[Beume et~al\mbox{.}(2007)]%
        {DBLP:journals/eor/BeumeNE07}
\bibfield{author}{\bibinfo{person}{Nicola Beume}, \bibinfo{person}{Boris
  Naujoks}, {and} \bibinfo{person}{Michael T.~M. Emmerich}.}
  \bibinfo{year}{2007}\natexlab{}.
\newblock \showarticletitle{{SMS-EMOA:} Multiobjective selection based on
  dominated hypervolume}.
\newblock \bibinfo{journal}{\emph{Eur. J. Oper. Res.}} \bibinfo{volume}{181},
  \bibinfo{number}{3} (\bibinfo{year}{2007}), \bibinfo{pages}{1653--1669}.
\newblock
\urldef\tempurl%
\url{https://doi.org/10.1016/J.EJOR.2006.08.008}
\showDOI{\tempurl}


\bibitem[C{\'{a}}ceres et~al\mbox{.}(2017)]%
        {DBLP:conf/ae/CaceresPFS17}
\bibfield{author}{\bibinfo{person}{Leslie~P{\'{e}}rez C{\'{a}}ceres},
  \bibinfo{person}{Federico Pagnozzi}, \bibinfo{person}{Alberto Franzin}, {and}
  \bibinfo{person}{Thomas St{\"{u}}tzle}.} \bibinfo{year}{2017}\natexlab{}.
\newblock \showarticletitle{Automatic Configuration of {GCC} Using Irace}. In
  \bibinfo{booktitle}{\emph{Artificial Evolution - 13th International
  Conference, {\'{E}}volution Artificielle, {EA} 2017, Paris, France, October
  25-27, 2017, Revised Selected Papers}} \emph{(\bibinfo{series}{Lecture Notes
  in Computer Science}, Vol.~\bibinfo{volume}{10764})},
  \bibfield{editor}{\bibinfo{person}{Evelyne Lutton}, \bibinfo{person}{Pierrick
  Legrand}, \bibinfo{person}{Pierre Parrend}, \bibinfo{person}{Nicolas
  Monmarch{\'{e}}}, {and} \bibinfo{person}{Marc Schoenauer}} (Eds.).
  \bibinfo{publisher}{Springer}, \bibinfo{pages}{202--216}.
\newblock
\urldef\tempurl%
\url{https://doi.org/10.1007/978-3-319-78133-4\_15}
\showDOI{\tempurl}


\bibitem[Chen et~al\mbox{.}(2021)]%
        {DBLP:conf/icse/0003XC021}
\bibfield{author}{\bibinfo{person}{Junjie Chen}, \bibinfo{person}{Ningxin Xu},
  \bibinfo{person}{Peiqi Chen}, {and} \bibinfo{person}{Hongyu Zhang}.}
  \bibinfo{year}{2021}\natexlab{}.
\newblock \showarticletitle{Efficient Compiler Autotuning via Bayesian
  Optimization}. In \bibinfo{booktitle}{\emph{43rd {IEEE/ACM} International
  Conference on Software Engineering, {ICSE} 2021, Madrid, Spain, 22-30 May
  2021}}. \bibinfo{publisher}{{IEEE}}, \bibinfo{pages}{1198--1209}.
\newblock
\urldef\tempurl%
\url{https://doi.org/10.1109/ICSE43902.2021.00110}
\showDOI{\tempurl}


\bibitem[Chen et~al\mbox{.}(2024)]%
        {DBLP:journals/corr/abs-2112-07303}
\bibfield{author}{\bibinfo{person}{Pengzhou Chen}, \bibinfo{person}{Tao Chen},
  {and} \bibinfo{person}{Miqing Li}.} \bibinfo{year}{2024}\natexlab{}.
\newblock \showarticletitle{{MMO:} Meta Multi-Objectivization for Software
  Configuration Tuning}.
\newblock \bibinfo{journal}{\emph{IEEE Transactions on Software Engineering}}
  (\bibinfo{year}{2024}).
\newblock


\bibitem[Chen(2022a)]%
        {DBLP:conf/wcre/Chen22}
\bibfield{author}{\bibinfo{person}{Tao Chen}.}
  \bibinfo{year}{2022}\natexlab{a}.
\newblock \showarticletitle{Lifelong Dynamic Optimization for Self-Adaptive
  Systems: Fact or Fiction?}. In \bibinfo{booktitle}{\emph{{IEEE} International
  Conference on Software Analysis, Evolution and Reengineering, {SANER} 2022,
  Honolulu, HI, USA, March 15-18, 2022}}. \bibinfo{publisher}{{IEEE}},
  \bibinfo{pages}{78--89}.
\newblock
\urldef\tempurl%
\url{https://doi.org/10.1109/SANER53432.2022.00022}
\showDOI{\tempurl}


\bibitem[Chen(2022b)]%
        {DBLP:conf/seams/Chen22}
\bibfield{author}{\bibinfo{person}{Tao Chen}.}
  \bibinfo{year}{2022}\natexlab{b}.
\newblock \showarticletitle{Planning Landscape Analysis for Self-Adaptive
  Systems}. In \bibinfo{booktitle}{\emph{International Symposium on Software
  Engineering for Adaptive and Self-Managing Systems, {SEAMS} 2022, Pittsburgh,
  PA, USA, May 22-24, 2022}}, \bibfield{editor}{\bibinfo{person}{Bradley~R.
  Schmerl}, \bibinfo{person}{Martina Maggio}, {and} \bibinfo{person}{Javier
  C{\'{a}}mara}} (Eds.). \bibinfo{publisher}{{ACM/IEEE}},
  \bibinfo{pages}{84--90}.
\newblock
\urldef\tempurl%
\url{https://doi.org/10.1145/3524844.3528060}
\showDOI{\tempurl}


\bibitem[Chen and Bahsoon(2017a)]%
        {DBLP:journals/tse/ChenB17}
\bibfield{author}{\bibinfo{person}{Tao Chen} {and} \bibinfo{person}{Rami
  Bahsoon}.} \bibinfo{year}{2017}\natexlab{a}.
\newblock \showarticletitle{Self-Adaptive and Online QoS Modeling for
  Cloud-Based Software Services}.
\newblock \bibinfo{journal}{\emph{{IEEE} Trans. Software Eng.}}
  \bibinfo{volume}{43}, \bibinfo{number}{5} (\bibinfo{year}{2017}),
  \bibinfo{pages}{453--475}.
\newblock
\urldef\tempurl%
\url{https://doi.org/10.1109/TSE.2016.2608826}
\showDOI{\tempurl}


\bibitem[Chen and Bahsoon(2017b)]%
        {DBLP:journals/tsc/ChenB17}
\bibfield{author}{\bibinfo{person}{Tao Chen} {and} \bibinfo{person}{Rami
  Bahsoon}.} \bibinfo{year}{2017}\natexlab{b}.
\newblock \showarticletitle{Self-Adaptive Trade-off Decision Making for
  Autoscaling Cloud-Based Services}.
\newblock \bibinfo{journal}{\emph{{IEEE} Trans. Serv. Comput.}}
  \bibinfo{volume}{10}, \bibinfo{number}{4} (\bibinfo{year}{2017}),
  \bibinfo{pages}{618--632}.
\newblock
\urldef\tempurl%
\url{https://doi.org/10.1109/TSC.2015.2499770}
\showDOI{\tempurl}


\bibitem[Chen et~al\mbox{.}(2018)]%
        {Chen2018FEMOSAA}
\bibfield{author}{\bibinfo{person}{Tao Chen}, \bibinfo{person}{Ke Li},
  \bibinfo{person}{Rami Bahsoon}, {and} \bibinfo{person}{Xin Yao}.}
  \bibinfo{year}{2018}\natexlab{}.
\newblock \showarticletitle{{FEMOSAA}: Feature Guided and Knee Driven
  Multi-Objective Optimization for Self-Adaptive Software}.
\newblock \bibinfo{journal}{\emph{ACM Transactions on Software Engineering and
  Methodology}} \bibinfo{volume}{27}, \bibinfo{number}{2}
  (\bibinfo{year}{2018}).
\newblock


\bibitem[Chen and Li(2021)]%
        {DBLP:conf/sigsoft/0001Chen21}
\bibfield{author}{\bibinfo{person}{Tao Chen} {and} \bibinfo{person}{Miqing
  Li}.} \bibinfo{year}{2021}\natexlab{}.
\newblock \showarticletitle{Multi-objectivizing software configuration tuning}.
  In \bibinfo{booktitle}{\emph{{ESEC/FSE} '21: 29th {ACM} Joint European
  Software Engineering Conference and Symposium on the Foundations of Software
  Engineering, Athens, Greece, August 23-28, 2021}},
  \bibfield{editor}{\bibinfo{person}{Diomidis Spinellis},
  \bibinfo{person}{Georgios Gousios}, \bibinfo{person}{Marsha Chechik}, {and}
  \bibinfo{person}{Massimiliano~Di Penta}} (Eds.). \bibinfo{publisher}{{ACM}},
  \bibinfo{pages}{453--465}.
\newblock
\urldef\tempurl%
\url{https://doi.org/10.1145/3468264.3468555}
\showDOI{\tempurl}


\bibitem[Chen and Li(2023a)]%
        {DBLP:journals/tosem/ChenL23a}
\bibfield{author}{\bibinfo{person}{Tao Chen} {and} \bibinfo{person}{Miqing
  Li}.} \bibinfo{year}{2023}\natexlab{a}.
\newblock \showarticletitle{Do Performance Aspirations Matter for Guiding
  Software Configuration Tuning? An Empirical Investigation under Dual
  Performance Objectives}.
\newblock \bibinfo{journal}{\emph{{ACM} Trans. Softw. Eng. Methodol.}}
  \bibinfo{volume}{32}, \bibinfo{number}{3} (\bibinfo{year}{2023}),
  \bibinfo{pages}{68:1--68:41}.
\newblock
\urldef\tempurl%
\url{https://doi.org/10.1145/3571853}
\showDOI{\tempurl}


\bibitem[Chen and Li(2023b)]%
        {DBLP:journals/tosem/ChenL23}
\bibfield{author}{\bibinfo{person}{Tao Chen} {and} \bibinfo{person}{Miqing
  Li}.} \bibinfo{year}{2023}\natexlab{b}.
\newblock \showarticletitle{The Weights Can Be Harmful: Pareto Search versus
  Weighted Search in Multi-objective Search-based Software Engineering}.
\newblock \bibinfo{journal}{\emph{{ACM} Trans. Softw. Eng. Methodol.}}
  \bibinfo{volume}{32}, \bibinfo{number}{1} (\bibinfo{year}{2023}),
  \bibinfo{pages}{5:1--5:40}.
\newblock
\urldef\tempurl%
\url{https://doi.org/10.1145/3514233}
\showDOI{\tempurl}


\bibitem[Chen et~al\mbox{.}(2019)]%
        {DBLP:journals/infsof/ChenLY19}
\bibfield{author}{\bibinfo{person}{Tao Chen}, \bibinfo{person}{Miqing Li},
  {and} \bibinfo{person}{Xin Yao}.} \bibinfo{year}{2019}\natexlab{}.
\newblock \showarticletitle{Standing on the shoulders of giants: Seeding
  search-based multi-objective optimization with prior knowledge for software
  service composition}.
\newblock \bibinfo{journal}{\emph{Inf. Softw. Technol.}}  \bibinfo{volume}{114}
  (\bibinfo{year}{2019}), \bibinfo{pages}{155--175}.
\newblock
\urldef\tempurl%
\url{https://doi.org/10.1016/j.infsof.2019.05.013}
\showDOI{\tempurl}


\bibitem[Deb et~al\mbox{.}(2002)]%
        {Deb2002}
\bibfield{author}{\bibinfo{person}{K. Deb}, \bibinfo{person}{A. Pratap},
  \bibinfo{person}{S. Agarwal}, {and} \bibinfo{person}{T. Meyarivan}.}
  \bibinfo{year}{2002}\natexlab{}.
\newblock \showarticletitle{A fast and elitist multiobjective genetic
  algorithm: NSGA-II}.
\newblock \bibinfo{journal}{\emph{IEEE Transactions on Evolutionary
  Computation}} \bibinfo{volume}{6}, \bibinfo{number}{2}
  (\bibinfo{year}{2002}), \bibinfo{pages}{182--197}.
\newblock


\bibitem[Derakhshanfar et~al\mbox{.}(2020)]%
        {derakhshanfar2020good}
\bibfield{author}{\bibinfo{person}{Pouria Derakhshanfar},
  \bibinfo{person}{Xavier Devroey}, \bibinfo{person}{Andy Zaidman},
  \bibinfo{person}{Arie van Deursen}, {and} \bibinfo{person}{Annibale
  Panichella}.} \bibinfo{year}{2020}\natexlab{}.
\newblock \showarticletitle{Good Things Come In Threes: Improving Search-based
  Crash Reproduction With Helper Objectives}. In \bibinfo{booktitle}{\emph{35th
  IEEE/ACM International Conference on Automated Software Engineering
  (ASE'20)}}.
\newblock


\bibitem[Ding et~al\mbox{.}(2015)]%
        {DBLP:conf/icpads/DingLQ15}
\bibfield{author}{\bibinfo{person}{Xiaoan Ding}, \bibinfo{person}{Yi Liu},
  {and} \bibinfo{person}{Depei Qian}.} \bibinfo{year}{2015}\natexlab{}.
\newblock \showarticletitle{JellyFish: Online Performance Tuning with Adaptive
  Configuration and Elastic Container in Hadoop Yarn}. In
  \bibinfo{booktitle}{\emph{21st {IEEE} International Conference on Parallel
  and Distributed Systems, {ICPADS} 2015, Melbourne, Australia, December 14-17,
  2015}}. \bibinfo{publisher}{{IEEE} Computer Society},
  \bibinfo{pages}{831--836}.
\newblock
\urldef\tempurl%
\url{https://doi.org/10.1109/ICPADS.2015.112}
\showDOI{\tempurl}


\bibitem[Durillo and Nebro(2011)]%
        {DBLP:journals/aes/DurilloN11}
\bibfield{author}{\bibinfo{person}{Juan~Jos{\'{e}} Durillo} {and}
  \bibinfo{person}{Antonio~J. Nebro}.} \bibinfo{year}{2011}\natexlab{}.
\newblock \showarticletitle{jMetal: {A} Java framework for multi-objective
  optimization}.
\newblock \bibinfo{journal}{\emph{Adv. Eng. Softw.}} \bibinfo{volume}{42},
  \bibinfo{number}{10} (\bibinfo{year}{2011}), \bibinfo{pages}{760--771}.
\newblock
\urldef\tempurl%
\url{https://doi.org/10.1016/j.advengsoft.2011.05.014}
\showDOI{\tempurl}


\bibitem[Fortin and Parizeau(2013)]%
        {DBLP:conf/gecco/FortinP13}
\bibfield{author}{\bibinfo{person}{F{\'{e}}lix{-}Antoine Fortin} {and}
  \bibinfo{person}{Marc Parizeau}.} \bibinfo{year}{2013}\natexlab{}.
\newblock \showarticletitle{Revisiting the {NSGA-II} crowding-distance
  computation}. In \bibinfo{booktitle}{\emph{Genetic and Evolutionary
  Computation Conference, {GECCO} '13, Amsterdam, The Netherlands, July 6-10,
  2013}}, \bibfield{editor}{\bibinfo{person}{Christian Blum} {and}
  \bibinfo{person}{Enrique Alba}} (Eds.). \bibinfo{publisher}{{ACM}},
  \bibinfo{pages}{623--630}.
\newblock
\urldef\tempurl%
\url{https://doi.org/10.1145/2463372.2463456}
\showDOI{\tempurl}


\bibitem[Gao et~al\mbox{.}(2021)]%
        {DBLP:conf/icse/GaoZ0LY21}
\bibfield{author}{\bibinfo{person}{Yanjie Gao}, \bibinfo{person}{Yonghao Zhu},
  \bibinfo{person}{Hongyu Zhang}, \bibinfo{person}{Haoxiang Lin}, {and}
  \bibinfo{person}{Mao Yang}.} \bibinfo{year}{2021}\natexlab{}.
\newblock \showarticletitle{Resource-Guided Configuration Space Reduction for
  Deep Learning Models}. In \bibinfo{booktitle}{\emph{43rd {IEEE/ACM}
  International Conference on Software Engineering, {ICSE} 2021, Madrid, Spain,
  22-30 May 2021}}. \bibinfo{publisher}{{IEEE}}, \bibinfo{pages}{175--187}.
\newblock
\urldef\tempurl%
\url{https://doi.org/10.1109/ICSE43902.2021.00028}
\showDOI{\tempurl}


\bibitem[Gong and Chen(2022)]%
        {DBLP:conf/msr/GongC22}
\bibfield{author}{\bibinfo{person}{Jingzhi Gong} {and} \bibinfo{person}{Tao
  Chen}.} \bibinfo{year}{2022}\natexlab{}.
\newblock \showarticletitle{Does Configuration Encoding Matter in Learning
  Software Performance? An Empirical Study on Encoding Schemes}. In
  \bibinfo{booktitle}{\emph{19th {IEEE/ACM} International Conference on Mining
  Software Repositories, {MSR} 2022, Pittsburgh, PA, USA, May 23-24, 2022}}.
  \bibinfo{publisher}{{ACM}}, \bibinfo{pages}{482--494}.
\newblock
\urldef\tempurl%
\url{https://doi.org/10.1145/3524842.3528431}
\showDOI{\tempurl}


\bibitem[Gong and Chen(2023)]%
        {DBLP:conf/sigsoft/GongChen2023}
\bibfield{author}{\bibinfo{person}{Jingzhi Gong} {and} \bibinfo{person}{Tao
  Chen}.} \bibinfo{year}{2023}\natexlab{}.
\newblock \showarticletitle{Predicting Software Performance with
  Divide-and-Learn}. In \bibinfo{booktitle}{\emph{Proceedings of the 31st {ACM}
  Joint European Software Engineering Conference and Symposium on the
  Foundations of Software Engineering, {ESEC/FSE} 2023, San Francisco, CA, USA,
  December 3-9, 2023}}, \bibfield{editor}{\bibinfo{person}{Satish Chandra},
  \bibinfo{person}{Kelly Blincoe}, {and} \bibinfo{person}{Paolo Tonella}}
  (Eds.). \bibinfo{publisher}{{ACM}}, \bibinfo{pages}{858--870}.
\newblock
\urldef\tempurl%
\url{https://doi.org/10.1145/3611643.3616334}
\showDOI{\tempurl}


\bibitem[Gong and Chen(2024)]%
        {DBLP:conf/sigsoft/0001L24}
\bibfield{author}{\bibinfo{person}{Jingzhi Gong} {and} \bibinfo{person}{Tao
  Chen}.} \bibinfo{year}{2024}\natexlab{}.
\newblock \showarticletitle{Predicting Configuration Performance in Multiple
  Environments with Sequential Meta-Learning}.
\newblock \bibinfo{journal}{\emph{{FSE}'24: Proceedings of the ACM on Software
  Engineering (PACMSE)}} \bibinfo{volume}{1}, \bibinfo{number}{FSE}.
\newblock
\urldef\tempurl%
\url{https://doi.org/10.1145/3643743}
\showDOI{\tempurl}


\bibitem[Han and Yu(2016)]%
        {DBLP:conf/esem/HanY16}
\bibfield{author}{\bibinfo{person}{Xue Han} {and} \bibinfo{person}{Tingting
  Yu}.} \bibinfo{year}{2016}\natexlab{}.
\newblock \showarticletitle{An Empirical Study on Performance Bugs for Highly
  Configurable Software Systems}. In \bibinfo{booktitle}{\emph{Proceedings of
  the 10th {ACM/IEEE} International Symposium on Empirical Software Engineering
  and Measurement, {ESEM} 2016, Ciudad Real, Spain, September 8-9, 2016}}.
  \bibinfo{publisher}{{ACM}}, \bibinfo{pages}{23:1--23:10}.
\newblock
\urldef\tempurl%
\url{https://doi.org/10.1145/2961111.2962602}
\showDOI{\tempurl}


\bibitem[Hutter et~al\mbox{.}(2011)]%
        {DBLP:conf/lion/HutterHL11}
\bibfield{author}{\bibinfo{person}{Frank Hutter}, \bibinfo{person}{Holger~H.
  Hoos}, {and} \bibinfo{person}{Kevin Leyton{-}Brown}.}
  \bibinfo{year}{2011}\natexlab{}.
\newblock \showarticletitle{Sequential Model-Based Optimization for General
  Algorithm Configuration}. In \bibinfo{booktitle}{\emph{LION5: Proc. of the
  5th International Conference Learning and Intelligent Optimization}}
  \emph{(\bibinfo{series}{Lecture Notes in Computer Science},
  Vol.~\bibinfo{volume}{6683})}. \bibinfo{publisher}{Springer},
  \bibinfo{pages}{507--523}.
\newblock


\bibitem[Hutter et~al\mbox{.}(2009)]%
        {DBLP:journals/jair/HutterHLS09}
\bibfield{author}{\bibinfo{person}{Frank Hutter}, \bibinfo{person}{Holger~H.
  Hoos}, \bibinfo{person}{Kevin Leyton{-}Brown}, {and} \bibinfo{person}{Thomas
  St{\"{u}}tzle}.} \bibinfo{year}{2009}\natexlab{}.
\newblock \showarticletitle{ParamILS: An Automatic Algorithm Configuration
  Framework}.
\newblock \bibinfo{journal}{\emph{J. Artif. Intell. Res.}}
  \bibinfo{volume}{36} (\bibinfo{year}{2009}), \bibinfo{pages}{267--306}.
\newblock
\urldef\tempurl%
\url{https://doi.org/10.1613/jair.2861}
\showDOI{\tempurl}


\bibitem[Jamshidi and Casale(2016)]%
        {DBLP:conf/mascots/JamshidiC16}
\bibfield{author}{\bibinfo{person}{Pooyan Jamshidi} {and}
  \bibinfo{person}{Giuliano Casale}.} \bibinfo{year}{2016}\natexlab{}.
\newblock \showarticletitle{An Uncertainty-Aware Approach to Optimal
  Configuration of Stream Processing Systems}. In
  \bibinfo{booktitle}{\emph{24th {IEEE} International Symposium on Modeling,
  Analysis and Simulation of Computer and Telecommunication Systems, {MASCOTS}
  2016, London, United Kingdom, September 19-21, 2016}}.
  \bibinfo{publisher}{{IEEE} Computer Society}, \bibinfo{pages}{39--48}.
\newblock


\bibitem[Jamshidi et~al\mbox{.}(2018)]%
        {DBLP:conf/sigsoft/JamshidiVKS18}
\bibfield{author}{\bibinfo{person}{Pooyan Jamshidi}, \bibinfo{person}{Miguel
  Velez}, \bibinfo{person}{Christian K{\"{a}}stner}, {and}
  \bibinfo{person}{Norbert Siegmund}.} \bibinfo{year}{2018}\natexlab{}.
\newblock \showarticletitle{Learning to sample: exploiting similarities across
  environments to learn performance models for configurable systems}. In
  \bibinfo{booktitle}{\emph{Proceedings of the 2018 {ACM} Joint Meeting on
  European Software Engineering Conference and Symposium on the Foundations of
  Software Engineering, {ESEC/SIGSOFT} {FSE} 2018, Lake Buena Vista, FL, USA,
  November 04-09, 2018}}, \bibfield{editor}{\bibinfo{person}{Gary~T. Leavens},
  \bibinfo{person}{Alessandro Garcia}, {and} \bibinfo{person}{Corina~S.
  Pasareanu}} (Eds.). \bibinfo{publisher}{{ACM}}, \bibinfo{pages}{71--82}.
\newblock
\urldef\tempurl%
\url{https://doi.org/10.1145/3236024.3236074}
\showDOI{\tempurl}


\bibitem[Li et~al\mbox{.}(2020a)]%
        {DBLP:conf/kbse/LiXCT20}
\bibfield{author}{\bibinfo{person}{Ke Li}, \bibinfo{person}{Zilin Xiang},
  \bibinfo{person}{Tao Chen}, {and} \bibinfo{person}{Kay~Chen Tan}.}
  \bibinfo{year}{2020}\natexlab{a}.
\newblock \showarticletitle{BiLO-CPDP: Bi-Level Programming for Automated Model
  Discovery in Cross-Project Defect Prediction}. In
  \bibinfo{booktitle}{\emph{35th {IEEE/ACM} International Conference on
  Automated Software Engineering, {ASE} 2020, Melbourne, Australia, September
  21-25, 2020}}. \bibinfo{publisher}{{IEEE}}, \bibinfo{pages}{573--584}.
\newblock
\urldef\tempurl%
\url{https://doi.org/10.1145/3324884.3416617}
\showDOI{\tempurl}


\bibitem[Li et~al\mbox{.}(2020b)]%
        {DBLP:conf/icse/LiX0WT20}
\bibfield{author}{\bibinfo{person}{Ke Li}, \bibinfo{person}{Zilin Xiang},
  \bibinfo{person}{Tao Chen}, \bibinfo{person}{Shuo Wang}, {and}
  \bibinfo{person}{Kay~Chen Tan}.} \bibinfo{year}{2020}\natexlab{b}.
\newblock \showarticletitle{Understanding the automated parameter optimization
  on transfer learning for cross-project defect prediction: an empirical
  study}. In \bibinfo{booktitle}{\emph{{ICSE} '20: 42nd International
  Conference on Software Engineering, Seoul, South Korea, 27 June - 19 July,
  2020}}, \bibfield{editor}{\bibinfo{person}{Gregg Rothermel} {and}
  \bibinfo{person}{Doo{-}Hwan Bae}} (Eds.). \bibinfo{publisher}{{ACM}},
  \bibinfo{pages}{566--577}.
\newblock
\urldef\tempurl%
\url{https://doi.org/10.1145/3377811.3380360}
\showDOI{\tempurl}


\bibitem[Li et~al\mbox{.}(2022)]%
        {DBLP:journals/tse/LiCY22}
\bibfield{author}{\bibinfo{person}{Miqing Li}, \bibinfo{person}{Tao Chen},
  {and} \bibinfo{person}{Xin Yao}.} \bibinfo{year}{2022}\natexlab{}.
\newblock \showarticletitle{How to Evaluate Solutions in Pareto-Based
  Search-Based Software Engineering: {A} Critical Review and Methodological
  Guidance}.
\newblock \bibinfo{journal}{\emph{{IEEE} Trans. Software Eng.}}
  \bibinfo{volume}{48}, \bibinfo{number}{5} (\bibinfo{year}{2022}),
  \bibinfo{pages}{1771--1799}.
\newblock
\urldef\tempurl%
\url{https://doi.org/10.1109/TSE.2020.3036108}
\showDOI{\tempurl}


\bibitem[Li et~al\mbox{.}(2014)]%
        {DBLP:conf/hpdc/LiZMTZBF14}
\bibfield{author}{\bibinfo{person}{Min Li}, \bibinfo{person}{Liangzhao Zeng},
  \bibinfo{person}{Shicong Meng}, \bibinfo{person}{Jian Tan},
  \bibinfo{person}{Li Zhang}, \bibinfo{person}{Ali~Raza Butt}, {and}
  \bibinfo{person}{Nicholas~C. Fuller}.} \bibinfo{year}{2014}\natexlab{}.
\newblock \showarticletitle{{MRONLINE:} MapReduce online performance tuning}.
  In \bibinfo{booktitle}{\emph{The 23rd International Symposium on
  High-Performance Parallel and Distributed Computing, HPDC'14, Vancouver, BC,
  Canada - June 23 - 27, 2014}}, \bibfield{editor}{\bibinfo{person}{Beth
  Plale}, \bibinfo{person}{Matei Ripeanu}, \bibinfo{person}{Franck Cappello},
  {and} \bibinfo{person}{Dongyan Xu}} (Eds.). \bibinfo{publisher}{{ACM}},
  \bibinfo{pages}{165--176}.
\newblock
\urldef\tempurl%
\url{https://doi.org/10.1145/2600212.2600229}
\showDOI{\tempurl}


\bibitem[L{\'o}pez-Ib{\'a}nez et~al\mbox{.}(2016)]%
        {lopez2016irace}
\bibfield{author}{\bibinfo{person}{Manuel L{\'o}pez-Ib{\'a}nez},
  \bibinfo{person}{J{\'e}r{\'e}mie Dubois-Lacoste},
  \bibinfo{person}{Leslie~P{\'e}rez C{\'a}ceres}, \bibinfo{person}{Mauro
  Birattari}, {and} \bibinfo{person}{Thomas St{\"u}tzle}.}
  \bibinfo{year}{2016}\natexlab{}.
\newblock \showarticletitle{The irace package: Iterated racing for automatic
  algorithm configuration}.
\newblock \bibinfo{journal}{\emph{Operations Research Perspectives}}
  \bibinfo{volume}{3} (\bibinfo{year}{2016}), \bibinfo{pages}{43--58}.
\newblock


\bibitem[Lukasiewycz et~al\mbox{.}(2011)]%
        {DBLP:conf/gecco/LukasiewyczGRT11}
\bibfield{author}{\bibinfo{person}{Martin Lukasiewycz},
  \bibinfo{person}{Michael Gla{\ss}}, \bibinfo{person}{Felix Reimann}, {and}
  \bibinfo{person}{J{\"{u}}rgen Teich}.} \bibinfo{year}{2011}\natexlab{}.
\newblock \showarticletitle{Opt4J: a modular framework for meta-heuristic
  optimization}. In \bibinfo{booktitle}{\emph{13th Annual Genetic and
  Evolutionary Computation Conference, {GECCO} 2011, Proceedings, Dublin,
  Ireland, July 12-16, 2011}}, \bibfield{editor}{\bibinfo{person}{Natalio
  Krasnogor} {and} \bibinfo{person}{Pier~Luca Lanzi}} (Eds.).
  \bibinfo{publisher}{{ACM}}, \bibinfo{pages}{1723--1730}.
\newblock
\urldef\tempurl%
\url{https://doi.org/10.1145/2001576.2001808}
\showDOI{\tempurl}


\bibitem[Mendes et~al\mbox{.}(2020)]%
        {DBLP:conf/mascots/MendesCRG20}
\bibfield{author}{\bibinfo{person}{Pedro Mendes}, \bibinfo{person}{Maria
  Casimiro}, \bibinfo{person}{Paolo Romano}, {and} \bibinfo{person}{David
  Garlan}.} \bibinfo{year}{2020}\natexlab{}.
\newblock \showarticletitle{TrimTuner: Efficient Optimization of Machine
  Learning Jobs in the Cloud via Sub-Sampling}. In
  \bibinfo{booktitle}{\emph{28th International Symposium on Modeling, Analysis,
  and Simulation of Computer and Telecommunication Systems, {MASCOTS} 2020,
  Nice, France, November 17-19, 2020}}. \bibinfo{publisher}{{IEEE}},
  \bibinfo{pages}{1--8}.
\newblock
\urldef\tempurl%
\url{https://doi.org/10.1109/MASCOTS50786.2020.9285971}
\showDOI{\tempurl}


\bibitem[Mkaouer et~al\mbox{.}(2014)]%
        {DBLP:conf/ssbse/MkaouerKBC14}
\bibfield{author}{\bibinfo{person}{Mohamed~Wiem Mkaouer},
  \bibinfo{person}{Marouane Kessentini}, \bibinfo{person}{Slim Bechikh}, {and}
  \bibinfo{person}{Mel~{\'{O}} Cinn{\'{e}}ide}.}
  \bibinfo{year}{2014}\natexlab{}.
\newblock \showarticletitle{A Robust Multi-objective Approach for Software
  Refactoring under Uncertainty}. In \bibinfo{booktitle}{\emph{Search-Based
  Software Engineering - 6th International Symposium, {SSBSE} 2014, Fortaleza,
  Brazil, August 26-29, 2014. Proceedings}} \emph{(\bibinfo{series}{Lecture
  Notes in Computer Science}, Vol.~\bibinfo{volume}{8636})},
  \bibfield{editor}{\bibinfo{person}{Claire~Le Goues} {and}
  \bibinfo{person}{Shin Yoo}} (Eds.). \bibinfo{publisher}{Springer},
  \bibinfo{pages}{168--183}.
\newblock
\urldef\tempurl%
\url{https://doi.org/10.1007/978-3-319-09940-8\_12}
\showDOI{\tempurl}


\bibitem[Nair et~al\mbox{.}(2020)]%
        {nair2018finding}
\bibfield{author}{\bibinfo{person}{Vivek Nair}, \bibinfo{person}{Zhe Yu},
  \bibinfo{person}{Tim Menzies}, \bibinfo{person}{Norbert Siegmund}, {and}
  \bibinfo{person}{Sven Apel}.} \bibinfo{year}{2020}\natexlab{}.
\newblock \showarticletitle{Finding faster configurations using FLASH}.
\newblock \bibinfo{journal}{\emph{IEEE Transactions on Software Engineering}}
  \bibinfo{volume}{46}, \bibinfo{number}{7} (\bibinfo{year}{2020}).
\newblock


\bibitem[Scott and Knott(1974)]%
        {scott1974cluster}
\bibfield{author}{\bibinfo{person}{Andrew~Jhon Scott} {and} \bibinfo{person}{M
  Knott}.} \bibinfo{year}{1974}\natexlab{}.
\newblock \showarticletitle{A cluster analysis method for grouping means in the
  analysis of variance}.
\newblock \bibinfo{journal}{\emph{Biometrics}} (\bibinfo{year}{1974}),
  \bibinfo{pages}{507--512}.
\newblock


\bibitem[Shahbazian et~al\mbox{.}(2020)]%
        {DBLP:conf/sigsoft/ShahbazianKBM20}
\bibfield{author}{\bibinfo{person}{Arman Shahbazian}, \bibinfo{person}{Suhrid
  Karthik}, \bibinfo{person}{Yuriy Brun}, {and} \bibinfo{person}{Nenad
  Medvidovic}.} \bibinfo{year}{2020}\natexlab{}.
\newblock \showarticletitle{eQual: informing early design decisions}. In
  \bibinfo{booktitle}{\emph{{ESEC/FSE} '20: 28th {ACM} Joint European Software
  Engineering Conference and Symposium on the Foundations of Software
  Engineering, Virtual Event, USA, November 8-13, 2020}},
  \bibfield{editor}{\bibinfo{person}{Prem Devanbu}, \bibinfo{person}{Myra~B.
  Cohen}, {and} \bibinfo{person}{Thomas Zimmermann}} (Eds.).
  \bibinfo{publisher}{{ACM}}, \bibinfo{pages}{1039--1051}.
\newblock
\urldef\tempurl%
\url{https://doi.org/10.1145/3368089.3409749}
\showDOI{\tempurl}


\bibitem[Siegmund et~al\mbox{.}(2012)]%
        {DBLP:conf/icse/SiegmundKKABRS12}
\bibfield{author}{\bibinfo{person}{Norbert Siegmund},
  \bibinfo{person}{Sergiy~S. Kolesnikov}, \bibinfo{person}{Christian
  K{\"{a}}stner}, \bibinfo{person}{Sven Apel}, \bibinfo{person}{Don~S. Batory},
  \bibinfo{person}{Marko Rosenm{\"{u}}ller}, {and} \bibinfo{person}{Gunter
  Saake}.} \bibinfo{year}{2012}\natexlab{}.
\newblock \showarticletitle{Predicting performance via automated
  feature-interaction detection}. In \bibinfo{booktitle}{\emph{34th
  International Conference on Software Engineering, {ICSE} 2012, June 2-9,
  2012, Zurich, Switzerland}}, \bibfield{editor}{\bibinfo{person}{Martin
  Glinz}, \bibinfo{person}{Gail~C. Murphy}, {and} \bibinfo{person}{Mauro
  Pezz{\`{e}}}} (Eds.). \bibinfo{publisher}{{IEEE} Computer Society},
  \bibinfo{pages}{167--177}.
\newblock
\urldef\tempurl%
\url{https://doi.org/10.1109/ICSE.2012.6227196}
\showDOI{\tempurl}


\bibitem[Silva{-}Mu{\~{n}}oz et~al\mbox{.}(2021)]%
        {DBLP:journals/peerj-cs/Silva-MunozFB21}
\bibfield{author}{\bibinfo{person}{Mois{\'{e}}s Silva{-}Mu{\~{n}}oz},
  \bibinfo{person}{Alberto Franzin}, {and} \bibinfo{person}{Hugues Bersini}.}
  \bibinfo{year}{2021}\natexlab{}.
\newblock \showarticletitle{Automatic configuration of the Cassandra database
  using irace}.
\newblock \bibinfo{journal}{\emph{PeerJ Comput. Sci.}}  \bibinfo{volume}{7}
  (\bibinfo{year}{2021}), \bibinfo{pages}{e634}.
\newblock
\urldef\tempurl%
\url{https://doi.org/10.7717/peerj-cs.634}
\showDOI{\tempurl}


\bibitem[Soltani et~al\mbox{.}(2018)]%
        {DBLP:conf/ssbse/SoltaniDPDZD18}
\bibfield{author}{\bibinfo{person}{Mozhan Soltani}, \bibinfo{person}{Pouria
  Derakhshanfar}, \bibinfo{person}{Annibale Panichella},
  \bibinfo{person}{Xavier Devroey}, \bibinfo{person}{Andy Zaidman}, {and}
  \bibinfo{person}{Arie van Deursen}.} \bibinfo{year}{2018}\natexlab{}.
\newblock \showarticletitle{Single-objective Versus Multi-objectivized
  Optimization for Evolutionary Crash Reproduction}. In
  \bibinfo{booktitle}{\emph{Search-Based Software Engineering - 10th
  International Symposium, {SSBSE} 2018, Montpellier, France, September 8-9,
  2018, Proceedings}} \emph{(\bibinfo{series}{Lecture Notes in Computer
  Science}, Vol.~\bibinfo{volume}{11036})},
  \bibfield{editor}{\bibinfo{person}{Thelma~Elita Colanzi} {and}
  \bibinfo{person}{Phil McMinn}} (Eds.). \bibinfo{publisher}{Springer},
  \bibinfo{pages}{325--340}.
\newblock
\urldef\tempurl%
\url{https://doi.org/10.1007/978-3-319-99241-9\_18}
\showDOI{\tempurl}


\bibitem[Valov et~al\mbox{.}(2017)]%
        {DBLP:conf/wosp/ValovPGFC17}
\bibfield{author}{\bibinfo{person}{Pavel Valov},
  \bibinfo{person}{Jean{-}Christophe Petkovich}, \bibinfo{person}{Jianmei Guo},
  \bibinfo{person}{Sebastian Fischmeister}, {and} \bibinfo{person}{Krzysztof
  Czarnecki}.} \bibinfo{year}{2017}\natexlab{}.
\newblock \showarticletitle{Transferring Performance Prediction Models Across
  Different Hardware Platforms}. In \bibinfo{booktitle}{\emph{Proceedings of
  the 8th {ACM/SPEC} on International Conference on Performance Engineering,
  {ICPE} 2017, L'Aquila, Italy, April 22-26, 2017}},
  \bibfield{editor}{\bibinfo{person}{Walter Binder}, \bibinfo{person}{Vittorio
  Cortellessa}, \bibinfo{person}{Anne Koziolek}, \bibinfo{person}{Evgenia
  Smirni}, {and} \bibinfo{person}{Meikel Poess}} (Eds.).
  \bibinfo{publisher}{{ACM}}, \bibinfo{pages}{39--50}.
\newblock
\urldef\tempurl%
\url{https://doi.org/10.1145/3030207.3030216}
\showDOI{\tempurl}


\bibitem[Vargha and Delaney(2000)]%
        {Vargha2000ACA}
\bibfield{author}{\bibinfo{person}{Andr{\'a}s Vargha} {and}
  \bibinfo{person}{Harold~D. Delaney}.} \bibinfo{year}{2000}\natexlab{}.
\newblock \showarticletitle{A Critique and Improvement of the CL Common
  Language Effect Size Statistics of McGraw and Wong}.
\newblock


\bibitem[Wilcoxon(1945)]%
        {Wilcoxon1945IndividualCB}
\bibfield{author}{\bibinfo{person}{Frank Wilcoxon}.}
  \bibinfo{year}{1945}\natexlab{}.
\newblock \showarticletitle{Individual Comparisons by Ranking Methods}.
\newblock


\bibitem[Yang et~al\mbox{.}(2013)]%
        {DBLP:journals/tec/YangLLZ13}
\bibfield{author}{\bibinfo{person}{Shengxiang Yang}, \bibinfo{person}{Miqing
  Li}, \bibinfo{person}{Xiaohui Liu}, {and} \bibinfo{person}{Jinhua Zheng}.}
  \bibinfo{year}{2013}\natexlab{}.
\newblock \showarticletitle{A Grid-Based Evolutionary Algorithm for
  Many-Objective Optimization}.
\newblock \bibinfo{journal}{\emph{{IEEE} Trans. Evol. Comput.}}
  \bibinfo{volume}{17}, \bibinfo{number}{5} (\bibinfo{year}{2013}),
  \bibinfo{pages}{721--736}.
\newblock
\urldef\tempurl%
\url{https://doi.org/10.1109/TEVC.2012.2227145}
\showDOI{\tempurl}


\bibitem[Yuan and Banzhaf(2020)]%
        {DBLP:journals/tse/YuanB20}
\bibfield{author}{\bibinfo{person}{Yuan Yuan} {and} \bibinfo{person}{Wolfgang
  Banzhaf}.} \bibinfo{year}{2020}\natexlab{}.
\newblock \showarticletitle{{ARJA:} Automated Repair of Java Programs via
  Multi-Objective Genetic Programming}.
\newblock \bibinfo{journal}{\emph{{IEEE} Trans. Software Eng.}}
  \bibinfo{volume}{46}, \bibinfo{number}{10} (\bibinfo{year}{2020}),
  \bibinfo{pages}{1040--1067}.
\newblock
\urldef\tempurl%
\url{https://doi.org/10.1109/TSE.2018.2874648}
\showDOI{\tempurl}


\bibitem[Zhu et~al\mbox{.}(2017)]%
        {DBLP:conf/cloud/ZhuLGBMLSY17}
\bibfield{author}{\bibinfo{person}{Yuqing Zhu}, \bibinfo{person}{Jianxun Liu},
  \bibinfo{person}{Mengying Guo}, \bibinfo{person}{Yungang Bao},
  \bibinfo{person}{Wenlong Ma}, \bibinfo{person}{Zhuoyue Liu},
  \bibinfo{person}{Kunpeng Song}, {and} \bibinfo{person}{Yingchun Yang}.}
  \bibinfo{year}{2017}\natexlab{}.
\newblock \showarticletitle{BestConfig: tapping the performance potential of
  systems via automatic configuration tuning}. In
  \bibinfo{booktitle}{\emph{Proceedings of the 2017 Symposium on Cloud
  Computing, SoCC 2017, Santa Clara, CA, USA, September 24-27, 2017}}.
  \bibinfo{publisher}{{ACM}}, \bibinfo{pages}{338--350}.
\newblock
\urldef\tempurl%
\url{https://doi.org/10.1145/3127479.3128605}
\showDOI{\tempurl}


\bibitem[Zitzler and K{\"{u}}nzli(2004)]%
        {DBLP:conf/ppsn/ZitzlerK04}
\bibfield{author}{\bibinfo{person}{Eckart Zitzler} {and} \bibinfo{person}{Simon
  K{\"{u}}nzli}.} \bibinfo{year}{2004}\natexlab{}.
\newblock \showarticletitle{Indicator-Based Selection in Multiobjective
  Search}. In \bibinfo{booktitle}{\emph{Parallel Problem Solving from Nature -
  {PPSN} VIII, 8th International Conference, Birmingham, UK, September 18-22,
  2004, Proceedings}} \emph{(\bibinfo{series}{Lecture Notes in Computer
  Science}, Vol.~\bibinfo{volume}{3242})},
  \bibfield{editor}{\bibinfo{person}{Xin Yao}, \bibinfo{person}{Edmund~K.
  Burke}, \bibinfo{person}{Jos{\'{e}}~Antonio Lozano}, \bibinfo{person}{Jim
  Smith}, \bibinfo{person}{Juan Juli{\'{a}}n~Merelo Guerv{\'{o}}s},
  \bibinfo{person}{John~A. Bullinaria}, \bibinfo{person}{Jonathan~E. Rowe},
  \bibinfo{person}{Peter Ti{\~{n}}o}, \bibinfo{person}{Ata Kab{\'{a}}n}, {and}
  \bibinfo{person}{Hans{-}Paul Schwefel}} (Eds.).
  \bibinfo{publisher}{Springer}, \bibinfo{pages}{832--842}.
\newblock
\urldef\tempurl%
\url{https://doi.org/10.1007/978-3-540-30217-9\_84}
\showDOI{\tempurl}


\bibitem[Zuluaga et~al\mbox{.}(2016)]%
        {DBLP:journals/jmlr/ZuluagaKP16}
\bibfield{author}{\bibinfo{person}{Marcela Zuluaga}, \bibinfo{person}{Andreas
  Krause}, {and} \bibinfo{person}{Markus P{\"{u}}schel}.}
  \bibinfo{year}{2016}\natexlab{}.
\newblock \showarticletitle{e-PAL: An Active Learning Approach to the
  Multi-Objective Optimization Problem}.
\newblock \bibinfo{journal}{\emph{J. Mach. Learn. Res.}}  \bibinfo{volume}{17}
  (\bibinfo{year}{2016}), \bibinfo{pages}{104:1--104:32}.
\newblock
\urldef\tempurl%
\url{http://jmlr.org/papers/v17/15-047.html}
\showURL{%
\tempurl}


\end{thebibliography}

%\printbibliography

\end{document}